\documentclass[a4paper, 11pt,twoside,openright]{book}

\usepackage{graphicx}

\usepackage{amsmath, amsthm} 
\usepackage{amsfonts}
\usepackage{amssymb,graphics,psfrag}
\usepackage[bottom]{footmisc}
\usepackage{hyperref}
\usepackage{mathrsfs}
\usepackage{cite}
\usepackage{setspace}
\usepackage{color}

\usepackage{tikz}

\usepackage{fancyhdr}
\renewcommand{\chaptermark}[1]{%
\markboth{\chaptername
\ \thechapter.\ #1}{}}

\makeatletter
\def\cleardoublepage{\clearpage\if@twoside \ifodd\c@page\else
\hbox{}
\vspace*{\fill}
\begin{center}
\end{center}
\vspace{\fill}
\thispagestyle{empty}
\newpage
\if@twocolumn\hbox{}\newpage\fi\fi\fi}
\makeatother

\fancypagestyle{plain}{%
\fancyhf{} 
\fancyfoot[C]{\bfseries \thepage} 

}

\DeclareFontFamily{OT1}{rsfs}{} \DeclareFontShape{OT1}{rsfs}{m}{n}{
<-7> rsfs5 <7-10> rsfs7 <10-> rsfs10}{}
\DeclareMathAlphabet{\mycal}{OT1}{rsfs}{m}{n}
\def\scri{{\mycal I}}

\newcommand{\half}{\frac{1}{2}}
\newcommand{\dd}{\partial}
\renewcommand{\d}{\partial}

\newcommand{\bdelta}{\slash\hspace{-0.185cm}{\delta}}

\def\cL{\mathcal{L}}
\def\cX{\mathcal{X}}
\def\cY{\mathcal{Y}}

\newcommand{\CSeneque}{\LARGE{C}\normalsize}		
\newcommand{\NSeneque}{\LARGE{N}\normalsize}	

\newcommand{\ffrac}[2]{\raisebox{.5pt}%
  {\footnotesize$\displaystyle\frac{#1}{#2}$}\kern1pt}
\newcommand{\dover}[2]{\ffrac{\dd #1}{\dd #2}}

\newtheorem{prop}{Proposition}[chapter]

\newtheorem{theorem}[prop]{Theorem}

\newcommand*\xbar[1]{%
  \hbox{%
    \vbox{%
      \hrule height 0.5pt 
      \kern0.3ex
      \hbox{%
        \kern-0.0em
        \ensuremath{#1}%
        \kern-0.0em
      }%
    }%
  }%
}

\makeatletter
  \renewcommand\chapter{\if@openright\cleardoublepage\else\clearpage\fi      
                      \thispagestyle{plain}
			   \global\@topnum\z@
                      \@afterindentfalse
                      \secdef\@chapter\@schapter}
  \def\@chapter[#1]#2{\ifnum \c@secnumdepth >\m@ne
                           \refstepcounter{chapter}%
                           \typeout{\@chapapp\space\thechapter.}%
                           \addcontentsline{toc}{chapter}%
                                     {\protect\numberline{\thechapter}#1}%
                      \else
                        \addcontentsline{toc}{chapter}{#1}%
                      \fi
                      \chaptermark{#1}%
                      \addtocontents{lof}{\protect\addvspace{10\p@}}%
                      \addtocontents{lot}{\protect\addvspace{10\p@}}%
                      \if@twocolumn
                        \@topnewpage[\@makechapterhead{#2}]%
                      \else
                        \@makechapterhead{#2}%
                        \@afterheading
                      \fi}
  \def\@makechapterhead#1{%
    \vspace*{50\p@}%
    {\parindent \z@ \raggedright \normalfont
      \ifnum \c@secnumdepth >\m@ne
          \flushright\LARGE\scshape \@chapapp\space \thechapter              
          \par\nobreak
          \vskip 8\p@                                                        
          \hrule\vskip 6.9\p@                                                
      \fi
      \interlinepenalty\@M
      \Huge \normalfont \bfseries #1\par\nobreak                             
      \vskip 40\p@}}
  
  \def\@schapter#1{\if@twocolumn
                     \@topnewpage[\@makeschapterhead{#1}]%
                   \else
                     \@makeschapterhead{#1}%
                     \@afterheading
                   \fi}
  \def\@makeschapterhead#1{%
    \vspace*{50\p@}%
    {\parindent \z@ \raggedright
      \normalfont
      \interlinepenalty\@M
      \Huge \bfseries  #1\par\nobreak
      \vskip 40\p@
    }}
\makeatother

\makeatletter
\renewcommand\appendix{\par
  \setcounter{chapter}{0}%
  \setcounter{section}{0}%
  \gdef\@chapapp{\appendixname}%
  \gdef\thechapter{\@Alph\c@chapter}
  \renewcommand\chaptername{Annexe}
}
\makeatother

\makeatletter
    \renewcommand\part{%
        \if@openright
            \cleardoublepage
        \else
            \clearpage
        \fi
        \thispagestyle{empty}%
        \if@twocolumn
            \onecolumn
            \@tempswatrue
        \else
            \@tempswafalse
        \fi
        \null\vfil
        \secdef\@part\@spart}

        \def\@part[#1]#2{%
            \ifnum \c@secnumdepth >-2\relax
                \refstepcounter{part}%
                \addcontentsline{toc}{part}{\thepart\hspace{1em}#1}%
            \else
                \addcontentsline{toc}{part}{#1}%
            \fi
            \markboth{}{}%
            {\centering
                \interlinepenalty \@M
            \normalfont
            \ifnum \c@secnumdepth >-2\relax
                \huge\bfseries \partname\nobreakspace\thepart
                \par
                \vskip 10\p@
            \fi
            \hrule\vskip 10\p@
            \Huge \bfseries #2\par}%
        \@endpart}
        
        \def\@spart#1{%
            {\centering
                \interlinepenalty \@M
                \normalfont
                \Huge \bfseries #1\par}%
        \@endpart}
        \def\@endpart{\vfil\newpage
            \if@twoside
                \if@openright
                    \null
                    \thispagestyle{empty}%
                    \newpage
                \fi
            \fi
            \if@tempswa
                \twocolumn
            \fi}
\makeatother

\makeatletter
    \let\appendixOriginal\appendix
    \renewcommand{\appendix}{%
        \clearpage{
            \pagestyle{empty}
            \cleardoublepage
        }
        \renewcommand\chaptername{\appendName}
        \appendixOriginal
    }

\makeatletter
    \let\backmatterOriginal\backmatter
    \renewcommand{\backmatter}{%
        \clearpage{
            \pagestyle{empty}
            \cleardoublepage
        }
        \backmatterOriginal
    }

    {\cleardoublepage\thispagestyle{empty}\null\vfill\begin{center}%
    \bfseries\abstractname\end{center}}%
    {\vfill\null}

\begin{document}
\thispagestyle{empty}

\begin{titlepage}
	\begin{center}
\vspace{-1cm}
    UNIVERSITE LIBRE DE BRUXELLES \\
    Facult\'e des Sciences \\
    Physique Math\'ematique des Interactions Fondamentales\\
    \vfill \vfill \vfill\vfill \vfill\vfill
    {\huge Conformal symmetries of gravity\\ from asymptotic methods:\\ further developments\\[2em]}
\vfill
\textsc{Pierre-Henry Lambert}
	    \emph{}  
    \vfill \vfill \vfill \vfill \vfill\vfill
    Th\`ese de doctorat pr\'esent\'ee en vue de l'obtention\\ du grade acad\'emique de Docteur en Sciences\\ [2em]
	Directeur: Prof. Glenn Barnich \\
     Ann\'ee acad\'emique 2013-2014\\
     \vfill \vfill
    \phantom{PH}
  \end{center}

\end{titlepage}

\frontmatter

\thispagestyle{empty}
\newpage
\thispagestyle{empty}
\null \vspace {\stretch{1}}
\begin{flushright}
\textit{ \NSeneque on quia difficilia sunt non audemus,\\
 sed quia non audemus difficilia sunt}.
\vspace{1cm}\\
\textit{ \CSeneque e n'est pas parce que les choses\\sont difficiles que nous n'osons pas,\\
c'est parce que nous n'osons pas\\qu'elles sont difficiles.}
\vspace{1cm}\\
\textsc{S\'en\`eque}, Lettres \`a Lucilius.
\end{flushright}
\vspace{1cm}
\null \vspace {\stretch {1}}

\newpage
\chapter*{Acknowledgements}

The results presented in this thesis have been realized in the group of theoretical physics of fundamental interactions at Universit\'e Libre de Bruxelles.
This intellectual marathon of four years would have never been completed without interacting with a lot a people. In particular, I want to thank:\\

\noindent Glenn Barnich, the director of this doctoral thesis, for having accepted to be my guide during these four years, for his constant disponibility and for his patience. His experience, insight and knownledge, and his constant attention are the seeds of this thesis. I also want to thank him warmly for the contagious motivation that he is equipped with, and also for his (always) judicious advice.\\

\noindent Laura Donnay,  for the infinitely many discussions about physics (in various dimensions) and life in general, for her enthusiasm and good mood, and for her friendship. Moreover, her remarks and comments, and her proofreading of this thesis were greatly appreciated. Thanks a lot for all of the laughs and good times, and good luck for the end of the PhD!\\

\noindent The other PhD students of the same year: Gustavo Lucena G\'omez (my officemate in the noisy box), Diego Redigolo, and Micha Moskovic, for the great time spent during these four years.\\

\noindent Marco Fazzi, Eduardo Conde Pena, Andrea Mezzalira, Andrea Marzolla and Ignacio Cortese Mombelli, for all the sport activities performed and enjoyed: from horizontal ones (running races, football matches, roller parade) to vertical ones (stairs race, climbing); and also Rakibur Rhaman for the nice chess games played together.\\

\noindent Geoffrey Comp\`ere, Pujian Mao, Waldemar Schulgin for discussions about physics, and for projects in progress. I also want to thank all the other members of the group of theoretical physics of fundamental interactions, for the nice atmosphere they contributed to create.\\

\noindent The members of the jury of my thesis (alphabetically listed): Glenn Barnich, Xavier Bekaert, Geoffrey Comp\`ere, Thomas Hambye and Philippe Spindel, for having accepted to be part of the jury, and also for the attention to this thesis.\\

\thispagestyle{plain}

\noindent St\'ephane-Olivier Guillaume, for his support and also for the shared feeling of doing a PhD in physics.\\

\noindent My parents and my three brothers, for their unshaleable and constant support during all my studies and in particular during my thesis.\\

\thispagestyle{plain}


\newpage
\tableofcontents

\mainmatter

\newpage
\chapter*{Prologue}
\addcontentsline{toc}{chapter}{Prologue}
\thispagestyle{plain}
The aim of theoretical physics is to understand and provide a mathematical description of the fundamental laws of Nature. At present, the understanding of physical phenomena encountered in Nature is based on two rather different pillars. On the one hand, there is Einstein's theory of general relativity, providing a description of the gravitational interaction using the classical theory of fields. It is founded on elegant geometric concepts drawn from thinking of spacetime as a curved manifold. On the other hand, there is the standard model of particle physics, describing the quantum interactions of elementary particles of spins zero, one half and one.\\

There are several reasons for which the theory of general relativity cannot be the final theory in
order to describe the gravitational interaction. On the one side, the theory predicts the existence of
singularities but does not resolve them. This is clearly an internal evidence that general relativity
theory is incomplete. On the other side, general relativity is a classical theory but Nature is fundamentally
quantum, and at present a quantum theory of gravity is still missing.
This is an external piece of evidence that general relativity cannot be the final theory concerning gravity.

\subsection*{The search for quantum gravity}

The theory of general relativity predicts the existence of particular solutions, known as black holes. Black holes are surrounded by a surface,  the event horizon, such that in classical general relativity nothing inside the event horizon can escape from it. In the context of quantum fields theory in curved spacetimes however, it has been shown by Hawking \cite{Hawking:1974sw,Hawking:1976ra} in the early seventies that black holes behave as thermodynamic systems, emit radiation at a certain temperature and finally evaporate. Black holes solutions are completely classified by their mass, angular momentum and charge, and furthermore their dynamics is governed by a few laws \cite{Bardeen:1973gs}, analogous to those of thermodynamics.\\
\newpage

Black holes have an entropy $S$, known as the Bekenstein-Hawking entropy, given by the simple expression (in Planck units)
\begin{eqnarray}\notag
S=\dfrac{A}{4},
\end{eqnarray}
where $A$ is the area of the horizon. This formula is macroscopic and universal. Macroscopic, because it only depends on the geometric quantity that is the area of the horizon of the black hole; universal, because it is valid for any black hole independently of its shape, and is also valid for cosmological horizons such as those found for de Sitter or Rindler spacetimes.
The appearance of an area in the entropy is somehow counterintuitive, since the entropy of common physical systems is usually proportional to the volume of the region where the physical system evolves.\\

A natural question one can wonder about is whether this macroscopic relation for the black hole entropy can be understood and described from a microscopic (i.e. statistical) point of view, as being the logarithm of the number of quantum states associated with the black hole.
A positive answer to this question
would
allow  to understand and to identify the microscopic degrees of freedom of the gravitational interaction and, as a consequence, it would represent a promising step in the search for a quantum theory of gravity.\\

\thispagestyle{plain}

In 1995, progress was made by Strominger and Vafa\cite{Strominger:1996sh} towards understanding the microscopic origin of the Bekenstein-Hawking entropy formula. Their microscopic derivation heavily depends on string theory and supersymmetry and only applies to a restricted class of black holes (extremal black holes) that does not include the most basic black hole solution (the Schwarzschild one) and furthermore does not explain the case of cosmological horizons. 
This microscopic derivation relies on the holographic duality; more precisely, the microstates are counted by the conformal field theory that describes the black hole in string theory in the weak-coupling regime.\\

This intertwining between gravity and conformal field theory can be seen as an illustration of the holographic principle \cite{Susskind,'thooft}, according to which information contained inside a spacetime region can be described by data located at the boundary of that region.
The holographic principle is an extremely fruitful idea that pervades modern research in high energy physics. At present, the most concrete realization of it is the celebrated AdS-CFT correspondence, proposed by Maldacena \cite{Maldacena} in 1998. It is the best understood holographic duality, between
superconformal field theory in the limit of a large number of colors from one side and supergravity theory embedded in string theory on the other side.
Beyond its AdS-CFT realization, holography is argued to be a key ingredient for gravity in general \cite{'thooft}.\\

At first sight, it thus seems that precise detail of the string theory are needed in order to provide a microscopic derivation of the entropy of black holes.
 Actually the derivation can be performed without the help of string theory, at least in some cases.
In fact, the aforementioned relation between gravity and conformal field theory had already been anticipated more than ten years before by Brown and Henneaux \cite{Brown:1986nw}, coming from asymptotic computations in gravity.
More precisely,  it was shown in 1986 that the asymptotic symmetry algebra of gravity  in three dimensions with a negative cosmological constant consists of the conformal algebra in two dimensions (two copies of the Witt algebra). Moreover, it was also shown that the charges associated with the asymptotic symmetries form a representation of the symmetry algebra, with a classical central extension (the Viraroso algebra). In other words, this result shows that in three dimensions, a consistent theory of quantum gravity defined around anti-de Sitter space should be holographically dual to a two dimensional conformal field theory.\\

This purely classical observation was used by Strominger in 1998 \cite{Strominger:1998eq} to give a completely non-stringy microscopic derivation of the macroscopic entropy law for black holes whose near horizon geometry is locally anti-de Sitter, by combining the value of the classical Virasoro central extension of Brown and Henneaux, with a statistical formula for the entropy of bi-dimensional conformal field theories, due to Cardy \cite{Cardy:1986ie}. The important point in Strominger's derivation is that neither string theory nor supersymmetry is required.\\

Besides the fact that  the appearance of a classical central extension in the algebra of asymptotic charges is a sign that the entropy of black holes can be understood microscopically, the result of Brown and Henneaux illustrates another physical phenomenon, known as symmetry enhancement.

\thispagestyle{plain}

\subsection*{Symmetry enhancement phenomenon}

The phenomenon of symmetry enhancement occurs in gauge theories containing gravity, when the symmetry algebra at the boundary of the spacetime differs from the rigid symmetry algebra of the bulk theory.
In this case, the asymptotic symmetry algebra may become infinite-dimensional. Furthermore, the asymptotic symmetry algebra becomes the global symmetry algebra of the dual theory.\\

The role of symmetries cannot be overestimated in theoretical physics. Symmetries allow, for instance, to make complicated problems tractable, to classify solutions and to describe fundamental laws of Nature. In non-linear fields theories such as general relativity, the benefits of symmetries are particularly important in the search of new solutions and in the building of conserved quantities.

\subsubsection*{bms$_4^{global}$: appearance of supertranslations}
In gravitational theories, the symmetry enhancement phenomenon was first observed in the sixties in the case of asymptotically flat four dimensional spacetimes at null infinity with a vanishing cosmological constant. In this setup, it was shown by Bondi, Metzner and Sachs \cite{Bondi:1962px,Sachs:1962wk,Sachs2} that  the asymptotic symmetry algebra was not the usual Poincar\'e algebra (the symmetry algebra of flat Minkowski space) but instead an infinite dimensional extension of it, which is called the bms$_4^{global}$ algebra below. The structure of the bms$_4^{global}$ algebra and associated group with respect to the Poincar\'e group is the following,
\begin{align*}
\text{Poincar\'e}:&\hspace{1cm}\text{translations} \ltimes \text{Lorentz transformations},\\
\text{bms$_4^{global}$}:&\hspace{1cm}\text{supertranslations} \ltimes \text{Lorentz transformations}.
\end{align*}
When the elements of the symmetry algebra are expanded in globally well-defined modes (as explicitly assumed in the original papers \cite{Bondi:1962px,Sachs:1962wk,Sachs2}), the algebra then consists of the semi-direct sum between the usual infinitesimal Lorentz transformations and an infinite dimensional enhancement of the infinitesimal Poincar\'e translations to supertranslations\footnote{The prefixe ``super" in this context has nothing to do with supersymmetry.}, that are characterized by arbitrary smooth functions over the 2-sphere.\\

\thispagestyle{plain}

\subsubsection*{Asymptotically adS$_3$: appearance of Virasoro algebra}
The enhancement phenomenon is of particular interest when the asymptotic symmetry algebra is realized by the Virasoro algebra, the infinite-dimensional algebra of conformal field theory in two dimensions,
\begin{eqnarray*}
\{\mathcal{Q}_m,\mathcal{Q}_n\}=(m-n)\mathcal{Q}_{n+m}+\frac{c}{12}m(m^2-1)\delta_{m+n,0}.
\end{eqnarray*}
This case is interesting because it gives the perspective that gravity can be analyzed by the powerful  techniques \cite{BPZ} available in these theories.
This symmetry enhancement was first observed in 1986 by Brown and Henneaux \cite{Brown:1986nw} in the case of gravity with a negative cosmological constant, when expanded around anti-de Sitter space in three dimensions with suitable boundary conditions.
As already mentioned in the first part of this prologue, this symmetry enhancement to a Virasoro algebra relates gravity and conformal field theory and, in particular, allows for a kind of symmetry based counting \cite{Strominger:1998eq} of the microscopic degrees of freedom of the bulk three dimensional gravity, thereby exhibiting its quantum properties.

\subsubsection*{bms$_4^{local}$: appearance of superrotations}

A generalization of the work of Brown and Henneaux to four spacetime dimensions was performed in the case of 
asymptotically flat spacetimes (therefore going beyond the standard AdS-CFT correspondence) by Barnich and Troessaert \cite{Barnich:2009se,Barnich:2010eb} at null infinity in 2010.
The novel feature of this analysis is that local singularities are allowed in the symmetry generators (more precisely: meromorphic functions). This has the effect of enhancing the Lorentz factor in the bms$_4^{global}$ algebra to the non-centrally extended Virasoro algebra. The generators of this Virasoro algebra are called superrotations \cite{Barnich:2011ct}, and both factors  in the bms$_4^{local}$ algebra (i.e. superrotations and supertranslations) are infinite dimensional,\\
\begin{align*}
\text{bms$_4^{global}$}:&\hspace{1cm}\text{supertranslations} \ltimes \text{Lorentz transformations},\\
\text{bms$_4^{local}$}:&\hspace{1cm}\text{supertranslations} \ltimes \text{superrotations}.\\
\end{align*}

\section*{New results of this thesis}

\thispagestyle{plain}

In this thesis, the symmetry structure of gravity at null infinity is studied further, in the case of pure gravity in four dimensions, and also in the case of Einstein-Yang-Mills theory in $d$ dimensions with and without a cosmological constant.\\

Firstly, it is shown that the enhancement from Lorentz to Virasoro algebra also occurs for asymptotically flat spacetimes  defined in the sense of Newman-Unti \cite{newman:891}. Their definition corresponds to a gauge that differs from the one by BMS through a different choice of radial coordinate.
As a first application, the transformation laws of the Newman-Penrose coefficients characterizing the solution space of the Newman-Unti approach are worked out, focusing on the inhomogeneous terms that contain the information about central extensions of the theory. These transformations laws make the conformal structure particularly transparent, and constitute the main original result of the thesis.\\

Secondly, asymptotic symmetries of the Einstein-Yang-Mills system with or
without cosmological constant are explicitly worked out in a unified manner in $d$ dimensions. In
agreement with a recent conjecture \cite{Strominger:2013lka}, one finds a Virasoro-Kac-Moody type algebra
not only in three dimensions but also in the four dimensional asymptotically
flat case.\\


\newpage
\thispagestyle{plain}

During the realization of this thesis, the following research papers and proceedings have been published:\\
\begin{enumerate}
\item 
G.~Barnich and P.-H. Lambert, ``{A Note on the Newman-Unti Group and the BMS
  Charge Algebra in Terms of Newman-Penrose Coefficients},'' {\em Adv. Math.
  Phys.} {\bfseries 16} (2012) 197385,
  \href{http://arxiv.org/abs/1102.0589}{{\ttfamily arXiv:1102.0589 [gr-qc]}}.
\url{http://dx.doi.org/10.1155/2012/197385}, see also: J.\ Phys.\ Conf.\ Ser.\  {\bf 410}, 012142 (2013) \url{http://iopscience.iop.org/1742-6596/410/1/012142/}.\\
\item
G.~Barnich and P.-H. Lambert, ``{Asymptotic symmetries at null infinity
and local conformal properties of spin coefficients}'', Tomsk State Pedagogical University Bulletin, 2012, 13 (128),
  \href{http://arxiv.org/abs/1301.5754}{{\ttfamily  	arXiv:1301.5754 [gr-qc]}}.\\
\item
G.~Barnich and P.~-H.~Lambert,
  ``Einstein-Yang-Mills theory: Asymptotic symmetries,''
  Phys.\ Rev.\ D {\bf 88}, 103006 (2013)
 \href{http://arxiv.org/abs/arXiv:1310.2698}{{\ttfamily arXiv:1310.2698 [hep-th]}}.
\url{http://dx.doi.org/10.1103/PhysRevD.88.103006}\\
\item
 P.~-H.~Lambert,
  ``Introduction to Black Hole Evaporation,''
  PoS Modave {\bf 2013}, 001 (2013)
 \href{http://arxiv.org/abs/arXiv:1310.8312}{{\ttfamily arXiv:1310.8312 [gr-qc]}}.
\url{http://pos.sissa.it/archive/conferences/201/001/Modave\%202013_001.pdf}.\\
\end{enumerate}

\section*{Outline of the thesis}
Besides the present prologue, this thesis contains two main parts and a few appendices.\\

The first part of this thesis is devoted to the presentation of asymptotic methods (symmetries, solution space and surface charges) applied to gravity in the case of the BMS gauge in three and four spacetime dimensions. Most results presented in this first part are not new, but the emphasis is put on the fact that BMS gauge can describe in a unified way the case of asymptotically flat, anti-de Sitter and de Sitter spacetimes.
More precisely, the first chapter consists of an introduction to the BMS gauge, from the very beginning in the sixties up to the recent literature on the subject. The second and third chapters present asymptotic symmetries, solution space and surface charges of gravity in the BMS gauge with and without a cosmological constant in four and three dimensions, respectively.\\

The second part of this thesis contains the original contributions. Chapter four is devoted to the asymptotic symmetry analysis in the case of asymptotically flat spacetimes in the Newman-Unti gauge, while chapter five presents the asymptotic symmetries analysis in the case of Einstein-Yang-Mills system in $d$ dimensions with and without cosmological constant. Chapter six contains a presentation of the equations of motion for the Einstein-Maxwell set-up.\\

These two parts of the thesis are supplemented by appendices. In appendix A, some properties of conformal Killing vectors in $d$ dimensions are proven. These properties are used in the first part of the thesis.
Appendix B is devoted to attempting deriving the charged rotating BTZ black hole through the Newman-Janis' trick applied to the static charged BTZ black hole.
In appendix C, the relation between the bms boundary conditions and those obtained by acting with the Minkowski Killing vectors on the Kerr metric is investigated.

\thispagestyle{plain}

\newpage 
\thispagestyle{empty}
\part{BMS gauge - a guided tour}
\thispagestyle{empty}

\newpage

\chapter{Introduction to the BMS gauge}

One of the main research areas in general relativity in the sixties was the problem of gravitational radiation.
For instance, the following questions concerning the nature of gravitational waves were investigated in the case of asymptotically flat spacetime in four dimensions \cite{Sachs:1962wk}: What is the relation between gravitational radiation and the mass loss of the radiative source? What is the precise mechanism of the mass loss phenomenon? The asymptotic symmetry group leaves invariant the form of the boundary conditions for asymptotically flat spacetimes: what is the structure of this group?
In order to investigate these physically interesting questions, a metric ansatz representing asymptotically flat spacetimes in four dimensions was constructed\footnote{Bondi, Metzner and van der Burgh \cite{Bondi:1962px} in the axially symmetric case, and Sachs \cite{Sachs:1962wk} when the latter condition is removed.}, based on geometric considerations \cite{d'Inverno:1992rk}. This general ansatz for the metric field is called the BMS metric\footnote{From Bondi, Metzner and Sachs. Throughout this thesis, capital ``BMS" refers to the metric ansatz, while small ``bms" refers to the associated asymptotic symmetry algebra.}.\\

Asymptotically flat spacetimes defined with the use of the BMS metric is, by definition, coordinate dependent.  A natural question one can wonder is whether asymptotically flat spacetimes can be studied without referring to any particular coordinate system. The answer to this question is positive and was given by Penrose \cite{PhysRevLett.10.66,penrose:1964} who introduced conformal compactification in order to describe asymptotic regions of spacetimes in a geometric way. Moreover,
 infinity is brought to a finite distance through a conformal factor. However, both approaches to asymptotically flat spacetimes can be shown to be equivalent (see \cite{Wald}, section (11.1) and also \cite{Tamburino:1966}).\\

At the level of the symmetries that preserve asymptotically flat spacetimes at null infinity, the interesting result obtained in the sixties by Bondi, Metzner and Sachs is that the symmetry algebra is not the Poincar\'e algebra, but an infinite extension of it, called the bms algebra. The main difference with respect to the Poincar\'e algebra is that the translations are replaced by arbitrary angle-dependent translations, called supertranslations.\\

Recently, it has been pointed out by Barnich and Troessaert \cite{Barnich:2009se} that besides the usual difference between BMS and Poincar\'e translations, another difference between the two algebras can be considered. Indeed,  if one admits local singularities (meromorphic functions) in the asymptotic Killing vector, then the Lorentz part of the bms algebra becomes enhanced to two copies of the Virasoro algebra. In \cite{Barnich:2011ct}, the original bms algebra (i.e. supertranslations and Lorentz transformations) is called bms$^{global}$, the enhanced algebra is called bms$^{local}$. Throughout this thesis, this nomenclature will be used. Note that the Virasoro enhancement of the bms algebra was conjectured in 2003 by Banks\footnote{Footnote 17, page 29 of \cite{Banks}.}.\\

In four spacetime dimensions, the BMS ansatz can be generalized by allowing for a conformal factor in front on the angular part of the metric\footnote{This is related to the ambiguity in the Penrose conformal compactification scheme.}. In this case, both the symmetry algebra and the equations of motion change. This extension of the BMS gauge was considered in detail in \cite{Barnich:2010eb}, and the associated symmetry algebra is called bms extended\footnote{``Extended" thus refers to the presence of Weyl transformations in the symmetry algebra, and not to the enhancement of the Lorentz subalgebra to the Virasoro subalgebra; contrary to the nomenclature used in \cite{Strominger:2013jfa},\cite{He:2014laa} and related papers, in which  ``extended bms" algebra refers to the Virasoro enhanced symmetry algebra.}: it is the direct sum between the Weyl  transformations with bms$^{local}$.\\

The symmetries of asymptotically flat spacetimes at null infinity can be worked out in other spacetime dimensions than four.
In three dimensions for instance, the bms algebra was introduced by  Ashtekar, Bicak and Schmidt in \cite{Ashtekar}. In higher dimensions, a first computation of bms$_n$ algebra was computed by Barnich and Comp\`ere in \cite{Barnich:2006av}. \\

In three spacetime dimensions, the relation between asymptotically anti-de Sitter and asymptotically flat spaces has been studied in detail in \cite{Barnich:2012aw,Hernan}, where there is a nice observation that the BMS gauge can be adapted to describe asymptotically anti-de Sitter spaces as well, in addition to the description of asymptotically flat spaces. This adaptation of the BMS metric to include the presence of a non-vanishing cosmological constant has previously been considered, for instance in \cite{Stephani:2003tm}.\\

It is worth stressing that the BMS gauge can be used to describe the cases of a non-vanishing cosmological constant (positive or negative) in any dimensions, even though the BMS gauge was first build in order to describe asymptotically flat spaces in four dimensions.\\

An interesting property of the BMS ansatz is the fact that the gauge is completely fixed\footnote{Contrary to the Brown-Henneaux gauge fixing for asymptotically anti-de Sitter space in three dimensions, see section 2 of \cite{Barnich:2011ct} for a nice discussion about the difference between complete versus asymptotic gauge fixing.}. In three spacetime dimensions, one can also observes that the perturbative expansion of the solution (in the pure gravity case) stops at next to leading order, while for on-shell asymptotic symmetry generators, the expansion stops at next-to-next to leading order. Furthemore, the associated surface charges computed at any radial coordinate appear to be the same\cite{P-HLaura}.

\chapter{BMS gauge in four dimensions}
In this chapter, the BMS gauge in four dimensions is reviewed. Except for the presentation, the example of section (2.2.3) and the proof of \eqref{4:75} in section (2.4), the results described in this chapter are not new. This chapter is mainly based on \cite{Barnich:2010eb,Barnich:2011ct,Barnich:2011mi} and also on the original BMS papers \cite{Bondi:1962px,Sachs:1962wk}. The three first sections concern the BMS gauge in the case of asymptotically flat spacetimes; the last two sections present the case of (anti-)de Sitter spacetimes and introduce the Newman-Unti gauge, respectively.

\section{BMS$_4$ metric and bms$_4$ symmetry algebra}
In four spacetime dimensions with coordinates $(u,r,x^A)\footnote{$x^A$ represents the two angular coordinates.}$, the BMS metric ansatz representing asymptotically flat spacetimes depends on seven functions $(V, \beta,U^A,g_{AB})$ and is given (off-shell) by
\begin{eqnarray}\label{3:3.1}
g_{\mu\nu}=
\begin{pmatrix}
\dfrac{V}{r}e^{2\beta}+g_{AB}U^AU^B	&-e^{2\beta}	&-g_{BC}U^C\\
-e^{2\beta}&0	&0\\
-g_{AC}U^C&0&g_{AB}
\end{pmatrix},
\end{eqnarray}
with the following boundary conditions
\begin{align}
\beta&=\mathcal O(r^{-2}),~~U^A=\mathcal O(r^{-2}),\label{3:3.2}\\
g_{AB}&=r^2\bar\gamma_{AB}+\mathcal O(r),~~det(g_{AB})=r^4det(\bar\gamma_{AB}),\label{3:3.3}
\end{align}
where $\bar\gamma_{AB}$ is the metric of the two-sphere (for instance with coordinates $x^A=(\theta,\phi): \bar\gamma_{AB}dx^Adx^B=d\theta^2+\sin^2\theta d\phi^2$). The covariant derivative, Laplacian and Ricci scalar  associated with the two-sphere are denoted by $\bar D_A,\bar\Delta,\phantom a^{(2)}\bar R$, respectively.\\
The gauge fixing condition of the function $V$ is actually a consequence of the boundary conditions \eqref{3:3.2}, due to the equations of motion. See section 2.2, equation \eqref{3:51} for a detailed explanation. However in this section, the asymptotic symmetries analysis is performed in an off-shell way. The gauge fixing condition for the function $V$ is given to be
\begin{eqnarray}\label{bms4:3.4}
\dfrac{V}{r}&=-1+\mathcal O(r^{-1}).
\end{eqnarray}
In the asymptotic symmetry analysis (see below), asking that the gauge transformations leave invariant this boundary condition will not impose any new constraint on the symmetry parameter.

\subsection{Asymptotic symmetries and algebra}
By definition, asymptotic symmetries are gauge transformations that preserve the boundary conditions of the field\footnote{This is not the only way to define asymptotic symmetries, see \cite{Barnich:2001jy} section (1.3.1).}. For gravitational theories, the gauge transformations are given by the Lie derivative acting of the metric, $\delta_{\xi}g_{\mu\nu}=-\mathcal L_{\xi}g_{\mu\nu}$. Explicitly, asymptotic symmetries of the BMS gauge \eqref{3:3.1} are solutions to
\begin{align}
\mathcal L_\xi g_{\mu\nu}&=-\delta g_{\mu\nu}\notag\\
&=
\begin{pmatrix}
\mathcal O(r^{-1})&\mathcal O(r^{-2})&\mathcal O(r^{-2})\\
\mathcal O(r^{-2})&0&0\\
\mathcal O(r^{-2})&0&\mathcal O(r)
\end{pmatrix}\label{3:3.4},\\
\notag\\
\mathcal L_\xi (det g_{AB})&=0\notag\\
&=(det g_{AB}) (g^{DC}\mathcal L_\xi g_{DC})
\rightarrow g^{AB}\mathcal L_\xi g_{AB}=0.\label{3:3.5}
\end{align}

The explicit form of the asymptotic symmetry parameter $\xi$ is obtained in two steps:
\begin{itemize}
\item
First, the exact\footnote{Exact, as opposite to asymptotic.} equations of \eqref{3:3.4} and \eqref{3:3.5} are solved,
\begin{eqnarray}\label{chuiscreve}
\mathcal L_\xi g_{rr}=0,~~\mathcal L_\xi g_{rA}=0,~~g^{AB}\mathcal L_\xi g_{AB}=0.
\end{eqnarray}
The three first equations in \eqref{chuiscreve} give differential equations for the components of the gauge parameters ($\xi^u,\xi^A$) with respect to the radial coordinate, and produce three integration functions ($f,Y^A$) with respect to $r$. The last equation in \eqref{chuiscreve} is an algebraic equation for $\xi^r$.
More precisely,
\begin{align}
\partial_r\xi^u&=0\hspace{0.5cm}\Rightarrow\xi^u=f(u,x^A),\\
\partial_r\xi^A&=(\partial_B f)e^{2\beta}g^{AB}\hspace{0.5cm}\Rightarrow
\xi^A=Y^A(u,x^B)-\partial_B f\int^\infty_{r'}dr' e^{2\beta}g^{AB},\\
\xi^r&=-\frac{r}{2}\left(\bar D_A\xi^A-U^C\partial_C f\right),
\end{align}
with $f(u, x^C)$ and $Y^A(u,x^C)$ arbitrary integration functions.
\item
The second step in finding the asymptotic symmetry parameter $\xi$ is to solve the remaining equations in  \eqref{3:3.4}, i.e.
\begin{align}\label{3:3.10}
\mathcal L_\xi g_{ur}=\mathcal O(r^{-2}),~\mathcal L_\xi g_{uA}=\mathcal O(r^{-2}),~
\mathcal L_\xi g_{AB}=\mathcal O(r),~\mathcal L_\xi g_{uu}=\mathcal O(r^{-1}).
\end{align}
These equations give relations between integration functions $f,Y^A$. Solutions to equations \eqref{3:3.10} are, respectively,
\begin{align}
\partial_u f&=\dfrac{1}{2}\bar D_A Y^A,~~
\partial_u Y^A=0,\label{BMS4:4:12}\\
Y^C\partial_C \bar\gamma_{AB}&+\bar\gamma_{CB}\partial_AY^C+\bar\gamma_{AC}\partial_BY^C=\bar D_CY^C\bar\gamma_{AB}.\label{4:3:conf}
\end{align}
Equation \eqref{4:3:conf} is the conformal Killing equation for $Y^A$ with respect to the metric $\bar\gamma_{AB}$.
The last equation of \eqref{3:3.10} does not give any new relation, due to the following property, satisfied by any conformal Killing vector of the two-sphere,
\begin{align}\label{3:12}
\bar D_A\bar D_BY_C~&=~\dfrac{1}{2}~(\bar\gamma_{CA}\bar D_B\psi+\bar\gamma_{CB}\bar D_A\psi-\bar\gamma_{AB} \bar D_C\psi)-\notag\\
&-\bar\gamma_{AB}Y_C+\bar\gamma_{AC}Y_D,
\end{align}
where $\psi$ is the conformal factor of the conformal Killing vector, i.e. $\psi=\bar D_CY^C$. See equation \eqref{A:4} of appendix A for a proof of this formula and for a generalization in $d$ dimensions. The first equation of \eqref{BMS4:4:12} can be integrated with respect to $u$ so that $\xi^u=T(x^A)+\dfrac{u}{2}\bar D_CY^C$.
\end{itemize}

Therefore, the most general gauge parameter leaving invariant the form of the BMS gauge in four dimensions in the case of vanishing cosmological constant can be written, in components, as
\begin{align}
\xi^u&=T(x^A)+\dfrac{u}{2}\bar D_CY^C\label{3:14},\\
\xi^A&=Y^A(x^C)-\partial_B f\int_{r'}^\infty e^{2\beta}g^{AB}dr',\\
\xi^r&=-\frac{r}{2}\left(\bar D_A\xi^A-U^C\partial_C f\right).\label{3:16}
\end{align}
and is characterized by three functions of the angles: $T(x^A)$  and $Y^A(x^C)$, a conformal Killing vector of the two-sphere. Note that this vector field $\xi$ explicitly depends on the metric components, through $\beta,g^{AB}$ and $U^C$.
When performing an expansion in inverse power of $r$, the gauge parameter \eqref{3:14}-\eqref{3:16} becomes asymptotically 
\begin{align}\label{3:3.17}
\xi=\left(T+\dfrac{u}{2}\bar D_CY^C\right)\partial_u+Y^A\partial_A-\dfrac{1}{2}\left(r\bar D_CY^C-\bar\Delta f\right)\partial_r+\mathcal O(r^{-1}).
\end{align}

At future null infinity\footnote{Null infinity is denoted by $\mathcal I$, pronounced scri (for ``SCRipt Infinity"). As noted by Valiente Kroon in \cite{ValienteKroon:2001pc}, the phonetic transliteration of scri to polish incidentally means ``boundary".} (with coordinates $(u,x^A)$ and by taking $\lim_{r\to\infty}$) the vector field \eqref{3:3.17} reduces to
\begin{eqnarray}\label{3:18}
\bar\xi=\left(T+\dfrac{u}{2}\bar D_CY^C\right)\partial_u+Y^A\partial_A,
\end{eqnarray}
 and does no longer depend on the metric components, but is still parametrized by three functions of the angles ($T$ and $Y^A$). One can also consider the $T$ and $Y^A$ generators separately,
\begin{eqnarray}
\bar\xi_T=T\partial_u,\hspace{1cm}
\bar\xi_Y=\left(\dfrac{u}{2}\bar D_CY^C\right)\partial_u+Y^A\partial_A.
\end{eqnarray}
The two types of generators in the BMS asymptotic symmetries are thus the supertranslations (translations depending on the angles $x^C$), generated by $\bar\xi_T$, and the conformal transformations (global or local, see later) generated by $\bar\xi_Y$.\\

The algebra of a field-independent gauge parameter is found by considering the commutator of two variations acting on the metric field, $[\delta_{\bar\xi_1},\delta_{\bar\xi_2}]g_{\mu\nu}=\delta_{[\bar\xi_1,\bar\xi_2]g_{\mu\nu}}$ with the Lie bracket defined by
\begin{eqnarray}
[\bar\xi_1,\bar\xi_2]&= \bar\xi_1^\sigma\partial_\sigma\bar\xi_2-\bar\xi_2^\sigma\partial_\sigma\bar\xi_1.
\end{eqnarray}
In the case of the vector field \eqref{3:18}, let $\hat f$ denote the $u$ component of the Lie bracket, and let $\hat Y^A$ denote the angular components,
\begin{eqnarray}\label{3:20}
[\bar\xi_1,\bar\xi_2]=[\bar\xi_1,\bar\xi_2]^u\partial_u+[\bar\xi_1,\bar\xi_2]^A\partial_A\equiv \hat f\partial_u+\hat Y^A\partial_A.
\end{eqnarray}
A computation shows that these components of the Lie bracket are
\begin{align}
\hat Y^A&=Y_1^C\partial_CY_2^A-Y_2^C\partial_CY_1^A,\label{3:21}\\
\hat T&=Y_1^C\partial_C T_2-Y_2^C\partial_C T_1+\dfrac{1}{2}(T_1\bar D_CY^C_2-T_2\bar D_CY^C_1),\label{3:22}\\
\hat f&=\hat T+\dfrac{1}{2}\bar D_C \hat Y^C.\label{3:23}
\end{align}
These relations define the bms$_4$ symmetry algebra. Relations  \eqref{3:20}-\eqref{3:22} can also be written more compactly as
\begin{eqnarray}\label{3:24}
[(T_1,Y_1),(T_2,Y_2)]&=(\hat T,\hat Y^A),
\end{eqnarray}
with $\hat T,\hat Y^A$ defined in \eqref{3:21} and \eqref{3:23}. The symmetry algebra can be developped in modes by choosing suitable generators basis and the resulting symmetry algebra then depends on the space of integration functions ($T,Y^A$) that one considers. As observed by Barnich and Troessaert in \cite{Barnich:2009se,Barnich:2010eb}, there is not only one but actually two possible choices for the mode expansion of the elements of the symmetry algebra:
\begin{enumerate}
\item The algebra elements are required to be globally well-defined. In this case, the functions $T,Y^A$ are developped in spherical harmonics and the resulting symmetry algebra is bms$_4^{global}$. This is the original choice for the mode expansions since the sixties.
\item The algebra elements are allowed to be meromorphic functions (i.e. to admit pole singularities). In this case the functions $T,Y^A$ are developped in Laurent series and the associated symmetry algebra is bms$_4^{local}$. This is the new approach to the BMS symmetry \cite{Barnich:2009se,Barnich:2010eb,Barnich:2011mi}.
\end{enumerate}
In both cases (global and local), the BMS metric ansatz can be extended to take into account a conformal factor in front of the angular part of the metric, see \eqref{3:extended} below.\\

Before entering into the detail of the two different mode developments of the symmetry algebra, let us digress briefly about the asymptotic symmetry vector field \eqref{3:14}-\eqref{3:16},
\begin{align}\label{3:25}
\xi=&\left(T+\dfrac{u}{2}\bar D_CY^C\right)\partial_u+
\left(Y^A(x^C)-\partial_B f\int_{r'}^\infty e^{2\beta}g^{AB}dr'\right)\partial_A-\notag\\
&-\frac{r}{2}\left(\bar D_A\xi^A-U^C\partial_C f\right)\partial_r.
\end{align}

\subsection{Realization of the symmetry algebra in the bulk}

The vector field \eqref{3:25} is defined everywhere in the bulk of the spacetime (i.e. for all coordinates $u,r,x^A$) and, as mentioned before, is field dependent through the functions $\beta,g^{AB},U^A$ that are present in $\xi^\phi$ and $\xi^r$.\\

The algebra of this field-dependent gauge parameter can be found in the same way as before, by considering the commutator of two variations acting on the metric, $[\delta_{\xi_1},\delta_{\xi_2}]g_{\mu\nu}=\delta_{[\xi_1,\xi_2]_M}g_{\mu\nu}$ but now with the modified Lie bracket in the right hand side, defined by
\begin{eqnarray}\label{bms4:m}
[\xi_1,\xi_2]_M= \xi_1^\sigma\partial_\sigma\xi_2-\xi_2^\sigma\partial_\sigma\xi_1+\delta_{\xi_1}\xi_2(g)-\delta_{\xi_2}\xi_1(g),
\end{eqnarray}
where the two last terms take into account the fact that the gauge parameters are field dependent, so that the second variation in the commutators affects both the metric and the first variation,
\begin{align}
\delta_{\xi_1(g)}(\delta_{\xi_2(g)}g_{\mu\nu})&=\delta_{\xi_1}(-\mathcal L_{\xi_2(g)}g_{\mu\nu})=\mathcal L_{\xi_2(g)}\mathcal L_{\xi_1(g)}g_{\mu\nu}-\mathcal L_{\delta_{\xi_1}\xi_2(g)}g_{\mu\nu}\notag\\
&=\mathcal L_{\xi_2(g)}\mathcal L_{\xi_1(g)}g_{\mu\nu}+\delta_{(\delta_{\xi_1}\xi_2(g))}g_{\mu\nu}.\\
[\delta_{\xi_1(g)},\delta_{\xi_2(g)}]g_{\mu\nu}&=\mathcal L_{\xi_2(g)}\mathcal L_{\xi_1(g)}g_{\mu\nu}+\delta_{(\delta_{\xi_1}\xi_2(g))}g_{\mu\nu}-(1\leftrightarrow2)\notag\\
&=\delta_{[\xi_1,\xi_2]}g_{\mu\nu}+(\delta_{(\delta_{\xi_1}\xi_2(g))}-\delta_{(\delta_{\xi_2}\xi_1(g))})g_{\mu\nu},\notag\\
&=\delta_{([\xi_1,\xi_2]+\delta_{\xi_1}\xi_2(g)-\delta_{\xi_2}\xi_1(g))}g_{\mu\nu}.
\end{align}
hence the result \eqref{bms4:m}.
With this modified bracket, one can show \cite{Barnich:2010eb} that the vector field $\xi$ defined in the bulk of the spacetime by \eqref{3:25} forms a representation of the bms$_4$ symmetry algebra \eqref{3:24},
\begin{align}\label{3:27}
[\xi_1,\xi_2]_M=&\left(\hat T+\dfrac{u}{2}\bar D_C\hat Y^C\right)\partial_u+
\left(\hat Y^A(x^C)-\partial_B \hat f\int_{r'}^\infty e^{2\beta}g^{AB}dr'\right)\partial_A-\notag\\
&-\frac{r}{2}\left(\bar D_A\hat \xi^A-U^C\partial_C \hat f\right)\partial_r,
\end{align}
with $\hat \xi^A$ defined by $\xi^A$ in which $T,f$ are replaced by $\hat f,\hat Y^A$. The proof of this representation of the asymptotic symmetry algebra in the bulk is given in \cite{Barnich:2010eb} and consists in showing that the modified Lie bracket satisfies the same initial conditions and differential equations as the vector field \eqref{3:25}.\\

This result shows that the bms$_4$ symmetry algebra, even though it has been defined at infinity with the vector field $\bar\xi$ of \eqref{3:18}, is actually represented everywhere in the bulk of the spacetime throught the modified bracket.\\

The different mode expansions of the bms$_4$ algebra are now considered. 

\subsection{Global bms$_4$}
In this case, the functions $T,Y^A$ are developped in spherical harmonics on the two-sphere. The resulting bms$_4^{global}$ symmetry algebra consists of the semi-direct sum between smooth functions on the two-sphere (supertranslations) with global conformal Killing vectors. The latter is isomorphic to the Lorentz algebra. This is the first approach of bms$_4$ symmetries, see \cite{Bondi:1962px,Sachs:1962wk,Sachs2}.

\subsection{Local bms$_4$}

In this new approach, advocated by Barnich and Troessaert \cite{Barnich:2009se,Barnich:2010eb,Barnich:2011ct,Barnich:2011mi}, the integration constants are allowed to be meromorphic functions (i.e. to admit poles singularities) and in particular the functions $T,Y^A$ are now developped in Laurent series.\\

More precisely, let us introduce complex coordinates $(z,\bar z)$ by the stereographic relations
\begin{eqnarray}\label{zz}
 z=e^{i\phi}\cot{\frac {\theta}{2}},\hspace{1cm}\bar z=e^{-i\phi}\cot{\frac {\theta}{2}},
\end{eqnarray}
so that the leading of the angular part of the metric becomes conformally flat\footnote{Recall that conformal Killing vectors are invariant under conformal rescalings of the metric.}, $\bar\gamma_{AB}dx^Adx^B=\frac{4}{(1+z\bar z)^2}dzd\bar z$, and the conformal Killing equation \eqref{3:12} is solved for factorized vector fields $Y^z=Y^z(z)\equiv Y$ and $ Y^{\bar z}= Y^{\bar z}(\bar z)\equiv \bar Y$, so that the symmetry generator \eqref{3:18} becomes
\begin{eqnarray}\label{3:}
\bar\xi_{(T,Y,\bar Y)}=\left(T(z,\bar z)+\dfrac{u}{2} (\partial_zY+\partial_{\bar z}\bar Y)-\dfrac{u}{1+z\bar z}(\bar z Y+z\bar Y)\right)\partial_u+\notag\\
+Y(z)\partial_z+\bar Y(\bar z)\partial_{\bar z}.
\end{eqnarray}
Let the generators be expanded  in modes in the following way,
\begin{align}\label{3:29}
l_n&=\{\bar\xi_{T,Y,\bar Y}:~~ T=0,~~Y=-z^{n+1},~~\bar Y=0 \},\notag\\
\bar l_n&=\{\bar\xi_{T,Y,\bar Y}:~~ T=0,~~Y=0, ~~\bar Y=-\bar z^{n+1} \},\\
T_{m,n}&=\{\bar\xi_{T,Y,\bar Y}:~~ T=2\dfrac{z^m\bar z^n}{1+z\bar z},~~Y=0,~~\bar Y=0 \}\notag.
\end{align}
The non-vanishing commutation relations of the bms$_4^{local}$ symmetry algebra, derived from \eqref{3:24} with mode expansion \eqref{3:29}, read
\begin{align}
[l_m,l_n]&=(m-n)l_{m+n},~~[l_l,T_{m,n}]=\left(\dfrac{l+1}{2}-m\right)T_{m+l,n},\label{bms4:chaptitre2:fin1}\\
[\bar l_m,\bar l_n]&=(m-n)\bar l_{m+n},~~[\bar l_l,T_{m,n}]=\left(\dfrac{l+1}{2}-n\right)T_{m,n+l}.\label{bms4:chaptitre2:fin2}
\end{align}
Two copies of the Witt algebra are recognized in the first commuation relation of \eqref{bms4:chaptitre2:fin1} and \eqref{bms4:chaptitre2:fin1}.
The bms$_4^{local}$ symmetry algebra consists of the semi-direct sum between the supertranslations and the local conformal transformations. The latter were called superrotations in \cite{Barnich:2011ct}, and contain the Lorentz algebra as a subalgebra\footnote{When the indice $n$ takes the values $(-1,0,1)$ in $l_n$.}. The commutation relations between the superrotations

\subsection{Extended bms$_4$}
 In \cite{Barnich:2010eb}, a generalization of the BMS metric ansatz was considered, in which an arbitrary conformal factor is present in front of the angular part of the metric. The gauge fixing conditions \eqref{3:3.3}  are replaced by
\begin{eqnarray}\label{3:extended}
g_{AB}=e^{2\varphi}r^2\bar\gamma_{AB}+\mathcal O(r),~~det(g_{AB})=r^4e^{4\varphi}det(\bar\gamma_{AB}),
\end{eqnarray}
with $\varphi$ an arbitrary function of ($u,x^A$). This conformal factor in front the two-sphere is important, for instance in the case of the Robinson-Trautman solutions. The symmetries and solutions to equations of motion of this extended BMS gauge were studied in detail in \cite{Barnich:2010eb}. In particular, the vector field \eqref{3:25} is replaced by
\begin{align}
\xi=&e^{\varphi}\left(T+\int_0^u\dfrac{\bar D_CY^C}{2}e^{-\varphi} du'\right)\partial_u+
\left(Y^A(x^C)-\partial_B f\int_{r'}^\infty e^{2\beta}g^{AB}dr'\right)\partial_A-\notag\\
&-\frac{r}{2}\left(\bar D_A\xi^A-U^C\partial_C \xi^u+2\xi^u\partial_u\varphi\right)\partial_r,
\end{align}
with $Y^A$ a conformal Killing vector of $e^{2\varphi}\bar\gamma_{AB}$ and also of $\bar\gamma_{AB}$.

\section{BMS$_4$ equations of motion}
All the results described above were derived off-shell. In this section, the equations of motion are first considered and then solved for the metric ansatz \eqref{3:3.1} with boundary conditions \eqref{3:3.2} and \eqref{3:3.3}\footnote{Recall that no boundary condition is needed to be imposed on the function $V$ in order to solve to equations of motion.}.\\

\subsection{Integration procedure}
Bianchi identities for the gravitational field written in the BMS gauge allow to solve the field equations in a very convenient way. First, a relation between the equations of motion is derived, that is valid in $d$ dimensions for any metric field\footnote{The original proof in \cite{Bondi:1962px} was only given for $d=4$, without cosmological constant and without gauge field.}. Then, the form of the BMS metric in four dimensions is explicitly used to obtain a suitable hierarchy of the field equations that simplifies their resolution. The justification of using this hierarchy will become clear a posteriori; at this stage it is only a practical tool.\\

The Bianchi identitites in $d$ dimensions can always be written as
\begin{eqnarray}\label{3:33}
g^{\alpha\epsilon}\left(R_{\mu\alpha,\epsilon}-\dfrac{1}{2}R_{\alpha\epsilon,\mu}-\Gamma^{\sigma}_{\alpha\epsilon}R_{\mu\sigma}\right)=0.\label{eq:bianchiBMS4}
\end{eqnarray}
Indeed, one has
\begin{align*}
0=
g^{\alpha\epsilon}G_{\mu\alpha;\epsilon}&=g^{\alpha\epsilon}(R_{\mu\alpha}-\dfrac{1}{2}Rg_{\mu\alpha})_{;\epsilon}\\
&=g^{\alpha\epsilon}(R_{\mu\alpha;\epsilon}-\dfrac{1}{2}R_{\alpha\epsilon;\mu})\\
&=g^{\alpha\epsilon}(R_{\mu\alpha,\epsilon}-\dfrac{1}{2}R_{\alpha\epsilon,\mu}-\Gamma^\sigma_{\alpha\epsilon}R_{\mu\sigma}).
\end{align*}

Let us now focus on $d=4$. Bianchi identities \eqref{3:33} imply that for the BMS metric one has the
\begin{theorem}
If the six equations of motion $R_{rr}=0,R_{rA}=0,R_{AB}=0$ are satisfied, then the four relations $R_{ur}=0$ and $R_{A u}=\dfrac{\text{fct}(u,x^A)}{r^2}$, $R_{uu}=\dfrac{\text{fct}(u,x^A)}{r^2}$ are valid.
\end{theorem}
The proof goes as follow. Suppose that the six following equations of motion (called the main equations) are satisfied,
\begin{eqnarray}
R_{rr}=0,\hspace{1cm}R_{rA}=0,\hspace{1cm}R_{AB}=0,\label{eq:mainBMS4}
\end{eqnarray}
then the Bianchi identities \eqref{eq:bianchiBMS4} applied to the BMS metric \eqref{3:3.1} reduce to the following relations
\begin{align}
\mu=r:\hspace{1cm}&-g^{\alpha\epsilon}\Gamma^u_{\alpha\epsilon}R_{ur}=0,\label{eq:bianchi1BMS4}\\
\mu=A:\hspace{1cm}&g^{ur}(R_{A u,r}-R_{ur,A})-g^{\alpha\epsilon}\Gamma^u_{\alpha\epsilon}R_{uA}=0,\label{eq:bianchi2BMS4}\\
\mu=u:\hspace{1cm}&
g^{ur}R_{uu,r}+g^{rr}R_{ur,r}+g^{rA}(R_{ur,A}+R_{uA,r})+g^{AB}R_{uA,B}-\notag\\
&-g^{\alpha\epsilon}\Gamma^u_{\alpha\epsilon}R_{uu}-g^{\alpha\epsilon}\Gamma^r_{\alpha\epsilon}R_{ur}-g^{\alpha\epsilon}\Gamma^A_{\alpha\epsilon}R_{A u}=0.\label{eq:bianchi3BMS4}
\end{align}
From the BMS metric \eqref{3:3.1} one has $g^{\alpha\epsilon}\Gamma^u_{\alpha\epsilon}=g^{AB}\Gamma^u_{AB}=2\frac{e^{-2\beta}}{r}$ and thus equation \eqref{eq:bianchi1BMS4} implies that
\begin{eqnarray}
R_{ur}=0.\label{eq:trivial}
\end{eqnarray}
This means that if the six main equations \eqref{eq:mainBMS4} are satisfied, then the equation of motion $R_{ur}$ is automatically satisfied, as a direct consequence of the Bianchi identities. For this reason, equation $R_{ur}=0$ is called the trivial equation.\\
This result in relation \eqref{eq:bianchi2BMS4} implies
\begin{align}
-e^{-2\beta}\left(\partial_r+\dfrac{2}{r}\right)R_{Au}&=0\notag\\
-\dfrac{e^{-2\beta}}{r^2}\partial_r(r^2R_{Au})&=0\hspace{1cm}\Rightarrow\hspace{1cm}
R_{A u}=\dfrac{\text{fct}(u,x^A)}{r^2}\label{eq:supppl1},
\end{align}
and means that if the six main equations \eqref{eq:mainBMS4} are satisfied, then the equation of motion $R_{uA}=0$ is automatically satisfied at all orders, except the one of $r^{-2}$. In other words, after the main equations \eqref{eq:mainBMS4} have been solved, solving $R_{Au}=0$ is equivalent to solve the same equation but only at order $r^{-2}$.\\
Finally, using $R_{ur}=0$ and $R_{u\phi}=0$ in \eqref{eq:bianchi3BMS4} yields
\begin{align}
-e^{-2\beta}\left(\partial_r+\dfrac{2}{r}\right)R_{uu}&=0\notag\\
-\dfrac{e^{-2\beta}}{r^2}\partial_r(r^2R_{uu})&=0\hspace{1cm}\Rightarrow\hspace{1cm}
R_{uu}=\dfrac{\text{fct}(u,x^A)}{r^2}\label{eq:supppl2},
\end{align}
and, again, means that if the six main equations are satisfied, and also the equation $R_{uA}=0$, then the equation of motion $R_{uu}=0$ is automatically satisfied at all orders, except at order $r^{-2}$. This concludes the proof.\\

In summary, the Bianchi identities \eqref{eq:bianchiBMS4} therefore allow to organize the integration procedure of the equations of motion for the BMS metric field in a very convenient way. Once the six equations $R_{rr}=0, R_{rA}=0, R_{AB}=0$ (the main equations) are solved, then the equation $R_{ur}=0$ (the trivial equation) is automatically satisfied, and furthermore the last three equations of motion $R_{Au}=0, R_{uu}=0$ (the supplementary equations) are automatically solved at all orders for $r$ except at order $r^{-2}$, which remains to be solved.\\

As will be seen below, the main equations produce arbitrary integration functions (with respect to $r$). Some of these functions will be set to zero (in order to preserve the fall-off conditions that were imposed on the BMS metric), and the time evolution of some of the remaining constants will be constrained by the supplementary equations.
The only free data (apart from integration constants) needed to characterize a solution of the equations of motion is the angular part of the metric, $g_{AB}$.\\

\subsubsection{Comment on the main equations and Bondi-Sachs classification}

The main equations $R_{rr}=0$ and $R_{rA}=0$ give differential equations with respect to $r$ for the functions $\beta$ and $U^A,$ respectively. These differential equations can be solved in term of $g_{AB}$.\\

The last main equation, $R_{AB}=0$, contains three equations\footnote{$R_{AB}$ is symmetric.} and can be splitted in a trace part ($g^{AB}R_{AB}$) and a traceless part ($g^{DA}R_{AB}$\footnote{A priori these traceless part equations represent four different equations, because each of the indices $(B,D)$ takes the values $(\theta,\phi)$. However, in the case when $D=B=\theta$ and when $D=B=\phi$, these two equations are not independent with respect to the trace part equation $g^{AB}R_{AB}$. In other words, the only linearly independent equations in the traceless part equation $g^{DA}R_{AB}$ is when $B=\theta$ and $D=\phi$ or when $B=\phi$ and $D=\theta$.}). The trace part equation gives rise to a differential equation with respect to $r$ for the function $V/r$, while the traceless part equation fixes the $u$ dependence of the initial data $g_{AB}$.
When expanded in inverse power of $r$, there is one subleading of the free data $g_{AB}$ whose time derivative is not determined by the traceless equation, and it is the first subleading of $g_{AB}$ (denoted $C_{AB}$, see  \eqref{3:47} below).\\

The equations of motion can be named as follows, according to the original papers \cite{Bondi:1962px,Sachs:1962wk}.
\begin{itemize}
\item Main equations:
\begin{itemize}
\item Hypersurface equations: $R_{rr}=0,~~R_{rA}=0,~~g^{AB}R_{AB}=0;$
\item Standard equations: $g^{DA}R_{AB}=0;$
\end{itemize}
\item Trivial equation: $R_{ur}=0;$
\item Supplementary equations: $R_{uA}=0, R_{uu}=0.$\\
\end{itemize}

In the next subsection, the Einstein's field equations are solved for the BMS metric ansatz \eqref{3:3.1}.

\subsection{General solution}

Following \cite{Barnich:2010eb}, let us introduce some auxiliary quantities (for computational convenience) $k_{AB},K_{AB},l_{AB},L_{AB},n_A$, defined by the relations
\begin{align}
k_{AB}&=\dfrac{1}{2}\partial_rg_{AB},\hspace{1cm}k^A_{\phantom AB}=g^{AC}k_{CB},\hspace{1cm}k^A_{\phantom AB}=\dfrac{\delta^A_{\phantom AB}}{r}+\dfrac{K^A_{\phantom AB}(r)}{r^2}\label{BMS4aux1},\\
l_{AB}&=\dfrac{1}{2}\partial_ug_{AB},\hspace{1cm}l^A_{\phantom AB}=g^{AC}l_{CB},\hspace{1cm}l^A_{\phantom AB}=\dfrac{L^A_{\phantom AB}(r)}{r}\label{BMS4aux2},\\
n_A&=\dfrac{1}{2}e^{-2\beta}g_{AB}\partial_rU^B.\label{BMS4aux3}
\end{align}
\noindent The main equations (hypersurface and standard) are
\begin{align}
R_{rr}=0:\hspace{0.5cm}&\partial_r\beta=\dfrac{-1}{2r}+\dfrac{r}{16}(\partial_r g^{BC})(\partial_rg_{BC})
=\dfrac{1}{4r^3}K^A_{\phantom AB}K^B_{\phantom BA}\label{3:44}\\
R_{Ar}=0:\hspace{0.5cm}&\partial_r(r^2n_A)=r^2\left(\partial_r-\dfrac{2}{r}\right)\partial_A\beta-\phantom{D}^{(2)}D_BK^B_{\phantom BA}\label{3:45}\\
g^{AB}R_{AB}=0:\hspace{0.5cm}&\partial_r V=e^{2\beta}r^2\left[\Delta\beta+\partial^A\beta\partial_A\beta-\dfrac{1}{2}\phantom{ }^{(2)}R\right]-\notag\\
&\phantom{\partial r V =}-\dfrac{r^2}{2}\left(\partial_r+\dfrac{4}{r}\right)\phantom{ }^{(2)}D_AU^A,\label{3:46}\\
g^{AD}R_{AB}=0:\hspace{0.5cm}&\partial_r(r~l^D_{\phantom DB})+r(l^C_{\phantom DB}k^D_{\phantom DC}-l^D_{\phantom DC}k^C_{\phantom CB})=\notag\\
&=re^{2\beta}\left({\phantom a}^{(2)}D_B\partial^c\beta+\partial^D\beta\partial_B\beta+n^Dn_B-\dfrac{1}{2}\phantom{a}^{(2)}R^D_{\phantom DB}\right)-\notag\\
&-\dfrac{1}{2r}\partial_r\left(\dfrac{r^2}{2}(D^DU_B+D_BU^D)+rVk^D_{\phantom DB}\right)-\notag\\
&-\dfrac{r}{2}\left(k_{BC}D^CU^D-k^{CD}D_BU_C-\phantom{a}^{(2)}D_C(k^D_{\phantom DB}U^C)\right),\label{3:46bis}
\end{align}
where $\phantom{ }^{(2)}D_A$ and $\phantom{ }^{(2)}R$ respectively refer to the covariant derivative and Ricci scalar with respect to the two-dimensional metric $g_{AB}$. 
These equations can be integrated, up to integration constants $N^A(u,x^C)$ (for equation \eqref{3:45}) and $M(u,x^C)$ (for equation \eqref{3:46}). Note that there is no integration constant coming for equation \eqref{3:44}, because the fall-off condition for the beta function reduces it to be zero, $\beta=\mathcal O(r^{-2})$ see \eqref{3:3.2}.\\

Before integrating these equations, let us assume (following the approach of \cite{Bondi:1962px,Sachs:1962wk}) that the angular part of the metric can be expanded in inverse power of $r$, as follow
\begin{eqnarray}
g_{AB}=r^2\bar\gamma_{AB}+rC_{AB}+D_{AB}+\dfrac{1}{4}\bar\gamma_{AB}C^{EF}C_{EF}+\dfrac{E_{AB}}{r}+\mathcal O(r^{-2}),\label{3:47}
\end{eqnarray}
with $C_{AB},D_{AB},E_{AB}$ some functions of $(u,x^C)$ and where indices on these symmetric tensors are raised with the metric $\bar\gamma^{AB}$\footnote{Note that $\bar\gamma_{AB}$ is independent of $u$.}. The determinant condition \eqref{3:3.3} implies
 $C^A_{\phantom AA}=0, D^A_{\phantom AA}=0, E^A_{\phantom AA}=0$. Defining $g^{BC}$ such that $g^{BC}g_{AB}=\delta_A^{\phantom AC}$ gives
\begin{align}\label{bms4:gab}
g^{AB}=\dfrac{\bar\gamma^{AB}}{r^2}-\dfrac{C^{AB}}{r^3}-\dfrac{D^{AB}}{r^4}&+\dfrac{C^{AE}C^B_{\phantom BE}}{2r^4}-\dfrac{E^{AB}}{r^5}+\notag\\
&+\dfrac{C^A_{\phantom AE}D^{BE}+D^A_{\phantom AE}C^{BE}}{r^5}+\mathcal O(r^{-6}).
\end{align}
These expansions on the auxiliary quantities \eqref{BMS4aux1}-\eqref{BMS4aux3} imply
\begin{align}
k_{AB}&=r\bar\gamma_{AB}+\dfrac{C_{AB}}{2}-\dfrac{E_{AB}}{2r^2}+\mathcal O(r^{-3}),\\
l_{AB}&=\dfrac{r}{2}\dot C_{AB}+\dfrac{\dot D_{AB}}{2}+\dfrac{\dot C_{AE}C_B^{\phantom BE}+ C_{AE}\dot C_B^{\phantom BE}}{4}+\dfrac{\dot E_{AB}}{2r}+\mathcal O(r^{-2}),\\
k^D_{\phantom DB}&=\dfrac{\delta^D_{\phantom DB}}{r}-\dfrac{C^D_{\phantom DB}}{2r^2}-\dfrac{D^D_{\phantom DB}}{r^3}-\dfrac{3}{2r^4}E^D_{\phantom DB}+\dfrac{1}{r^4}D_{BE}C^{DE}+\notag\\
&+\dfrac{1}{2r^4}C_{BE}D^{DE}+\dfrac{1}{4r^4}C_{BE}C^{EF}C_F^{\phantom FD}+\mathcal O(r^{-5}),\\
l^D_{\phantom DB}&=\dfrac{\dot C^D_{\phantom DB}}{2r}+\dfrac{1}{2r^2}\left(\dot D^D_{\phantom DB}+\dfrac{1}{2}C_{BE}\dot C^{ED}-\dfrac{1}{2}\dot C_{BE} C^{DE} \right)+\notag\\
&+\dfrac{1}{2r^3}\left(\dot E^D_{\phantom DB}-\dot C_{BE} D^{DE}+\dot D_{BE}C^{DE}-\dfrac{1}{2}C_{BE}\dot C^{EF} C^D_{\phantom DF}\right)+\mathcal O(r^{-4}),\label{4:3:5228}
\end{align}
where a dot denotes $u$ derivative, and where the identity for traceless matrix in two dimensions $\frac{1}{2}\delta^A_{\phantom AB}C^{EF}C_{EF}=C^{AF}C_{BF}$ has been used in order to get the last relation of \eqref{4:3:5228}.
Even though a perturbative expansion in inverse power of $r$ has been performed, the integration of the field equations is still exact and closed: in principle, the solution can be known at every order in $r$.\\

Under these assumptions, the integration of the main equations \eqref{3:44}, \eqref{3:45} and \eqref{3:46} gives
\begin{align}
\beta&=\int_r^\infty dr'\dfrac{1}{4r'^3}K^A_{\phantom AB}K^B_{\phantom BA}
=-\dfrac{1}{32}\dfrac{C^A_{\phantom AB}C^B_{\phantom BA}}{r^2}+\mathcal O(r^{-3}),\label{3:49}\\
U^B&=-\dfrac{1}{2r^2} \phantom{D}^{(2)}D_C C^{BC}-\notag\\
&-\dfrac{2}{3r^3}\left(N^B(u,x^C)-\dfrac{1}{2}C^{AB}D_C C_A^{\phantom AC}+\left(\ln r+\dfrac{1}{3}\right)\bar D_CD^{CA}\right)+\mathcal O(r^{-4}),\label{3:50}\\
\dfrac{V}{r}&=-1+\dfrac{2M(u,x^C)}{r}+\mathcal O(r^{-2})\label{3:51}.
\end{align}
Note that the fall-off condition for the function $V$ is not needed during the integration procedure. Equation \eqref{3:51} is only a consequence of the gauge fixing conditions of functions $\beta,U^A,g_{AB}$ and of the dynamics of the system.\\
The integration provides integration constants (with respect to $r$) $M(u,x^C)$, called the Bondi mass aspect, and $N_A(u,x^C)$, the angular momentum aspect.\\

The two tensors $C_{AB}$ and $D_{AB}$ defined in the expansion of $g_{AB}$ in inverse power of $r$, \eqref{3:47} play the following role:
\begin{itemize}
\item
As first observed in \cite{Sachs:1962wk}, the presence of $\bar D_AD^{AB}$ and $D^{AB}$ are related to logarithmic terms in the solution and to singularities on the two-sphere, respectively. Indeed as can be seen from \eqref{3:50}, imposing $\bar D_AD^{AB}=0$ kills the logarithmic terms in the solution, and the globally well-defined solutions to $\bar D_AD^{AB}=0$ are $D^{AB}=0$. To see this, let us use stereographic coordinates $(z,\bar z)$, defined by \eqref{zz}. The trace condition $\bar\gamma_{AB}D^{AB}=0$ becomes $D^{\bar zz}=0$ so that equations $\bar D_AD^{AB}=0$ become explicitly
\begin{eqnarray}
\partial_zD^{zz}-\dfrac{4\bar z}{1+z\bar z}D^{zz}=0,\hspace{1cm}\partial_{\bar z}D^{\bar z\bar z}-\dfrac{4 z}{1+z\bar z}D^{\bar z\bar z}=0,
\end{eqnarray}
whose solutions are
\begin{eqnarray}
D^{zz}=f(\bar z)~(1+z\bar z)^4,\hspace{1cm}D^{\bar z\bar z}=g(z)~(1+z\bar z)^4,
\end{eqnarray}
with $f$ and $g$ arbitrary functions of their argument. The factor $(1+\bar zz)^4$ in the right hand side of these solutions is equal to $\frac{1}{\sin^8\theta}$ in $(\theta,\phi)$ coordinates, and the coordinate singularity at $\theta=0$ therefore becomes a physical singularity, except if the functions $f$ and $g$ are zero. In this case, the solutions become $D_{AB}=0$. This concludes the proof.\\
In view of the interesting symmetry enhancement that occurs when singularities are allowed on the two-sphere, the condition $\bar D_AD^{AB}=0$ is assumed to hold in this thesis, but not the further restriction $D_{AB}=0$.
\item
As will be seen later, the time derivative of the $C_{AB}$ tensors is responsible for the non-conservation of the mass of the system and is called the news tensors ($\partial_u C_{AB}=N_{AB}$). The tensor $C_{AB}$ is related to the presence of supertranslations.
\end{itemize}

The last main equation \eqref{3:46bis} gives the time evolution (contained in the auxiliary quantity $l^D_{\phantom DB}$) of the subleadings of the expansion of the free data $g_{AB}$, in terms of functions $\beta,U^A,\frac{V}{r}$ that are known functions at this stage of the integration procedure. However, the time derivative of the tensor $C_{AB}$ is not given by equation \eqref{3:46bis}. This can be seen by considering the left hand side of the equation of motion \eqref{3:46bis}, when taking into account the relations \eqref{4:3:5228},
\begin{align}
&\partial_r\left[\frac{1}{2}\partial_uC^D_{\phantom CB}+\mathcal O(r^{-1})\right]+r\left[\dfrac{1}{2r}\partial_u C^C_{\phantom DB}+\mathcal O(r^{-2})\right]\left[\dfrac{\delta^D_{\phantom DC}}{r}+\mathcal O(r^{-2})\right]-\notag\\
&-r\left[\dfrac{1}{2r}\partial_u C^D_{\phantom DC}+\mathcal O(r^{-2})\right]\left[\dfrac{\delta^C_{\phantom DB}}{r}+\mathcal O(r^{-2})\right]=\dots\label{4:3:56}
\end{align}
One immediatly sees from \eqref{4:3:56} that all the terms $\partial_u C^D_{\phantom DB}$ cancel, and therefore one concludes that the time evolution of the tensor $C_{AB}$ is not given by the dynamics of the system and must therefore be given as part of the initial data, as announced earlier.\\

The main equations are now completely solved. Following the general integration procedure described in previous section, one has that the trivial equation is satisfied as a consequence of the main equations. The only remaining equations to consider are the three supplementary equations. These equations are needed to be solved at order $\mathcal O(r^{-2}$)\footnote{Recall that all the remaining orders of these supplementary equations of motion are already satisfied, as a consequence of the main equations.}, and determine the $u$-dependence of the three integration constants $M(u,x^C),N^A(u,x^C)$.\\
Using the results obtained so far, the supplementary equations become
\begin{align}
R_{uA}=0:~~&\partial_uN^A=\partial_AM+\dfrac{1}{16}\partial_A(C_{BC}\partial_uC^{BC})-\dfrac{1}{4}(\partial_uC_{BA})\bar D_AC^{BC}\notag\\
&\phantom{\partial_uN^A=}-\dfrac{1}{4}\bar D_B(\bar D^B\bar D^C C_{CA}-\bar D_A\bar D_C C^{BC})\notag\\
&\phantom{\partial_uN^A=}-\dfrac{1}{4}\bar D_B(C^{CB}\partial_uC_{AC}-C_{AC}\partial_uC^{CB}),\label{3:53}\\
R_{uu}=0:~~&\partial_u M=\dfrac{1}{4}\bar D_A\bar D_B (\partial_u C^{AB})-\dfrac{1}{8}(\partial_uC^{AB})(\partial_u C_{AB})\label{3:52}.
\end{align}
As can be seen from equation \eqref{3:52}, the non-conservation of the mass is related to the presence of news (defined by $\partial_u C_{AB}$). When there is no news in the system, evolution equations \eqref{3:52} and \eqref{3:53} reduce to
\begin{align}
\partial_u N^A&=\partial_AM-\dfrac{1}{4}\bar D_B(\bar D^B\bar D^C C_{CA}-\bar D_A\bar D_C C^{BC}),\\
\partial_u M&=0.
\end{align}

The most general solution of the equations of motion for a metric in the BMS gauge \eqref{3:3.1} depends on the initial data $g_{AB}$ (assumed to be expanded in inverse power of $r$, see \eqref{3:47}) and is characterized by functions $\beta,U^A,\dfrac{V}{r}$ given explicitely by equations \eqref{3:49}-\eqref{3:51}, up to arbitrary integration functions $M,N^A$ subject to constraint (or evolution) equations \eqref{3:52},\eqref{3:53}. The time evolution of the first subleading in the expansion of $g_{AB}$ (i.e. the news, $\partial_u C_{AB}$) is not determined by the equations of motion, and must also be given as initial data to characterize the solution uniquely.\\
When there is no $u$ dependence in the initial data $g_{AB}$, then the traceless equations \eqref{3:46bis} become  constraint equations for the functions $\beta,U^A, V$ and, as a consequence, restrict the choice of integration functions $M$ and $N^A$. See the next subsection for an explicit example with no $u$ dependence in the choice of the initial data.

\subsection{Example of initial data: $g_{AB}=r^2\bar\gamma_{AB}$}
In this section, the general solution of the integration procedure described in detail in the previous section is applied in the simplest case of initial data one could think of: the case where the initial data is just the metric of the two-sphere, i.e. $g_{AB}=r^2\bar\gamma_{AB}$. Comparison between the form of $g_{AB}$ and the expansion \eqref{3:47}, and definitions \eqref{BMS4aux1}-\eqref{BMS4aux3}, gives that both tensors $C_{AB}$ and $D_{AB}$ are zero, and also that the auxiliary quantities  are
\begin{align}
k_{AB}&=r\bar\gamma_{AB},\hspace{1cm}l_{AB}=0,\\
k^A_{\phantom AB}&=\dfrac{\delta^A_{\phantom AB}}{r},\hspace{1cm}l^A_{\phantom AB}=0.
\end{align}
Solutions to the main equations \eqref{3:44} \eqref{3:46bis} are in this case
\begin{align}
\beta&=0,\\
U^A&=-\dfrac{2N^A(u,x^C)}{3r^3},\\
\dfrac{V}{r}&=-1+\dfrac{2M(u,x^C)}{r}-\dfrac{1}{r^2}\bar D_CU^C-\dfrac{3}{r^4}(\bar\gamma_{AB}N^AN^B).
\end{align}
Finally, the traceless equation \eqref{3:46bis} gives the following condition
\begin{align}\label{3:72}
&0=r\left(n^Dn_B-\dfrac{1}{2}\phantom{a}^{(2)}R^D_{\phantom DB}\right)\notag\\
&-\dfrac{1}{2r}\partial_r\left(\dfrac{r^2}{2}(D^DU_B+D_BU^D)+2rVk^D_{\phantom DB}\right)-\notag\\
&-\dfrac{r}{2}\left(k_{BC}D^CU^D-k^{CD}D_BU_C-\phantom{a}^{(2)}D_C(k^D_{\phantom DB}U^C)\right).
\end{align}
Equation \eqref{3:72} can then be interpreted as a  constraint equation for the function $U^A$. Secondly, this equation can be expanded in inverse power of $r$, and so one can solve this equation order by order in $r$. At order $\mathcal O(r^{-4})$ equation \eqref{3:72} reduces to
\begin{eqnarray}
-18 \dfrac{N^\phi N^\theta\sin^2\theta}{r^4}=0,
\end{eqnarray}
so either $N^\phi$ or $N^\theta$ must be zero. Combining this result with the equation $R_{\theta\theta}=0$ at order $\mathcal O(r^{-4})$,
\begin{eqnarray}
-\dfrac{9}{r^4}[(N^\theta)^2-(N^\phi)^2\sin^2\theta]=0,
\end{eqnarray}
forces the other component of $N^A$ to vanish as well.\\
The traceless part equations therefore  reduce $N^A$ to be zero.\\

The two supplementary equations give the following constraint on the integration function $M(u,x^C)$,
\begin{eqnarray}
\partial_A M(u,x^C)=0,\hspace{1cm}\partial_u M(u,x^C)=0.
\end{eqnarray}

The BMS gauge ansatz with initial data $g_{AB}=r^2\bar\gamma_{AB}$ therefore becomes on-shell
\begin{eqnarray}
ds^2=-\left(1-\dfrac{2M}{r}\right)du^2-2dudr+r^2\bar\gamma_{AB} dx^Adx^B,
\end{eqnarray}
with a constant parameter $M$. This is nothing but the Schwarzschild metric written in retarded null coordinates ($u,r,x^A$). This final metric is by no means surprising, since the choice of initial data is equivalent to requiring spherical symmetry, and the resulting metric can only be the Schwarzschild one, due to Birkoff's theorem.\\

This exercise of solving the equation of motion for the BMS gauge with the initial data $g_{AB}=r^2\bar\gamma_{AB}$ is nevertheless interesting. To the best, it is a new derivation of the Schwarzschild metric. To the worst, it is just a consistency check of the integration procedure that is associated with the bms gauge ansatz.

\subsection{News and Bondi-Sachs mass-loss formula}
In terms of stereographic coordinates \eqref{zz}, the  $C_{AB}$ tensor can be parametrized by $C_{zz}=c, C_{\bar z\bar z}=\bar c, C_{z\bar z}=0$.
 The evolution equation for the mass, equation \eqref{3:52}, becomes 
\begin{eqnarray}
\partial_u M=\partial_z^2(\partial_u \bar c)+\partial_{\bar z}^2(\partial_u c)-(\partial_u c)(\partial_u\bar c).\label{3:56}
\end{eqnarray}
Note that the two first terms are total derivative with respect to $(z,\bar z)$, and the last term can be written as a square. This equation can then be integrated over $(z,\bar z)$ to give
\begin{eqnarray}
\partial_u\int dzd\bar z\bar\gamma^{z\bar z}M=-\int  dzd\bar z \bar\gamma^{z\bar z}(\partial_u c)(\partial_u\bar c).
\end{eqnarray}
This is the famous Bondi-Sachs mass loss formula: the total energy of the system (i.e. when integrated over the angles) is decreasing. The physical interpretation of this formula is the following: the source emits waves and waves carry energy, so that the total mass of the source decreases when emitting waves.\\

In some recent considerations \cite{Strominger:2013jfa,He:2014laa} concerning the relation between the BMS symmetry and soft theorems, the evolution equation for the mass was integrated over $u$, instead of the angles. Writing $\Delta M(z,\bar z)=\int_{-\infty}^{+\infty}du (\partial_u M)$, equation \eqref{3:56} becomes
\begin{eqnarray}\label{4:rhs}
\Delta M(z,\bar z)=-\int_{-\infty}^{+\infty}du~~(\partial_u c)(\partial_u\bar c)+\int_{-\infty}^{+\infty}du\left(\partial_z^2(\partial_u \bar c)+\partial_{\bar z}^2(\partial_u c)\right).
\end{eqnarray}
There are two terms in the right hand side of \eqref{4:rhs}: a radiative term (the first one) and a term linear in the gravitational field (the second one). This second term was argued to be a soft graviton contribution to the local energy \cite{Strominger:2013jfa}. This term does not contribute to the total energy (it is a total derivative term), but gives only localized contributions. This term is important in the proof of the conservation of energy at every angles \cite{Strominger:2013jfa,He:2014laa}.\\

\subsection{News and supertranslations}

In view of the importance of the $C_{AB}$ tensor and its time derivative, it is interesting to see how the asymptotic symmetries act on it. In order to do this, the variation of $g_{AB}$ is computed on-shell, at order $\mathcal O(r)$
\begin{align}\label{3:59}
-\delta_{\xi} C_{AB}&=\left(\mathcal L_{\xi}g_{AB}\right)|_{\mathcal O(r)}\notag\\
&=\left(f\partial_u+\mathcal L_{Y^C}-\dfrac{\bar D_CY^C}{2}\right)C_{AB}+\left(\bar\gamma_{AB}\bar\Delta f-2\bar D_A\bar D_B f\right).
\end{align}
Note that the inhomogeneous term in the transformation law \eqref{3:59} is identically zero when the gauge parameter $\xi$ is an element of the Poincar\'e algebra. In other words, the inhomogeneous term is non-zero only in the case of supertranslations that are not pure translations, and in the case of superrotations.
In the globally well-defined case, comments about this inhomogeneous term can be made when the general transformation law of the $C_{AB}$ tensor is restricted to the case of supertranslations transformations (i.e. $T\neq0, Y=0$ in $\xi$).
In this case, equation \eqref{3:59} becomes
\begin{eqnarray}\label{3:60}
-\delta_TC_{AB}=T\partial_u C_{AB}+(\gamma_{AB}\bar\Delta T-2\bar D_A\bar D_B T).
\end{eqnarray}
The first term on the right hand side acts as a usual translation, the second term is the non-zero inhomogeneous term in the transformation law. This relation means that even if one starts with $C_{AB}=0$, then a supertranslation will produce a non-zero $C_{AB}$ tensor. In other word, there is a mixing between the $C_{AB}$ and the supertranslations.\\

In recent works relating BMS with soft theorems \cite{Strominger:2013jfa,He:2014laa}, equation \eqref{3:60} is the statement that the Minkowski vacuum spontaneously breaks supertranslation invariance.

\subsection{Magnetic part of Weyl tensor is zero:\\ no supertranslation for SPI - no superrotation for BMS}

The covariant derivative of the $D_{AB}$ tensor, defined in \eqref{3:47}, controls the presence of logarithmic terms in the solution of the equations of motion and the presence of singularities on the two-sphere, as can be seen from equation \eqref{3:50} and the discussion below.  The condition $\bar D_AD^{AB}=0$, which was assumed when solving the main equations of motion,  can be seen as an extra gauge fixing condition. It is therefore of interest to see if this condition restricts the form of the symmetry algebra. In order to see this, the variation of $g_{AB}$ is computed at order $\mathcal O(1)$. One finds
\begin{align}\label{BMS4:dab}
\delta_\xi D_{AB}&=0.
\end{align}
Since there is no inhomogeneous term in the transformation \eqref{BMS4:dab}, the condition $\bar D_A D^{AB}=0$ is therefore invariant under the asymptotic symmetry transformations. In other words, the gauge fixing condition $\bar D_A D^{AB}$  does not impose any new condition on $\xi$.\\

In some discussion about the  asymptotic symmetry group at spatial infinity\footnote{The so-called SPI group (for ``SPatial Infinity").}, supertranslations symmetry transformations can be reduced to ordinary translations in two different ways \cite{Henneaux:1985tv}: either by imposing parity conditions on the fields, either by requiring that the magnetic part of the Weyl tensor vanishes. It is interesting to observe that in the case of asymptotic symmetries at null infinity in the BMS gauge, the magnetic part of the Weyl tensor is nothing but the $D_{AB}$ tensor\footnote{See equation (4.38) of \cite{Barnich:2010eb}.}, and that $D_{AB}$ is zero in the case where singularities on the two-sphere are not allowed.\\

So at spatial infinity, the vanishing of the magnetic part of the Weyl tensor is related to the absence of supertranslation, while at null infinity its vanishing is related to absence of singularities on the two-sphere and, as a consequence, to the absence of superrotation.\\

\section{BMS$_4$ surface charges}

Properties of surface charges associated with asymptotic symmetries are of direct physical interest and allow, in some cases, for  insights on the microscopic degrees of freedom for the gravitational interaction. For instance, in the case of black holes whose geometry are locally $AdS_3$, the central extension appearing in the algebra of surface charges\cite{Brown:1986nw} is a crucial ingredient for a non-stringy microscopic derivation of the macroscopic Bekenstein-Hawking entropy\cite{Strominger:1998eq}. In general, properties of surface charges (such as their algebra and possible central extension of it) depend crucially on the question whether charges are (or not) integrable.\\

\subsection{Surface charges and non-integrability}
The surface charges associated with the BMS symmetry parameters $(T,Y^A)$ and with the on-shell solution were computed in \cite{Barnich:2011mi}, and are
\begin{align}
\bdelta \mathcal{Q}_T=\dfrac{1}{16\pi G}\int dx^Adx^B\sqrt{\bar\gamma}&
\left[4T\delta M+\dfrac{T}{2}(\partial_uC_{AB})\delta(C^{AB})\right],\\
\bdelta \mathcal{Q}_{Y^A}=\dfrac{1}{16\pi G}\int dx^Adx^B\sqrt{\bar\gamma}&
[2u\bar D_AY^A\delta M
+Y^A\delta\left(2N_A+\dfrac{1}{16}\partial_A(C^{EF}C_{EF})\right)\notag\\
&+u\bar D_AY^A(\partial_uC_{AB})\delta(C^{AB})].
\end{align}
These charges are non-integrable due to the presence of the news tensors $\partial_u C_{AB}$.

\subsection{Algebra of surface charges}
In the  integrable case, there exists some theorems (in both hamiltonian formalism \cite{Brown:1986ed} and covariant approach \cite{Barnich:2001jy,Barnich:2007bf}) that garantee that the surface charges always form a representation of the asymptotic symmetry algebra, up to a possible central extension (i.e. $\{ Q_{\xi_1},Q_{\xi_2}\}^*=Q_{[\xi_1,\xi_2]}+K_{\xi_1,\xi_2}$), where the Dirac bracket for the surface charges was defined by
\begin{eqnarray}
\{Q_{\xi_1},Q_{\xi_2}\}^*=-\delta_{\xi_2}Q_{\xi_1}.
\end{eqnarray}

In the non-integrable case, no such theorem exists, but there is however a recent proposition for the Dirac bracket in the literature. Indeed, in the case of charges associated with the BMS algebra in four spacetime dimensions at null infinity, it was shown in \cite{Barnich:2011mi} that the non-integrable surface charge $\bdelta\mathcal{Q}_\xi$ can be decomposed into a sum of an integrable piece $\delta Q_\xi$ and a non-integrable piece $\Theta_\xi$, so that the resulting expression for the surface charge can be cast in the following form $\bdelta \mathcal{Q}_\xi(\delta\chi,\chi)=\delta Q_\xi(\chi)+\Theta_\xi(\delta\chi,\chi)$\footnote{Up to an ambiguity in the definition of $\Theta$, and where $\chi$ denotes the free data of the theory, $\chi=\{g_{AB},\partial_uC_{AB},M,N^A\}.$}. It was furthermore proved that if the Dirac bracket is defined taking into account the non-integrable piece, than the (integrable part of the) charges realize the asymptotic symmetry algebra, up to a possible (field dependent) central extension. Explicitly \cite{Barnich:2011mi}, if the Dirac bracket is defined by\footnote{Note the indices in the variation in the two terms.}
\begin{eqnarray}\label{diracbracket}
\{Q_{\xi_1},Q_{\xi_2}\}^*[\chi]=-\delta_{\xi_2}Q_{\xi_1}[\chi]+\Theta_{\xi_2}[-\delta_{\xi_1}\chi,\chi],
\end{eqnarray}
then the charges satisfy
\begin{eqnarray}\label{chargealgebra}
\{Q_{\xi_1},Q_{\xi_2}\}^*[\chi]=Q_{[\xi_1,\xi_2]}[\chi]+K_{\xi_1,\xi_2}[\chi].
\end{eqnarray}
In \cite{Barnich:2011mi}, the field-dependent central extension appearing in the algebra of the integrable part of the charges associated with the BMS gauge in four dimensions was computed to be
\begin{align}
K_{\xi_1,\xi_2}=\dfrac{1}{16\pi G}\int dx^A&dx^B\sqrt{\bar\gamma}~~
[\left(f_1\partial_A F_2-f_2\partial_A f_1\right)\partial^A \bar R+\notag\\
&+C^{AB}\left(f_1\bar D_A\bar D_B\bar D_CY^C_2-f_2\bar D_A\bar D_B\bar D_CY^C_1\right)].
\end{align}

\subsection{Superrotations and holographic currents}
In \cite{Barnich:2013axa}, the charges were evaluated in the case of the superrotation symmetry parameter. However, it was shown that these charges were diverging because of the presence of pole singularities.
The way to cure this problem is not to integrate and to consider instead the associated hologaphic currents algebra, see \cite{Barnich:2013axa} for more detail.

\section{Anti-de Sitter and de Sitter cases}

In four dimensions, the BMS metric ansatz described in the three first sections of this chapter was only discussed in the case of a vanishing cosmological constant. Interestingly, the BMS gauge \eqref{3:3.1}
\begin{eqnarray}\label{bms4:386}
g_{\mu\nu}=
\begin{pmatrix}
\dfrac{V}{r}e^{2\beta}+g_{AB}U^AU^B	&-e^{2\beta}	&-g_{BC}U^C\\
-e^{2\beta}&0	&0\\
-g_{AC}U^C&0&g_{AB}
\end{pmatrix},
\end{eqnarray}
 can be used to describe asymptotically anti-de Sitter (see \cite{Henneaux:1985ey}, \cite{Hernan} for a discussion about the symmetries and equations of motion, respectively) and de Sitter cases as well.\\

To describe the case of a non-vanishing cosmological constant in a unified way, a parameter $\alpha$ is introduced and defined as being the sign of the cosmological constant and zero otherwise. When $\alpha\neq0$, the only difference with respect to the asymptotically flat case lies in the BMS function $\frac{V}{r}$, for which the (off-shell) gauge fixing condition \eqref{bms4:3.4} now becomes
\begin{eqnarray}
\dfrac{V}{r}=\alpha\dfrac{r^2}{l^2}+\mathcal O(1).
\end{eqnarray}
Of course, this modification in the gauge fixing  of the metric affects the symmetry algebra. As in the asymptotically flat case, the fall-off for the function $V$ is not needed during the integration of the equations of motion, as it is only a consequence of the remaining boundary conditions (for $\beta,U^A)$, due to the equations of motion. As shown in \cite{Henneaux:1985ey}, there is no symmetry enhancement for asymptotically anti-de Sitter in dimension higher than three. In the rest of this section, only the equations of motion will be considered.\\

The integration procedure of the equations of motion for the BMS gauge, described in section 3.2 in the asymptotically flat case, can also be used in the case of a non-vanishing cosmological constant. Indeed, Bianchi identities allow to have the same hierarchy between the equations of motion (main, trivial, supplementary), simplifying therefore the quest of solving them.\\

The generalization of \eqref{3:33} in the presence of a cosmological constant is the following (to be as general as possible, the proof is given in $d$ dimensions).
When a cosmological constant\footnote{With $\Lambda=\alpha\frac{(d-1)(d-2)}{2l^2}$, so that the equations of motion in $d$ dimensions are $R_{\mu\nu}=\alpha\frac{(d-1)}{l^2}g_{\mu\nu}$.} is present, the equations of motion read
\begin{align}
&G_{\mu\nu}+\Lambda g_{\mu\nu}=0~~
\Leftrightarrow~~ R_{\mu\nu}-\dfrac{2}{d-2}\Lambda g_{\mu\nu}=0.
\end{align}
Let the equations of motion be denoted by $E_{\mu\nu}=R_{\mu\nu}-\dfrac{2}{d-2}\Lambda g_{\mu\nu}$ in this case. The generalization of \eqref{3:33} in the case of non-zero cosmological constant is
\begin{eqnarray}\label{4:75}
g^{\alpha\epsilon}\left(E_{\mu\alpha,\epsilon}-\dfrac{1}{2}E_{\alpha\epsilon,\mu}-\Gamma^{\sigma}_{\alpha\epsilon}E_{\mu\sigma}\right)=0.
\end{eqnarray}
Note that
\begin{align}
g^{\alpha\epsilon}(G_{\mu\alpha}+\Lambda g_{\mu\nu})_{;\epsilon}&=g^{\alpha\epsilon}\left( R_{\mu\nu}-\dfrac{2}{d-2}\Lambda g_{\mu\nu}\right)_{;\epsilon}\\
&=
g^{\alpha\epsilon}(R_{\mu\alpha})_{;\epsilon}-\dfrac{2}{d-2} \Lambda_{;\mu}
\end{align}
with $\Lambda_{;\mu}=\Lambda_{,\mu}=\text{(cst)}_{,\mu}=0$. The rest of the proof is therefore exactly the same as for the flat case (equation \eqref{3:33}) and will not be reproduced.\\

The integration procedure starts with the main equations. Due to the form of \eqref{bms4:386}, the three first main equations are the same as in the flat case, $E_{rr}=R_{rr}=0$ and $E_{rA}=R_{rA}=0$. The solutions for $\beta$ and $U^A$ are thus exactly the same as in the asymptotically flat case, \eqref{3:49},\eqref{3:50}.\\
The last main equations (trace part and trace free part) now involve the cosmological constant, so that \eqref{3:46} and \eqref{3:46bis} are replaced by
\begin{align}
g^{AB}R_{AB}=0:\hspace{0.5cm}&\partial_r V=e^{2\beta}r^2\left[\Delta\beta+\partial^A\beta\partial_A\beta-\dfrac{1}{2}\phantom{D}^{(2)}R\right]-\notag\\
&\phantom{\partial r V =}-\dfrac{r^2}{2}\left(\partial_r+\dfrac{4}{r}\right)\phantom{a}^{(2)}D_AU^A+\dfrac{3\alpha r^2}{l^2}e^{2\beta},\label{bms4:392}\\
g^{AD}R_{AB}=0:\hspace{0.5cm}&\partial_r(r~l^D_{\phantom DB})+r(l^C_{\phantom DB}k^D_{\phantom DC}-l^D_{\phantom DC}k^C_{\phantom CB})=\notag\\
&=re^{2\beta}\left({\phantom a}^{(2)}D_B\partial^c\beta+\partial^D\beta\partial_B\beta+n^Dn_B-\dfrac{1}{2}\phantom{a}^{(2)}R^D_{\phantom DB}\right)-\notag\\
&-\dfrac{1}{2r}\partial_r\left(\dfrac{r^2}{2}(D^DU_B+D_BU^D)+rVk^D_{\phantom DB}\right)+\dfrac{3\alpha r}{2l^2}e^{2\beta}\delta^D_{\phantom DB}-\notag\\
&-\dfrac{r}{2}\left(k_{BC}D^CU^D-k^{CD}D_BU_C-\phantom{a}^{(2)}D_C(k^D_{\phantom DB}U^C)\right).\label{bms4:393}
\end{align}
Under the assumption of a perturbative expansion in inverse power of $r$ of $g_{AB}$, \eqref{3:47}, \eqref{bms4:gab}, the trace equation \eqref{bms4:392} can be integrated to give
\begin{eqnarray}
\dfrac{V}{r}=\alpha\dfrac{r^2}{l^2}-\left(1+\dfrac{3\alpha}{l^2}\dfrac{C_{AB}C^{AB}}{16}\right)+\dfrac{2M(u,x^C)}{r}+\mathcal O(r^{-2})\label{bms4:396},
\end{eqnarray}
with $M(u,\phi)$ an integration function.\\
\noindent Some comments are in order.
\begin{itemize}
\item
The fall-off condition for the function $V$ is not needed during the integration procedure, exactly as in the asymptotically flat case ($\alpha=0$).
\item
When $\alpha=0$, \eqref{bms4:396} reduces to \eqref{3:51} as it should.
\item
Besides the contribution to the leading term (order $\mathcal O(r^2)$), the presence of the cosmological constant also affects the first subleading term (order $\mathcal O(r)$) in $\frac{V}{r}$. This subleading contribution comes from the term $\frac{3\alpha r^2}{l^2}e^{2\beta}$ in \eqref{bms4:392}. 
\end{itemize}
The traceless equation \eqref{bms4:393} gives the time evolution of the initial data $g_{AB}$, \eqref{3:47}. Expanding in inverse power of $r$ up to order $\mathcal O(r^{-2})$, and using the solution for $\beta,U^A,\dfrac{V}{r}$ already obtained during the integration (equations \eqref{3:49},\eqref{3:50},\eqref{bms4:396}), the traceless equation becomes
\begin{eqnarray}
-\alpha\dfrac{C^D_{\phantom DB}}{2l^2}+\dfrac{\alpha}{r}\dfrac{D^D_{\phantom DB}}{2l^2}+\mathcal O(r^{-2})=0.
\end{eqnarray}
Solving this equation at order $\mathcal O(1)$ and $\mathcal O(r^{-1})$ gives
\begin{eqnarray}
C^D_{\phantom DB}=0,\hspace{1cm}D^D_{\phantom DB}=0.\label{dernierjour}
\end{eqnarray}
This result reduces $g_{AB},k_{AB},l_{AB},k^D_{\phantom DB},l^D_{\phantom DB}$ to
\begin{align}
g_{AB}&=r^2\bar\gamma_{AB}+\dfrac{E_{AB}}{r}+\mathcal O(r^{-2}),\label{bms4:}\\
k_{AB}&=r\bar\gamma_{AB}-\dfrac{E_{AB}}{2r^2}+\mathcal O(r^{-3}),\hspace{1cm}
l_{AB}=\dfrac{\dot E_{AB}}{2r}+\mathcal O(r^{-2}),\\
k^D_{\phantom DB}&=\dfrac{\delta^D_{\phantom DB}}{r}-\dfrac{3}{2r^4}E^D_{\phantom DB}+\mathcal O(r^{-5}),\hspace{1cm}
l^D_{\phantom DB}=
\dfrac{1}{2r^3}\dot E^D_{\phantom DB}+\mathcal O(r^{-4})\label{4:3:52}.
\end{align}
With these simplifications, the solutions for $\beta,U^A,\dfrac{V}{r}$ given by equations \eqref{3:49}, \eqref{3:50}, \eqref{bms4:396} become
\begin{align}
\beta&
=-\dfrac{3}{32}\dfrac{E_{AB}E^{BA}}{r^6}+\mathcal O(r^{-7}),\label{bms4:3102}\\
U^B&=
-\dfrac{2}{3r^3}N^B(u,x^C)+\dfrac{3}{4r^4}\bar D_BE^{AB}+\mathcal O(r^{-5}),\label{bms4:3103}\\
\dfrac{V}{r}&=\alpha\dfrac{r^2}{l^2}-1+\dfrac{2M(u,x^C)}{r}-\dfrac{1}{3r^2}\bar D_AN^A+\mathcal O(r^{-3})\label{bms4:3104}.
\end{align}
The last equations of motion to be solved are the supplementary equations. Due to the cosmological constant, there is one more term to consider in addition to those of asymptotically flat case \eqref{3:53},\eqref{3:52},
\begin{eqnarray}
R_{uA}-\dfrac{3\alpha}{l^2}g_{uA}=0,\hspace{1cm}R_{uu}-\dfrac{3\alpha}{l^2}g_{uu}=0.
\end{eqnarray}
Recall from the integration procedure that in the perturbative expansion in inverse power of $r$, the only order that remains to be solved is the one of $\mathcal O(r^{-2})$. Explicitly,
\begin{align}
\left(R_{uA}-\dfrac{3\alpha}{l^2}g_{uA}\right)=0:\hspace{1cm}&\partial_uN^A=\partial_AM+\dfrac{9\alpha}{4l^2}\bar D_CE^C_{\phantom CA},\label{bms4:3106}\\
\left(R_{uu}-\dfrac{3\alpha}{l^2}g_{uu}\right)=0:\hspace{1cm}&\partial_u M=\dfrac{\alpha}{2l^2}\bar D_AN^A.\label{bms4:3107}
\end{align}
In the absence of cosmological constat (i.e. when $\alpha=0$), one recovers the results \eqref{3:52} \eqref{3:53} when the $C_{AB}$ and $D_{AB}$ tensors are switched-off, as it should.\\

The general solution to $R_{\mu\nu}=\alpha\frac{3}{l^2}g_{\mu\nu}$ for the metric ansatz \eqref{bms4:386} with the boundary conditions $\beta=\mathcal O(r^{-2})=U^A$ is completely characterized by the initial data $g_{AB}$, up to integration functions $M(u,x^C),N^A(u,x^C)$. When $g_{AB}$ is assumed to be expanded in inverse power of $r$, the solution becomes asymptotically \eqref{bms4:}, \eqref{bms4:3102}-\eqref{bms4:3104} with integration functions $N^A,M$ satisfying the evolution equations \eqref{bms4:3106}\eqref{bms4:3107}, respectively. The first and the second subleading parts of $g_{AB}$ are zero \eqref{dernierjour} which implies in particular that there is no news in this system, contrary to the asymptotically flat case.\\

As an example, choosing the initial data to be $g_{AB}=r^2\bar\gamma_{AB}$ reduces the general solution described above to
\begin{eqnarray}
ds^2=\left(\alpha \dfrac{r^2}{l^2}-1+\dfrac{2M}{r}\right)du^2-2dudr+r^2\bar\gamma_{AB} dx^Adx^B,
\end{eqnarray}
in the same way as in section (3.2.3). This corresponds to the Schwarschild-(anti-)de Sitter black hole in four dimensions.

\section{BMS gauge versus Newman-Unti gauge}

As explained in chapter 1, the BMS gauge is the original way to describe  asymptotically flat spacetimes. However, there are other gauges which describe asymptotically flat spacetimes as well, in particular the Newman-Unti \cite{newman:891} gauge (based on the Newman-Penrose formalism \cite{newman:566}). It is therefore interesting to see if the two gauge are physically equivalent. \\

This question will be addressed in detail in chapter 4, but at this point, let us just stress that the problem in comparing these two gauges is that, besides a different gauge choice of the radial coordinate, the Newman-Unti gauge allows a conformal factor in front of the angular part of the metric. More precisely in the bms gauge, the radial coordinate is defined as being a luminosity distance, i.e. by being the $r$ that satisfies the gauge fixing condition of the determinant,
\begin{eqnarray}
r\hspace{1cm}\text{such that}\hspace{1cm}\text{det~}(g_{AB})=r^4\sin^2 \theta.
\end{eqnarray}
In the Newman-Unti gauge for asymptotically flat spacetimes, the $r$ coordinates is defined to be the affine parameter of the null geodesic congruence,
\begin{eqnarray}
r\hspace{1cm}\text{such that}\hspace{1cm}l^\mu_{\phantom\mu;\nu}l^\nu=0.
\end{eqnarray}

\newpage
\chapter{BMS gauge in three dimensions}

In this chapter, the BMS gauge in three dimensions is described in purely gravitational theories. For Einstein-Yang-Mills systems, see chapter five. For higher spins and supergravity considerations, see \cite{Gonzalez:2013oaa} and \cite{Barnich:2014cwa} respectively. In the cases of asymptotically flat and asymptotically anti-de Sitter, this chapter is based on \cite{Barnich:2010eb,Barnich:2012aw}. Apart from the presentation, there are two new observations in this chapter. The first one \cite{P-HLaura} is the fact that the surface charges associated to the bms$_3$ algebra can be computed for any value of the radial coordinate and are the same\footnote{This result is true for the three cases: flat and (anti-)de Sitter.}. The second observation is the fact that the BMS gauge covers all the Penrose diagram of de Sitter space.

\section{BMS$_3$ gauge: Flat, AdS and dS cases}
One of the nice features of the BMS gauge is the fact that it can describe spacetimes with or without a cosmological constant.
In order to provide a description in an unified way of these three cases (flat, AdS, dS), a parameter $\alpha$ is introduced and defined to be the sign of the cosmological constant (and zero in the flat case).\\
  
In three spacetime dimensions with coordinates $(u,r,\phi)$, the BMS gauge depends on three functions $(V, \beta,U)$ and is given (off-shell) by the metric
\begin{eqnarray}\label{5:1}
g_{\mu\nu}=
\begin{pmatrix}
\dfrac{V}{r}e^{2\beta}+r^2U^2	&-e^{2\beta}	&-r^2U\\
-e^{2\beta}&0	&0\\
-r^2U&0&r^2
\end{pmatrix},
\end{eqnarray}
supplemented by gauge fixing conditions on $U,\beta$. The fall-off for $U$ is
\begin{eqnarray}
U=\mathcal O(r^{-2}),\label{5:2}
\end{eqnarray}
while the fall-off for the beta function depends on the presence of the cosmological constant and is given by
\begin{align}\label{bms3:53}
\beta&=\mathcal O(r^{-1})\hspace{1cm}(\alpha=0),\\
\beta&=\mathcal O(r^{-2})\hspace{1cm}(\alpha\neq0).
\end{align}
As in the four dimensional case, the boundary condition on the function $V$ is a consequence of the other gauge fixing conditions (i.e. on $U$ and $\beta$), due to the equations of motion. However in an off-shell description, this fall-off is needed and given to be
\begin{eqnarray}\label{bms3:3.4}
\dfrac{V}{r}=\alpha\dfrac{r^2}{l^2}+\mathcal O(1).
\end{eqnarray}

In this BMS gauge \eqref{5:1}, flat, anti de Siter and de Sitter spaces are described by taking $U=0=\beta$ and $\dfrac{V}{r}=\alpha\dfrac{r^2}{l^2}-1$, so that the line element becomes
\begin{eqnarray}
ds^2=\left(\alpha\dfrac{r^2}{l^2}-1\right)du^2-2dudr+r^2d\phi^2.
\end{eqnarray}

As in the four dimensional case, one can consider an extended gauge fixing, by introducing a conformal factor $\varphi=\varphi(u,\phi)$ in the angular part of the metric,
\begin{eqnarray}
\begin{pmatrix}
\dfrac{V}{r}e^{2\beta}+e^{2\varphi}r^2U^2	&-e^{2\beta}	&-e^{2\varphi}r^2U\\
-e^{2\beta}&0	&0\\
-e^{2\varphi}r^2U&0&e^{2\varphi}r^2
\end{pmatrix}.
\end{eqnarray}
See \cite{Barnich:2010eb,Troessaert:2013fma} for a detailed analysis of this extended gauge in the asymptotically flat and anti-de Sitter cases, respectively.
In the rest of this chapter, this extended gauge fixing will not be considered.

\section{BMS$_3$ symmetry algebra}
The asymptotic symmetries are, by definition, the gauge transformations leaving invariant the gauge fixing of the metric field.
Explicitly, asymptotic symmetries are the $\xi^\mu$ solutions to
\begin{align}
\mathcal L_{\xi}g_{\mu\nu}&=-\delta g_{\mu\nu}\\
&=
\begin{pmatrix}
\mathcal O(1) &-\delta g_{ur}&\mathcal O (1)\\
-\delta g_{ur}&0&0\\
\mathcal O (1)&0&0
\end{pmatrix}.\label{bms3:6}
\end{align}
As in the four dimensional case, the explicit form of the asymptotic symmetry parameter $\xi$ is found in two steps:
\begin{itemize}
\item
First, the following equations of \eqref{bms3:6} are solved,
\begin{eqnarray}\label{bms3:7}
\mathcal L_{\xi}g_{rr}=0,~~\mathcal L_{\xi}g_{r\phi}=0,~~\mathcal L_{\xi}g_{\phi\phi}=0.
\end{eqnarray}
The two first equations of \eqref{bms3:7} give differential equations with respect to $r$ for the components of the gauge parameter ($\xi^u,\xi^\phi$) and can be integrated exactly, up to some integration functions (with respect to $r$). The third equation in \eqref{bms3:7} is an algebraic equation for $\xi^r$.
Explicitly,
\begin{align}
\partial_r\xi^u&=0,\hspace{1cm}\Rightarrow\xi^u=f(u,\phi),\label{bms3:8}\\
\partial_r\xi^\phi&=\dfrac{1}{r^2}e^{2\beta}\partial_\phi f\hspace{1cm}\Rightarrow\xi^\phi=Y(u,\phi)-\partial_\phi f\int_{r'}^{+\infty}dr' \dfrac{e^{2\beta}}{r'^2},\label{bms3:9}\\
\xi^r&=-r(\partial_\phi\xi^\phi-U\partial_\phi f),\label{bms3:10}
\end{align}
with $f(u,\phi), Y(u,\phi)$ arbitrary integration functions.
\item
The second step in finding the symmetry parameter, is to solve the remaining equations of \eqref{bms3:6}, i.e.
\begin{eqnarray}\label{bms3:414}
\mathcal L_{\xi}g_{u\phi}=\mathcal O(1),~~\mathcal L_{\xi}g_{uu}=\mathcal O(1),
\end{eqnarray}
while for the component $g_{ur}$ the right hand side depends on the gauge fixing condition on $\beta$,
\begin{align}
\mathcal L_{\xi}g_{ur}&=\mathcal O(r^{-1})\hspace{1cm}\text{(asymptotically flat case)},\label{bms3:415}\\
\mathcal L_{\xi}g_{ur}&=\mathcal O(r^{-2})\hspace{1cm}\text{(asymptotically adS, dS cases)}.\label{bms3:416}
\end{align}
The equations \eqref{bms3:415} and \eqref{bms3:416} give the same relation on the $u$ derivative of $f$,
\begin{eqnarray}\label{bms3:417}
\partial_uf=\partial_\phi Y.
\end{eqnarray}
The first equation of \eqref{bms3:414} implies
\begin{eqnarray}\label{bms3:418}
\partial_u Y=-\dfrac{\alpha}{l^2}\partial_\phi f,
\end{eqnarray}
while the last equation of \eqref{bms3:414} does not impose any new condition on the gauge parameter.\\
\end{itemize}

Therefore, the most general gauge parameter leaving invariant the form of the BMS metric ansatz in three dimensions \eqref{5:1} is given by \eqref{bms3:8}-\eqref{bms3:10} with integration functions $f,Y$ subject to \eqref{bms3:417},\eqref{bms3:418}.\\
Note that as in the four dimensional case, the gauge parameter $\xi$ depends explicitly on the metric throught the functions $U,\beta$ present in $\xi^\phi$ and $\xi^r$.\\

At future null infinity ($\lim_{r\to \infty}$ with coordinates $u,\phi$), the vector field $\xi$ defined by the components \eqref{bms3:8}-\eqref{bms3:10} reduces to
\begin{eqnarray}\label{5:bms}
\bar\xi=f\partial_u+Y\partial_\phi.
\end{eqnarray}
The algebra of this field independent gauge parameter is found in the same way as in the four dimensional case, i.e. by considering the commutator of two variations acting on the metric field. In the case of the vector field \eqref{5:bms}, let the $u$ and $\phi$ components of the Lie bracket be denoted by $\hat f,\hat Y$, respectively.
\begin{eqnarray}\label{bms3:algebra}
[\bar\xi_1,\bar\xi_2]\equiv\hat f\partial_u+\hat Y\partial_\phi.
\end{eqnarray}
A computation shows that these components of the Lie bracket associated with the gauge parameter \eqref{5:bms} are
\begin{align}
\hat Y&=Y_1\partial_\phi Y_2-Y_2\partial_\phi Y_1-\dfrac{\alpha}{l^2}\left(f_1\partial_\phi f_2-f_2\partial_\phi f_1\right),\label{bms3:421}\\
\hat f&=Y_1\partial_\phi f_2-Y_2\partial_\phi f_1+f_1\partial_\phi Y_2-f_2\partial_\phi Y_1.\label{bms3:422}
\end{align}
These relations define the symmetry algebra and can be written more compactly as
\begin{align}\label{bms3:al}
[(f_1,Y_1),(f_2,Y_2)]=(\hat f,\hat Y),
\end{align}
with $\hat f,\hat Y$ defined by \eqref{bms3:421} and \eqref{bms3:422}.\\
This result is true for the three cases (flat, anti-de Sitter, de Sitter) in a unified way. Let us stress that this does not mean that the algebra is the same  in the three cases! The explicit form of the right hand side of \eqref{bms3:421} depends on the parameter $\alpha$ explicitly, but also implicitly through relations \eqref{bms3:417}-\eqref{bms3:418}, so that the three symmetry algebras are of course not the same.

\subsection{Realization of the symmetry algebra in the bulk}
Note that the vector field, given in components by \eqref{bms3:8}-\eqref{bms3:10}
\begin{align}\label{bms3:424}
\xi=f\partial_u&-r\left(\partial_\phi \left[Y-\partial_\phi f\int_r^\infty dr'\dfrac{e^{2\beta}}{r'^2}\right]-U\partial_\phi f\right)\partial_r+\notag\\
&+\left(Y-\partial_\phi f\int_r^\infty dr'\dfrac{e^{2\beta}}{r'^2}\right)\partial_\phi,
\end{align}
is defined everywhere in the bulk of the spacetime (i.e. for all coordinates $u,r,\phi$) and is moreover field dependent.\\

As in the four dimensional case, the algebra of this field dependent gauge parameter can be found by considering the commutator of two variations acting on the metric field. This will give rise to the modified bracket,
\begin{eqnarray}\label{bms3:m}
[\xi_1,\xi_2]_M=[\xi_1,\xi_2]+\delta_{\xi_1}\xi_2(g)-\delta_{\xi_2}\xi_1(g),
\end{eqnarray}
 in which the two corrections terms are due to the fact that the second variation in the commutator acts not only on the metric, but also on the metric-dependent gauge parameter of the first variation. The proof of \eqref{bms3:m} is the same as in the four dimensional case\footnote{The proof given in section 2.1.2 does actually not depend on the spacetime dimensions.}.\\
When using the modified bracket, one can show that the bulk vector field \eqref{bms3:8}-\eqref{bms3:10} form a representation of the symmetry algebra,
\begin{align}
[\xi_1,\xi_2]_M=
\hat f\partial_u&-r\left(\partial_\phi \left[\hat Y-\partial_\phi \hat f\int_r^\infty dr'\dfrac{e^{2\beta}}{r'^2}\right]-U\partial_\phi\hat  f\right)\partial_r+\notag\\
&+\left(\hat Y-\partial_\phi\hat  f\int_r^\infty dr'\dfrac{e^{2\beta}}{r'^2}\right)\partial_\phi.
\end{align}
The proof of this representation result is given in \cite{Barnich:2010eb} in the asymptotically flat case ($\alpha=0$), and consists in showing that the modified Lie bracket satisfies the same initial conditions and differential equations as the vector field \eqref{bms3:424}.\\

This result means that the bms$_3$ symmetry algebra is represented (throught the modified bracket) everywhere in the bulk of the spacetime, even though it has been defined at infinity.

\subsection{Mode expansions}
In order to recover the usual asymptotic symmetry algebra, at least for anti-de Sitter and flat cases, let us make a mode expansion of the integration functions $f,T$ present in the gauge parameter.

\subsubsection{Asymptotically flat: $\alpha=0$}
In the asymptotically flat case, the relations \eqref{bms3:417} and \eqref{bms3:418} simplify to
\begin{eqnarray}
\partial_u f=\partial_\phi Y,\hspace{1cm}\partial_uY=0,
\end{eqnarray}
so that the first relation can be integrated to give $f=T(\phi)+u\partial_\phi Y$. The asymptotic symmetry parameter $\bar\xi$ \eqref{5:bms} of the BMS gauge in the asymptotically flat case is therefore
\begin{eqnarray}
\bar\xi_{T,Y}=(T(\phi)+u\partial_\phi Y)\partial_u+Y\partial_\phi.
\end{eqnarray}
Let the generators be expanded in the following way,
\begin{align}
l_n=\{\bar\xi_{T,Y}:\hspace{0.5cm}T=0,Y=e^{in\phi}\},\\
t_n=\{\bar\xi_{T,Y}:\hspace{0.5cm}T=e^{in\phi},Y=0\}.
\end{align}
The non-vanishing commutation relations \eqref{bms3:algebra} of the asymptotically flat algebra are 
\begin{eqnarray}
i[l_n,l_m]=(n-m)l_{m+n},\hspace{1cm}i[l_n,t_m]=(n-m)t_{n+m}.
\end{eqnarray}
This is the symmetry algebra of asymptotically flat spacetimes in three dimensions.

\subsubsection{Asymptotically anti-de Sitter: $\alpha=-1$}
In the asymptotically anti-de Sitter case, the relations \eqref{bms3:417} and \eqref{bms3:418} become
\begin{eqnarray}
\partial_u f=\partial_\phi Y,\hspace{1cm}\partial_uY=\dfrac{1}{l^2}\partial_\phi f.
\end{eqnarray}
These relations are equivalent to
\begin{eqnarray}
\left(\partial_u^2-\frac{1}{l^2}\partial_\phi\right)f=0,\hspace{1cm}\left(\partial_u^2-\frac{1}{l^2}\partial_\phi\right)Y=0,\label{finidans3jours}
\end{eqnarray}
and can be solved for $f,Y$ by introducing coordinates $x^\pm=\frac{u}{l}\pm\phi$. The wave equations \eqref{finidans3jours} become $\frac{4}{l^2}\partial_+\partial_-f=0$ and  $\frac{4}{l^2}\partial_+\partial_-Y=0$ and can be integrated for $f,Y$ in terms of two arbitrary functions $Y^+(x^+),Y^-(x^-)$,
\begin{eqnarray}
f=\frac{l}{2}\left(Y^+(x^+)+Y^-(x^-)\right),\hspace{1cm}Y=\dfrac{1}{2}\left(Y^+(x^+)-Y^-(x^-)\right).
\end{eqnarray}
 The symmetry parameter $\bar\xi$ \eqref{5:bms} of the BMS gauge in the asymptotically anti-de Sitter case therefore becomes
\begin{eqnarray}
\bar\xi_{Y^+,Y^-}=Y^+\partial_+~+~Y^-\partial_-.
\end{eqnarray}
Let the generators be expanded in the following way,
\begin{align}
l_n^+=\{\bar\xi_{Y^+,Y^-}:\hspace{0.5cm}Y^+=e^{inx^+},Y^-=0\},\\
l_n^-=\{\bar\xi_{Y^+,Y^-}:\hspace{0.5cm}Y^+=0,Y^-=e^{inx^-}\}.
\end{align}
The non-vanishing commutation relation \eqref{bms3:algebra} of the asymptotically anti-de Sitter algebra thus become two copies of the Witt algebra,
\begin{eqnarray}
i[l_n^+,l_m^+]=(n-m)l_{m+n}^+,\hspace{1cm}i[l_n^-,l_m^-]=(n-m)l_{m+n}^-.
\end{eqnarray}

\section{BMS$_3$ equations of motion}
In this section, the equations of motion are solved in BMS gauge \eqref{5:1}. In order to do so,
the equations of motion are organized in the same way as in four dimensions.
The reason for doing so is both physical and technical. Physical, because it allows to go back and forth between the four dimensional and three dimensional cases to compare the results concerning the physical quantities that are the news of the system, the mass, and the angular momentum. Technical, because this way of arranging the equations of motion is very efficient.\\

\subsection{Integration procedure}
The  relation between the Bianchi identities \eqref{4:75} holds in any dimensions and with or without a cosmological constant,
\begin{eqnarray}\label{bms3:339BMS3}
g^{\alpha\epsilon}\left(E_{\mu\alpha,\epsilon}-\dfrac{1}{2}E_{\alpha\epsilon,\mu}-\Gamma^{\sigma}_{\alpha\epsilon}E_{\mu\sigma}\right)=0,
\end{eqnarray}
with $E_{\mu\nu}\equiv R_{\mu\nu}-\frac{2\alpha}{l^2}g_{\mu\nu}$ in three dimensions.
When the three following equations of motion (called the main equations) are satisfied,
\begin{eqnarray}
E_{rr}=0,\hspace{1cm}E_{r\phi}=0,\hspace{1cm}E_{\phi\phi}=0,\label{bms3:440main}
\end{eqnarray}
then the three Bianchi identities \eqref{bms3:339BMS3} reduce to
\begin{align}
\mu=r:\hspace{1cm}&-g^{\alpha\epsilon}\Gamma^u_{\alpha\epsilon}E_{ur}=0,\label{eq:bianchi1BMS3}\\
\mu=\phi:\hspace{1cm}&g^{ur}(E_{\phi u,r}-E_{ur,\phi})-g^{\alpha\epsilon}\Gamma^u_{\alpha\epsilon}E_{u\phi}=0,\label{eq:bianchi2BMS3}\\
\mu=u:\hspace{1cm}&
g^{ur}E_{uu,r}+g^{rr}E_{ur,r}+g^{r\phi}(E_{ur,\phi}+E_{u\phi,r})+g^{\phi\phi}E_{u\phi,\phi}-\notag\\
&-g^{\alpha\epsilon}\Gamma^u_{\alpha\epsilon}E_{uu}-g^{\alpha\epsilon}\Gamma^r_{\alpha\epsilon}E_{ur}-g^{\alpha\epsilon}\Gamma^\phi_{\alpha\epsilon}E_{\phi u}=0.\label{eq:bianchi3BMS3}
\end{align}
From the metric \eqref{5:2} one has $g^{\alpha\epsilon}\Gamma^u_{\alpha\epsilon}=\frac{e^{-2\beta}}{r}$ and the Bianchi identity \eqref{bms3:339BMS3} then implies
\begin{eqnarray}
E_{ur}=0\label{eq:trivial}.
\end{eqnarray}
Equation of motion $E_{ur}$ is therefore called the trivial equation, because it is automatically satisfied as consequence of the main equations \eqref{bms3:440main} only.
This result in the Bianchi identity \eqref{eq:bianchi2BMS3} implies
\begin{align}
-e^{-2\beta}\left(\partial_r+\dfrac{1}{r}\right)E_{\phi u}&=0\notag\\
-\dfrac{e^{-2\beta}}{r}\partial_r(rE_{\phi u})&=0\hspace{1cm}\Rightarrow\hspace{1cm}
E_{\phi u}=\dfrac{\text{f}(u,\phi)}{r}\label{eq:supppl1},
\end{align}
where $f(u,\phi)$ is an arbitrary function of its arguments.
When the main equations are satisfied, the function $E_{u\phi}$ contains only terms of order $\mathcal O(r^{-1})$.\\
Finally, the last Bianchi identity \eqref{eq:bianchi3BMS3} becomes
\begin{align}
-e^{-2\beta}\left(\partial_r+\dfrac{1}{r}\right)E_{uu}&=0\notag\\
-\dfrac{e^{-2\beta}}{r}\partial_r(rE_{uu})&=0\hspace{1cm}\Rightarrow\hspace{1cm}
E_{uu}=\dfrac{\text{g}(u,\phi)}{r}\label{eq:supppl2},
\end{align}
where $g(u,\phi)$ is an arbitrary function of its arguments.
\\

In summary, the Bianchi identities \eqref{bms3:339BMS3} allow one to organize the equations of motion for the metric field in a very convenient way, similar to the four dimensional case. Indeed, once the main equations $E_{rr}=0, E_{r\phi}=0, E_{\phi\phi}=0$ are solved, then $E_{ur}=0$ is trivially satified and $E_{\phi u}$ is of the form $E_{u\phi}=\frac{f}{r}$, and solving $E_{u\phi}=0$ reduces $f$ to zero. Finally, equation $E_{uu}$ is of the form $E_{uu}=\frac{g}{r}$, and solving $E_{uu}=0$ restricts $g$ to be zero.

\subsection{General solution}
In this section, the Einstein's field equations of motion
\begin{eqnarray}\label{bms3:427}
R_{\mu\nu}-\dfrac{2\alpha}{l^2}g_{\mu\nu}=0,
\end{eqnarray}
are solved in a unified\footnote{Both for $\alpha=0$ and $\alpha\neq0$.} way for the BMS metric ansatz \eqref{5:1} with boundary conditions \eqref{5:2} and \eqref{bms3:53}. The boundary condition for the function $V$ is not needed in the integration procedure, as it is a consequence of the remaining boundary conditions.\\

Following the general procedure presented in section 3.3.1, the three main equations are first solved.
\begin{eqnarray}
R_{rr}=0:\hspace{1cm}\dfrac{2}{r}\partial_r\beta=0.
\end{eqnarray}
This equation implies that the function $\beta$ is zero, due to the boundary condition \eqref{bms3:53}. The next main equation to solve is
\begin{eqnarray}
R_{r\phi}=0:\hspace{1cm}\dfrac{1}{2r}\partial_r(r^3\partial_rU)=0.
\end{eqnarray}
Using the boundary condition \eqref{5:2}, this equation can be integrated twice to give $U=-\dfrac{N(u,\phi)}{2r^2}$ with $N(u,\phi)$ an integration function.\\
The last main equation depends on $\alpha$,
\begin{eqnarray}
R_{\phi\phi}-\dfrac{2\alpha}{l^2}g_{\phi\phi}=0:\hspace{1cm}\partial_r\left(\dfrac{V}{r}\right)=\dfrac{N^2}{2r^3}+\dfrac{2\alpha}{l^2}r.
\end{eqnarray}
Integration gives $\dfrac{V}{r}=\dfrac{\alpha r^2}{l^2}+M(u,\phi)-\dfrac{N^2}{4r^2}$, with $M(u,\phi)$ an integration function.\\
One can check that the trivial equation of motion $R_{ur}-\dfrac{2\alpha}{l^2}g_{ur}=0$ is satisfied, as a consequence of the main equations.\\
The last equations to be solved are the two supplementary equations. Using the previous results, one finds for $E_{u\phi}=0$,
\begin{eqnarray}\label{bms3:431}
R_{u\phi}-\dfrac{2\alpha}{l^2}g_{u\phi}=0:\hspace{1cm}\dfrac{\partial_\phi M}{2r}-\dfrac{\partial_u N}{2r}=0.
\end{eqnarray}
Recall that there are only terms of order $\mathcal O(r^{-1})$ in this equation, due to \eqref{eq:supppl1} as a consequence of the main equation. Equation \eqref{bms3:431} gives the time evolution of the function $N$,
\begin{eqnarray}\label{bms3:432}
\partial_uN=\partial_\phi M.
\end{eqnarray}
The last equation of motion to solve is
\begin{eqnarray}\label{bms3:431}
R_{uu}-\dfrac{2\alpha}{l^2}g_{uu}=0:\hspace{1cm}-\alpha\dfrac{\partial_\phi N}{2l^2r}-\dfrac{\partial_u M}{2r}=0.
\end{eqnarray}
This equation only involves terms of order $\mathcal O(r^{-1})$,  due to \eqref{eq:supppl2}, and depends on $\alpha$. This equation gives the time evolution of the function $M$,
\begin{eqnarray}\label{bms3:434}
\partial_u M=-\dfrac{\alpha}{l^2}\partial_\phi N.
\end{eqnarray}

The general solution of the equations of motion \eqref{bms3:427} in BMS gauge \eqref{5:1} is therefore
\begin{eqnarray}\label{bms3:435}
ds^2=\left(\alpha\dfrac{ r^2}{l^2}+M(u,\phi)\right) du^2-2dudr+N(u,\phi)dud\phi+r^2d\phi^2,
\end{eqnarray}
with functions $M,N$ satisfying the evolution equations \eqref{bms3:432},\eqref{bms3:434}. In the asymptotically flat case, this solution was derived in \cite{Barnich:2010eb}.

\subsection{Equations of motion and asymptotic symmetries}
The general solution to the equations of motion for the BMS gauge in three dimensions depends on two integration functions $M,N$. On the other side, the asymptotic symmetries associated with the BMS gauge also depend on two integration functions $f,Y$. The time evolution of all these integration functions are fixed, and are
\begin{align}
\partial_u f&=\partial_\phi Y,\hspace{1cm}\partial_u Y=-\dfrac{\alpha}{l^2}\partial_\phi f,\\
\partial_u N&=\partial_\phi M,\hspace{1cm}\partial_uM=-\dfrac{\alpha}{l^2}\partial_\phi N.\label{bms3:449}
\end{align}
These relations show the close relation between parameters characterizing the symmetry algebra (functions $f,Y$) and  those characterizing the general solution (functions $M,N$).

\subsection{On-shell expansion of symmetry generators}
On-shell, the symmetry generator  \eqref{bms3:424} can be written
\begin{align}\label{bms3:438}
\xi=f\partial_u&+\left(-r\partial_\phi Y+\partial_\phi^2 f-\dfrac{N}{2r}\partial_\phi f\right)\partial_r+\left(Y-\dfrac{\partial_\phi f}{r}\right)\partial_\phi.
\end{align}
Equation \eqref{bms3:438} shows that the perturbative expansion in inverse power of $r$ of the symmetry parameter associated to the BMS gauge stops at next-to-next to leading order.

\subsection{Action of asymptotic symmetries on solution space}
The solution \eqref{bms3:435} to the equations of motion for the BMS gauge is characterize by two functions $M,N$ subject to relations \eqref{bms3:449}. In the next section, the algebra of the surface charge will be computed. In order to do so, the variation of the data $M,N$ is needed and computed via the on-shell Lie derivative of the metric. More precisely,
\begin{eqnarray}
\mathcal L_{\xi}g_{uu}=-\delta g_{uu}=-\delta M,\hspace{1cm}\mathcal L_{\xi}g_{u\phi}=-\delta g_{u\phi}=-\dfrac{\delta N}{2}.
\end{eqnarray}
The evaluation of these Lie derivatives on-shell with $\xi$ given in \eqref{bms3:438} is
\begin{align}
-\delta M&=(f\partial_u+Y\partial_\phi+2\partial_\phi Y)M-2\partial_\phi^3Y-2\dfrac{\alpha}{l^2}N\partial_\phi f,\\
-\delta N&=(f\partial_u+Y\partial_\phi+2\partial_\phi Y)N-2\partial_\phi^3f+2M\partial_\phi f.
\end{align}
In these transformation laws, the field independent part of the inhomogeneous term is responsible for the presence of a central extension in the algebra of the surface charges.

\section{BMS$_3$ surface charges}
In this section, the surface charges associated to the BMS symmetry generator are computed. To perform the computation, the assumption of asymptotic linearity is not made but it is the variation of the charge that is really computed, and then shown to be integrable. The charge is computed for the three regimes (flat, anti-de Sitter, de Sitter) and is shown \cite{P-HLaura} to be independent or the $r$ coordinate.

\subsection{Computation of the surface charges}
The covariant formalism of Barnich-Brandt-Comp\`ere \cite{Barnich:2001jy,Barnich:2007bf} is used in order to compute the surface associated with the on-shell solution \eqref{bms3:435}
\begin{eqnarray}\label{bms3:sol}
ds^2=\left(\alpha\dfrac{r^2}{l^2}+M(u,\phi)\right) du^2-2dudr+N(u,\phi)dud\phi+r^2d\phi^2,
\end{eqnarray}
and associated to the on-shell asymptotic symmetry vector field \eqref{bms3:424}.\\
More precisely \cite{Barnich:2001jy,Barnich:2007bf}, to any vector field $\xi$ is associated a 1-form $\bdelta \mathcal Q_\xi$ depending on a solution $g$, and its variation $\delta g$, and is defined
\footnote{
The form $k_\xi$ in \eqref{bms3:kxi} can be defined in terms of a symplectic form by $dk_\xi=W$ with the symplectic form $W$ defined by $W=\frac{1}{2}I_{\delta\phi}^n\left(\frac{\delta L}{\delta\phi}\delta\phi\right)$, where $I_{\delta\phi}^n$ is the homotopy operator acting on an $n$ form. This expression of $k_\xi$ differs from the one proposed by Iyer-Wald \cite{Iyer:1994ys}, $dk_\xi^{IW}=\omega^{IW}$, with the symplectic form $\omega^{IW}$ defined by $\omega^{IW}=\delta (I_{\delta\phi}^nL)$. The two different symplectic forms differ by a total derivative term, $W=\omega^{IW}+dE$ with $E=\frac{1}{2}(I_{\delta\phi}^{n-1}I_{\delta\phi}^nL)$. This $E$ term plays no role in the case of exact symmetries but is however relevant in the asymptotic context. See \cite{Barnich:2007bf} appendix A5 for more detail.
} 
in three spacetime dimensions by
\begin{align}
\bdelta\mathcal Q_{\xi}(g,\delta g)~=~&\dfrac{1}{16\pi G}\int r d\phi ~~k_\xi,\label{bms3:bdQ}\\
&\notag\\
k_\xi~=~&\xi^r\left(D^u h-D_\sigma h^{u\sigma}+D^rh^u_{\phantom ur}-D^uh^r_{\phantom rr}\right)-\notag\\
-&\xi^u\left(D^r h-D_\sigma h^{r\sigma}-D^rh^u_{\phantom uu}+D^uh^r_{\phantom ru}\right)+\notag\\
+&\xi^\phi\left(D^rh^u_{\phantom u\phi}-D^uh^r_{\phantom u\phi}\right)+\dfrac{1}{2}h\left(D^r\xi^u-D^u\xi ^r\right)+\notag\\
+&\dfrac{1}{2}h^{r\sigma}\left(D^u\xi_\sigma-D_\sigma\xi^u\right)-\dfrac{1}{2}h^{u\sigma}\left(D^r\xi_\sigma-D_\sigma\xi^r\right),\label{bms3:kxi}
\end{align}
where $h_{\mu\nu}=\delta g_{\mu\nu},h^\mu_{\phantom\mu\nu}=g^{\mu\sigma}\delta g_{\sigma\nu}, h=h^\sigma_{\phantom\sigma\sigma}$.\\
In the case of the solution $g$ given by \eqref{bms3:sol}, the only non-vanishing $h_{\mu\nu}$ are $h_{uu}=\delta M$ and $h_{u\phi}= \dfrac{\delta N}{2}$. Evaluating $k_\xi$ in \eqref{bms3:kxi} gives
\begin{align}
k_\xi&=f\dfrac{\delta M}{r}-f\dfrac{\partial_\phi\delta N}{2r^2}+\left(Y-\dfrac{\partial_\phi f}{r}\right)\dfrac{\delta N}{2r}+\delta N\dfrac{Y}{2r},\\
&=\dfrac{1}{r}\left(f\delta M+Y\delta N\right)-\dfrac{1}{2r^2}\partial_\phi\left(f\delta N\right),\label{bms3:4:27}
\end{align}
so that \eqref{bms3:bdQ} is
\begin{eqnarray}\label{bms3:4:28}
\bdelta\mathcal Q_{\xi}(g,\delta g)~=~\dfrac{1}{16\pi G}\int d\phi\left(f\delta M+Y\delta N\right).
\end{eqnarray}
Note that the second term of $k_{\xi}$ in \eqref{bms3:4:27} is a total derivative and vanishes when integrated along the $\phi$ coordinate. Because the right-hand side of \eqref{bms3:4:28} is made of $\delta$ exact terms\footnote{Recall that the variation $\delta$ only acts on the field and not on the gauge parameters, when these latter are field independent.}, the charge associated to the BMS gauge in three dimensions is integrable, and reads
\begin{eqnarray}\label{bms3:4:29}
\mathcal Q_{\xi}~=~\dfrac{1}{16\pi G}\int d\phi\left(f M+Y N\right).
\end{eqnarray}
\noindent Some remarks concerning this result for the charge are in order.
\begin{itemize}
\item The final result for the charge, equation \eqref{bms3:4:29} is $r$ independent \cite{P-HLaura}. This means that one can evaluate the charge on a closed circle at any radius $r$ and get the same result.
\item The result \eqref{bms3:4:29} is valid for asymptotically flat, anti-de Sitter and de Sitter cases. Indeed, the only difference between these three cases in the BMS symmetry parameter $\xi$  shows up in the relations between the integration functions $f,Y$, but not in the expression \eqref{bms3:4:29} for the charge.
\item Equation \eqref{bms3:4:29} makes explicit the isomorphism between the integration functions of the asymptotic symmetries ($f,Y$) and the integration functions of the solution to the equations of motion ($M,N$). More precisely, the charge $Q_\xi$ in \eqref{bms3:4:29} provides an inner product between the space of solutions and the asymptotic symmetries.
\end{itemize}

\subsection{Algebra of surface charges and central extension}
The algebra of surface charges can be computed with the Poisson bracket given by
\begin{eqnarray}
\{\mathcal Q_{\xi_1},\mathcal Q_{\xi_2}\}=-\delta_{\xi_2}\mathcal Q_{\xi_1}.
\end{eqnarray}
One finds
\begin{align}
-\delta_{\xi_2}\mathcal Q_{\xi_1}&=\dfrac{1}{16\pi G}\int d\phi\left[f_1(-\delta_2 M)+Y_1(-\delta_2 N)\right],\notag\\
&=\dfrac{1}{16\pi G}\int d\phi\left(\hat f M+\hat Y N\right)-\dfrac{1}{8\pi G}\int d\phi (f_1\partial_\phi^3Y_2+Y_1 \partial_\phi^3f_2),\label{bms3:462}
\end{align}
with $\hat f,\hat Y$ defined in \eqref{bms3:421} and \eqref{bms3:422}. This can be rewritten as
\begin{eqnarray}\label{bms3:463}
\{\mathcal Q_{\xi_1},\mathcal Q_{\xi_2}\}=\mathcal Q_{[\xi_1,\xi_2]}+\mathcal K_{\xi_1,\xi_2},
\end{eqnarray}
where $\mathcal K_{\xi_1,\xi_2}$ is by definition the second term of \eqref{bms3:462}. The algebra of the surface charges associated with the BMS gauge is centrally extended, in the three cases (flat, anti-de Sitter, de Sitter). In the asymptotically flat and anti-de Sitter cases, the same mode expansion as in section 3.3 can be made.

\subsubsection{Mode expansion in asymptotically flat case}
In the asymptotically flat case, let the charges associated with the generators $l_n,t_n$ be denoted by
\begin{align}
L_n=\{\mathcal Q_{T,Y}:\hspace{0.5cm}T=0,Y=e^{in\phi}\},\\
T_n=\{\mathcal Q_{T,Y}:\hspace{0.5cm}T=e^{in\phi},Y=0\}.
\end{align}
The charge algebra \eqref{bms3:463} becomes
\begin{align}
i\{L_n,L_m\}&=(n-m)L_{m+n},\\
i\{L_n,T_m\}&=(n-m)T_{m+n}+\dfrac{1}{4G}n^3\delta_{m+n,0},\label{bms3:467}
\end{align}
where $\delta_{k,0}=\frac{1}{2\pi}\int d\phi~e^{ik\phi}$.
From \eqref{bms3:467}, the classical central charge is identified\footnote{Usually, $i\{L_n,T_m\}=(n-m)T_{m+n}+\frac{c}{12}n^3\delta_{n+m,0}$.} to be $c=\frac{3}{G}$.

\subsubsection{Mode expansion in asymptotically anti-de Sitter case}
In the asymptotically anti-de Sitter case, with the same notation as in section 4.3, let the charges associated with the generators $l_n,t_n$ be denoted by,
\begin{align}
L_n^+=\{\mathcal Q_{Y^+,Y^-}:\hspace{0.5cm}Y^+=e^{inx^+},Y^-=0\},\\
L_n^-=\{\mathcal Q_{Y^+,Y^-}:\hspace{0.5cm}Y^+=0,Y^-=e^{inx^-}\}.
\end{align}
The charge algebra \eqref{bms3:463} becomes two copies of the (classical) Virasoro algebra,
\begin{align}
i\{L_n^+,L_m^+\}=(n-m)L_{m+n}^++\dfrac{c}{12}n^3\delta_{m+n,0},\\
i\{L_n^-,L_m^-\}=(n-m)L_{m+n}^-+\dfrac{c}{12}n^3\delta_{m+n,0}.
\end{align}
The two classical central charges are the same and equal to $c=\frac{3l}{2G}$.

\part{Original contributions}

\newpage
\chapter{The Newman-Unti gauge}

The symmetry algebra of asymptotically flat
    spacetimes at null infinity in four dimensions in the sense of
    Newman and Unti is revisited. As in the Bondi-Metzner-Sachs gauge,
    it is shown to be isomorphic to the direct sum of the abelian
    algebra of infinitesimal conformal rescalings with
    $\mathfrak{bms}_4$. The latter algebra is the semi-direct sum of
    infinitesimal supertranslations with the conformal Killing vectors
    of the Riemann sphere. Infinitesimal local conformal
    transformations can then consistently be included. We work out the
    local conformal properties of the relevant Newman-Penrose
    coefficients, construct the surface charges and derive their
    algebra.

\section{Motivation}
\label{sec:introduction}

The definitions of asymptotically flat four dimensional
space-times at null infinity by Bondi-Van der Burg-Metzner-Sachs
\cite{Bondi:1962px,Sachs:1962wk} (BMS) and Newman-Unti (NU)
\cite{newman:891} in 1962 merely differ by the choice of
the radial coordinate. Such a change of gauge should not affect
the asymptotic symmetry algebra if, as we contend, this concept is
to have a major physical significance.\\

The problem of comparing the symmetry algebra in both cases is that,
besides the difference in gauge, the very definitions of these
algebras are not the same. Indeed, NU allow the leading part of the
metric induced on Scri to undergo a conformal rescaling. When this
generalization is considered in the BMS setting, it turns out that the
symmetry algebra is the direct sum of the BMS algebra
$\mathfrak{bms}_4$ \cite{Sachs2} with the abelian algebra of
infinitesimal conformal rescalings \cite{Barnich:2009se},
\cite{Barnich:2010eb}. There are two novel and independent aspects in
this computation.

\begin{itemize}

\item The first concerns the fact that the BMS algebra in $4$
  dimension involves the conformal Killing vectors of the unit, or
  equivalently, the Riemann sphere and can consistently accommodate
  infinitesimal local conformal transformations. The symmetry algebra
  $\mathfrak{bms}_4$ then involves two commuting copies of the non
  centrally extended Virasoro algebra, called superrotations in
  \cite{Barnich:2011ct}, and simultaneously the supertranslations
  generators are expanded in Laurent series. The standard, globally
  well-defined symmetry algebra $\mathfrak{bms}^{ \rm glob}_4$ consists in
  restricting to the globally well-defined conformal Killing vectors
  of the sphere which correspond to infinitesimal Lorentz
  transformation, while the supertranslation generators are expanded
  into spherical harmonics.

  This local versus global versions of the symmetry algebra are of
  course not related to the BMS gauge choice, but will also occur in
  alternative characterizations of the asymptotic symmetry algebra
  where the conformal Killing vectors of the sphere play a
  role. Examples of this are the geometrical approach of Geroch
  \cite{Geroch:1977aa} based on Penrose's definition of null infinity
  \cite{PhysRevLett.10.66} and also, as we will explicitly discuss in
  this chapter, the asymptotic symmetries in the NU framework.

\item The second aspect is related to the modified Lie bracket that
  should be used when the vector fields parametrising infinitesimal
  diffeomorphisms depend explicitly on the metric. Indeed, when using
  the modified Lie bracket, the space-time vectors realize the
  asymptotic symmetry algebra everywhere in the bulk and furthermore,
  even on Scri, this bracket is needed to disentangle the algebra when
  conformal rescalings of the induced metric on Scri are
  allowed. Similarly, in the context of the AdS/CFT correspondence,
  this bracket allows one to realize the asymptotic symmetry algebra
  in the bulk and to disentangle the symmetry algebra at infinity when
  considering transformations that leave the Fefferman-Graham ansatz
  invariant only up to conformal rescaling of the boundary metric
  \cite{Imbimbo:1999bj}. From a mathematical point of view, the
  modified Lie bracket is the natural bracket of the Lie algebroid
  that is associated to any theory with gauge invariance
  \cite{Barnich:2010xq}.

\end{itemize}

What we will do in this chapter is to re-derive from scratch the
asymptotic symmetry algebra in the NU framework by focusing on metric
aspects and on the two novel features discussed above. As expected,
the symmetry algebra is again the direct sum of $\mathfrak{bms}_4$
with the abelian algebra of infinitesimal conformal rescalings of the
metric on Scri and thus coincides, as it should, with the generalized
symmetry algebra in the BMS approach. A related analysis of asymptotic
symmetries in the NU context from the point of view of Scri and
emphasizing global issues instead can be found in
\cite{0305-4470-11-1-012}, \cite{springerlink:10.1007/BF00669365}.\\

Even though the results presented here are not really surprising in
view of those in the BMS framework and the close relation between the
NU and BMS approaches, the exercise of working out the details is
justified because the NU framework is embedded in the context of the
widely used Newman-Penrose formalism \cite{newman:566} so that
explicit formulae in this context are directly relevant in many
applications, see e.g.~the  review articles
\cite{newman:1980xx}.\\

As a first application, we study the transformation properties of the
Newman-Penrose coefficients parametrizing the solution space in the NU
approach. Our main focus is on the inhomogeneous terms in the
transformation laws that contain the information on the central
extensions of the theory. We then discuss the associated surface
charges by following the analysis in the BMS gauge
\cite{Barnich:2011mi} and briefly compare with standard expressions
that can be found in the literature. The algebra of these charges is
derived and shown to involve field dependent central charges in
the case of $\mathfrak{bms}_4$ which vanish for $\mathfrak{bms}^{\rm
  glob}_4$.

\section{NU metric ansatz for asymptotically flat spacetimes}
\label{sec:from-bondi-van}

The metric ansatz of NU is based on a family of null hypersurfaces
labelled by the first coordinate, $x^0\equiv u={\rm const}$. The
second coordinate $x^1\equiv r$ is chosen as an affine parameter for
the null geodesic generators $l^\mu$ of these hypersurfaces, so that
$l^\mu=-\delta^\mu_r$. Up to a change of signature from $(+,-,-,-)$ to
$(-,+,+,+)$, a renumbering of the indices and the tetrad
transformation that makes $P_{NU}$ real, the line element considered
in chapter 4 of NU \cite{newman:891} can be written as
\begin{equation}
  \label{eq:1010}
  ds^2=Wdu^2-2dr du+
g_{AB}(dx^A-V^Adu)(dx^B-V^Bdu)\,,
\end{equation}
with associated inverse metric
\begin{equation}
g^{\mu\nu}=  \begin{pmatrix} 0 &
  -1 & 0 \\
  -1&
-W  & -V^B \\
0 & -V^A & g^{AB}
\end{pmatrix}\,,\label{eq:NU}
\end{equation}
where
\begin{equation}
g_{AB}dx^Adx^B=r^2\bar\gamma_{AB}dx^Adx^B
+r C_{AB}dx^Adx^B+o(r)\,,\label{eq:11}
\end{equation}
with $\bar\gamma_{AB}$ conformally flat. Below, we will use standard
stereographic coordinates $\zeta=\cot{\frac{\theta}{2}}e^{i\phi},\bar
\zeta$, $\bar\gamma_{AB}dx^Adx^B=e^{2{\tilde\varphi}} d\zeta
d\bar\zeta$, ${\tilde\varphi}={\tilde\varphi}(u,x)$.\\

In addition, the choice of origin for the affine parameter of the
null geodesics is fixed through the requirement that the
term proportional to $r^{-2}$ in the expansion of the spin
coefficient $-\rho=D_\rho l_\nu m^\rho\bar m^\nu$ is absent\footnote{$l_\mu$ and $m_\mu$ being two of the four complex null tetrad vectors.}.\\

When expressed in terms of the metric, one finds
\begin{equation}
\rho=-\frac{1}{4} g^{AB}g_{AB,r}=-\frac{1}{4}\d_r
\ln |g| =-r^{-1}+\frac{1}{4}C^A_A r^{-2}+o(r^{-2})\,, \label{eq:36}
\end{equation}
where $g={\rm det}\, g_{\rho\nu}$ and the index has been raised with
the inverse of $\bar\gamma_{AB}$.  The requirement is thus equivalent
to the condition
\begin{equation}
  \label{eq:38}
  C^A_A=0\,.
\end{equation}
In the following we denote by $\bar D_A$ the covariant derivative with
respect to $\bar \gamma_{AB}$ and by $\bar\Delta$ the associated
Laplacian and by $\bar R$ the scalar curvature. In complex coordinates
$\zeta,\bar\zeta$, $C_{\zeta\bar\zeta}=0$ and we define for later
convenience $C_{\zeta\zeta}=e^{2{\tilde\varphi}} c,C_{\bar\zeta\bar
  \zeta}=e^{2{\tilde\varphi}} \bar c$. Finally,
\begin{equation}
  \label{eq:12}
V^A=O(r^{-2}),\qquad  W=-2 r\d_u
{\tilde\varphi}+\bar\Delta {\tilde\varphi}+O(r^{-1})\,,
\end{equation}
where $\bar \Delta
{\tilde\varphi}=4 e^{-2{\tilde\varphi}}\d\bar\d{\tilde\varphi}$ with $\d=
\d_\zeta,\bar\d=\d_{\bar\zeta}$.\\

The more restrictive fall-off conditions in \cite{newman:891} are
relevant for integrating the field equations but play no role in the
discussion of the asymptotic symmetry algebra.

\section{Asymptotic symmetries in the NU approach}
\label{sec:asympt-symm-newm}

The infinitesimal NU transformations can be defined as those
infinitesimal transformations that leave the form \eqref{eq:NU} and
the fall-off conditions \eqref{eq:11}-\eqref{eq:12} invariant, up to a
rescaling of the conformal factor
$\delta{\tilde\varphi}(u,x^A)={\tilde\omega}(u,x^A)$. In other words, they
satisfy
\begin{equation}
  \label{eq:4aext1}
  \cL_\xi g^{uu}=0,\quad \cL_\xi g^{uA}=0,\quad \cL_\xi
  g^{ur}=0,
\end{equation}
\begin{equation}
  \label{eq:39}
  \d_r \Big[\frac{1}{\sqrt{|g|}}\d_\rho (\sqrt {|g|} \xi^\rho)\Big]=o(r^{-2})\,,
\end{equation}
\begin{equation}
\begin{gathered}
\cL_\xi g^{rA}=O( r^{-2}),\quad  \cL_\xi
g^{AB}=-2 {\tilde\omega} g^{AB}+O( r^{-3}),\\
\cL_\xi g^{rr} =2r\d_u{\tilde\omega} +2{\tilde\omega} \bar\Delta
{\tilde\varphi} -\bar\Delta  {\tilde\omega} + O( r^{-1})\,.\label{eq:4aext2}
\end{gathered}
\end{equation}
Equations \eqref{eq:4aext1} are equivalent to
\begin{equation}
  \label{eq:4aextbis}
 \d_r \xi^\nu=g^{\nu\rho}\d_\rho \xi^u\iff \left\{
\begin{array}{l}\d_r\xi^u=0\,,\\ \d_r \xi^A=\d_B\xi^u g^{BA}\,,\\
 \d_r\xi^r=-\d_u\xi^u-\d_A \xi^u V^A\,,
\end{array}\right.
\end{equation}
and are explicitly solved by
\begin{gather}
  \label{eq:26}
\left\{\begin{array}{l}
  \xi^u=f,\\
\xi^A=Y^A+I^A, \quad  I^A=- \d_B f\int_r^\infty dr^\prime
g^{AB},\\
\xi^r=-r \d_u f+Z+ J,\quad J=\d_A f \int_r^\infty dr^\prime V^A,
\end{array}\right.
\end{gather}
with $\d_r f=0=\d_r Y^A=\d_r Z$. Equation \eqref{eq:39} then implies
\begin{equation}
  \label{eq:9}
  Z=\half \bar\Delta f\,.
\end{equation}
The first equation of \eqref{eq:4aext2} requires $\d_u Y^A=0$, the
second that $Y^A$ is a conformal Killing vector of $\bar\gamma_{AB}$,
which amounts to
\begin{equation}
  \label{eq:21}
  Y^\zeta \equiv Y=Y(\zeta),\quad  Y^{\bar \zeta} \equiv \bar Y=\bar
  Y(\bar \zeta)\,,
\end{equation}
in the coordinates $\zeta,\bar\zeta$, and also that
\begin{equation}
  \d_u f =f\d_u {\tilde\varphi}+\half \tilde \psi\,,
\label{eq:44ter}
\end{equation}
with $\psi=\bar D_A Y^A $, or more
explicitly in
$\zeta,\bar\zeta$ coordinates, $\psi= \d Y+\bar
\d\bar Y+2 Y \d{\tilde\varphi} +2\bar Y \bar \d {\tilde\varphi}$, and
$\tilde \psi=\psi-2{\tilde\omega}$.
Finally, the last equation of \eqref{eq:4aext2} implies
\begin{equation}
  \label{eq:24}
  2(\d_u Z+Z\d_u{\tilde\varphi})=Y^A\d_A
  \bar\Delta{\tilde\varphi}+\psi\bar \Delta{\tilde\varphi}+2 \d_A
  f\bar\gamma^{AB} \d_B\d_u{\tilde\varphi}+f\bar\Delta \d_u{\tilde\varphi}
-\bar\Delta {\tilde\omega},
\end{equation}
which is identically satisfied when taking the previous relations into
account.\\

One approach is to consider that \eqref{eq:44ter} fixes
${\tilde\omega}$ in terms of $f$ and $Y$,
${\tilde\omega}=\half\psi+f\d_u{\tilde\varphi}-\d_u f$.  Consider Scri, the space $\scri$
with coordinates $u,\zeta,\bar\zeta$ and metric
\begin{equation}
ds^2_{\scri}=0du^2+e^{2{\tilde\varphi}}d\zeta d\bar \zeta\,.\label{eq:16}
\end{equation}
The NU algebra is then defined as the commutator algebra of the vector
fields
\begin{equation}
\bar \xi =f\dover{}{u}+Y^A\dover{}{x^A}\,,\label{eq:17}
\end{equation}
with $f=f(u,x^A)$ arbitrary and $Y^A(x)$ conformal Killing vectors of
a conformally flat metric in $2$ dimensions, or equivalently, the
algebra of conformal vector fields of the degenerate metric
\eqref{eq:16}.\\

This is not the symmetry algebra of asymptotically flat spacetimes in
the sense of NU however. Indeed, ${\tilde\varphi}$ is arbitrary, it can for
instance be considered as the finite ambiguity related to Penrose's
conformal approach
\cite{PhysRevLett.10.66,penrose:1964,Penrose:1965am} to null
infinity. One can then interpret ${\tilde\varphi}$ as part of the background
structure, or in other words, of the gauge fixing
\cite{Geroch:1977aa}, and compute the asymptotic symmetries for a
fixed choice of ${\tilde\varphi}$, i.e., ${\tilde\omega}=0$ in the formulae above, or
ask the more general question of how the asymptotic symmetries depend
on changes in ${\tilde\varphi}$ by an arbitrary infinitesimal amount
${\tilde\omega}$. In both cases, one has to consider \eqref{eq:44ter} as a
differential equation for $f$. As we now show, the symmetry algebra
will then be isomorphic to the trivially extended $\mathfrak{bms}_4$
algebra by the abelian algebra of infinitesimal conformal rescalings,
as it should, and as a consequence, the Poincar\'e algebra is embedded
therein in a natural way. Furthermore, there is a natural realization
of the asymptotic symmetry algebra on an asymptotically flat 4
dimensional bulk spacetime. Note also that, for ${\tilde\omega}=0$, equation
\eqref{eq:44ter} has been interpreted from the point of view of
Penrose's conformal approach to null infinity in
\cite{0305-4470-11-1-012} following \cite{Tamburino:1966} and related
to the preservation of null angles, which is the standard way
\cite{PhysRevLett.10.66,penrose:1964,Penrose:1974,springerlink:10.1007/BF00762453}
to recover the BMS algebra from geometrical data on Scri.\\

The general solution for \eqref{eq:44ter} reads
\begin{equation}
  \label{eq:14}
  f=e^{
    {\tilde\varphi}}\big[ \tilde T+
  \half\int_0^u du^\prime
  e^{- {\tilde\varphi}}\tilde \psi\big],\quad  \tilde T= \tilde T(\zeta,\bar \zeta)\,,
\end{equation}
and the general solution to equations
\eqref{eq:4aext1}-\eqref{eq:4aext2} defining the asymptotic symmetries
is given by $\xi^\rho$ as in \eqref{eq:26} where $Z$,$Y^A$,$f$ satisfy
\eqref{eq:9}, \eqref{eq:21}, \eqref{eq:14} with ${\tilde\omega}$
arbitrary. Asymptotic Killing vectors thus depend on
$Y^A,\tilde T,{\tilde\omega}$ and the metric,
$\xi=\xi[Y,\tilde T,{\tilde\omega};g]$.\\

For such metric dependent vector fields, consider on the one hand the
suitably modified Lie bracket taking the metric dependence of the
spacetime vectors into account,
\begin{equation}
  \label{eq:43}
  [\xi_1,\xi_2]_M=[\xi_1,\xi_2]-\delta^g_{
    \xi_1}\xi_2+\delta^g_{ \xi_2}\xi_1,
\end{equation}
where $\delta^g_{\xi_1}\xi_2$ denotes the variation in $\xi_2$ under
the variation of the metric induced by $\xi_1$, $\delta^g_{
  \xi_1}g_{\mu\nu}=\cL_{\xi_1}g_{\mu\nu}$.\\

Consider on the other hand the extended $\mathfrak{bms}_4$ algebra,
i.e., the semi-direct sum of the algebra of conformal Killing vectors
of the Riemann sphere with the abelian ideal of infinitesimal
supertranslations, trivially extended by infinitesimal conformal
rescalings of the conformally flat degenerate metric on Scri. More
explicitly, the commutation relations are given by\\
$[(Y_1,\tilde T_1,{\tilde\omega}_1),(Y_2,\tilde T_2,{\tilde\omega}_2)]=(\hat Y,\hat
{\tilde T},\hat{{\tilde\omega}})$ where
\begin{equation}
  \left\{\begin{array}{l}
      \label{eq:5}\hat Y^A= Y^B_1\d_B
Y^A_2-Y^B_2\d_B Y^A_1,\\
\hat {\tilde T}=Y^A_1\d_A
\tilde  T_2-Y^A_2\d_A \tilde T_1 +\half (\tilde T_1\d_AY^A_2-\tilde T_2\d_AY^A_1),\\
\hat{{\tilde\omega}}=0\,.
\end{array}\right.
\end{equation}
It thus follows that
\begin{theorem}
  The spacetime vectors $\xi[Y,\tilde T,{\tilde\omega};g]$ realize the extended
  $\mathfrak{bms}_4$ algebra in the modified Lie bracket,
\begin{equation}
  \Big[\xi[Y_1,\tilde T_1,{\tilde\omega}_1;g],\xi[Y_2,\tilde T_2,{\tilde\omega}_2;g]\Big]_M=
\xi[\hat Y,\hat {\tilde T},\hat{{\tilde\omega}};g]\,,\label{eq:1}
\end{equation}
in the bulk of an asymptotically flat spacetime in the sense of Newman
and Unti.
\end{theorem}
Note in particular that for two different choices of the conformal
factor ${\tilde\varphi}$ which is held fixed, ${\tilde\omega}=0$, the
asymptotic symmetry algebras are isomorphic to $\mathfrak{bms}_4$,
which is thus a gauge invariant statement.

\proof{The proof follows closely the one in \cite{Barnich:2010eb} for
  the BMS gauge. In order to be self-contained we recall the different
  steps here.  In a first stage, one shows that on $\scri$, the
  vectors fields $\bar \xi[Y,\tilde T,{\tilde\omega};\bar\gamma]$ given in
  \eqref{eq:17} with $f$ as in \eqref{eq:14} realize the extended
  $\mathfrak{bms}_4$ algebra in terms of the modified Lie
  bracket. Indeed, this is obvious for the $A$ components which do not
  depend on the metric so that the modified bracket reduces to the
  standard Lie bracket for these components. For the $u$ component,
  taking into account that
\[\delta^g_{\bar \xi_1} f_2= {\tilde\omega}_1 f_2+\half e^{{\tilde\varphi}}\int_0^udu^\prime
e^{-{\tilde\varphi}}[-{\tilde\omega}_1(\psi_2-2{\tilde\omega}_2)+
2Y^A_2\d_A {\tilde\omega}_1]\,,\] we have
$[\bar\xi_1,\bar\xi_2]^u_M|_{u=0}= e^{\tilde\varphi}|_{u=0}\hat
T$. Direct computation then shows that
$\d_u([\bar\xi_1,\bar\xi_2]^u_M)=\hat f \d_u{\tilde\varphi}+\half\bar
D_A \hat Y^A$ with $\hat f$ given by \eqref{eq:14} with
$\tilde T,Y,{\tilde\omega}$ replaced by their hatted counterparts, implying
the result for the $u$ component.

For the spacetime vectors, direct computation gives
$[\xi_1,\xi_2]^u_M= [\bar\xi_1,\bar\xi_2]^u_M=\hat f$.  Using the
defining property \eqref{eq:4aextbis}, one then finds that
$\d_r([\xi_1,\xi_2]^\rho_M)=g^{\rho\nu}\d_\nu\hat f$. For the $A$
components the result then follows from the one on $\scri$,
$\lim_{r\to\infty} [\xi_1,\xi_2]_M^A=\hat Y^A$. This is due to the
fact that $I^A$ goes to zero at infinity, that the non-vanishing term
at infinity does not involve the metric and that the correction term
in the bracket does not change the asymptotic behaviour. Finally, for
the $r$ component, we still need to check that the $r$ independent
component of $[\xi_1,\xi_2]_M^r$ is given by $\half \bar\Delta \hat
f$, which follows by direct computation. \qed}\\

For completeness, let us also stress here that, if one focuses on
local properties and expands the conformal Killing vectors $Y^A\d_A$
and the infinitesimal supertranslations $T$ in Laurent series,
\begin{equation}
l_n=-\zeta^{n+1}\frac{\d}{\d\zeta},\quad \bar l_n=-\bar
\zeta^{n+1}\frac{\d}{\d\bar \zeta},\quad n\in \mathbb Z\,,\label{eq:55}
\end{equation}
\begin{equation}
\tilde  T_{m,n}=\zeta^m\bar\zeta^n, 
\quad m,n\in\mathbb Z\,, \label{eq:15}
\end{equation}
the commutation relations for the complexified
$\mathfrak{bms}_4$ algebra read
\begin{equation}
\begin{gathered}
  \label{eq:37}
  [l_m,l_n]=(m-n)l_{m+n},\quad [\bar l_m,\bar l_n]=(m-n)\bar
  l_{m+n},\quad [l_m,\bar l_n]=0, \\
[l_l,T_{m,n}]=(\frac{l+1}{2}-m)T_{m+l,n},
\quad [\bar l_l,T_{m,n}]= (\frac{l+1}{2}-n)T_{m,n+l}. 
\end{gathered}
\end{equation}
The $\mathfrak{bms}_4$ algebra contains as subalgebra the Poincar\'e
algebra, which we identify with the algebra of exact Killing vectors
of the Minkowski metric equipped with the standard Lie bracket.  It is
spanned by the generators
\begin{equation}
  l_{-1},\,l_0,\,l_1,\quad \bar l_{-1},\, \bar l_0,\, \bar l_1,\quad
\tilde T_{0,0},\,\tilde T_{1,0},\,\tilde T_{0,1},\,\tilde T_{1,1}\,.\label{eq:61}
\end{equation}

Non trivial central extensions of the algebra \eqref{eq:37} have been
studied in \cite{Barnich:2011ct}: the computation of
$H^2(\mathfrak{bms}_4)$ reveals that there are only the standard ones
for the Virasoro algebra extending the first two commutation
relations.

\section{Explicit relation between the NU and the BMS gauges}
\label{sec:expl-relat-with}
~
The definition of asymptotically flat space-times in the
BMS approach \cite{Bondi:1962px}, \cite{Sachs:1962wk},
\cite{Sachs2} as reviewed in \cite{Barnich:2009se},
\cite{Barnich:2010eb}, amounts to replacing $g_{uu}=1/g^{uu}=-1$ by
\begin{equation}
g_{uu}=1/g^{uu}=-e^{2\beta},\qquad \beta=O(r^{-2})\label{eq:6}
\end{equation}
in \eqref{eq:1010} and \eqref{eq:NU} while imposing the additional
requirement that
\begin{equation}
{\rm det}\, g_{AB}=r^4{\rm det}\,\bar\gamma_{AB}\,.\label{eq:7}
\end{equation}

Both definitions then differ just by a choice of radial
coordinate. Indeed, replacing the radial coordinate by a function of
the $4$ coordinates preserves the zeros in \eqref{eq:10} and
\eqref{eq:NU} (see e.g.~the discussion in
\cite{0264-9381-20-19-302}). Furthermore, to first non trivial order
in $r$, the determinant condition leads to the same restriction
\eqref{eq:38} as the choice of the origin of the affine parameter. It
follows that the relation between the two radial coordinates does
not involve constant terms and is of the form
\begin{equation}
r^\prime=r +O(r^{-1})\label{eq:3}\,.
\end{equation}
More explicitly, starting from the NU approach,
BMS coordinates are obtained by defining the new
radial coordinates as \cite{Kroon:1999fk}
\begin{equation}
  \label{eq:2}
  r_{\rm BMS}=\big(\frac{{\rm det}\, g_{AB}}{{\rm det}\,\bar \gamma_{AB}}\big)^{\frac{1}{4}}\,.
\end{equation}
Conversely, starting from the BMS approach with radial coordinate $r$,
NU coordinates are obtained by changing the radial coordinate
to
\begin{equation}
  \label{eq:4}
   r_{\rm N} =r-\int^\infty_{ r} dr^\prime
  (e^{2\beta}-1)\,.
\end{equation}
These changes of coordinates only affect lower order terms in the
asymptotic expansion of the metric that play no role in the definition
of asymptotic symmetries and explains a posteriori why the asymptotic
symmetry algebras in both approaches are isomorphic.\\

We will now work out the explicit relation between the free data
characterizing asymptotic solution space in both approaches.  The
inverse metric in the BMS gauge (as discussed in
\cite{Barnich:2010eb}) is given by
\begin{equation}
g^{\mu\nu}_{BMS}=  \begin{pmatrix} 0 &
  -e^{-2\beta} & 0 \\
  -e^{-2\beta}&
-e^{-2\beta}\frac{V}{r}  & -e^{-2\beta}U^B \\
0 & -e^{-2\beta}U^A & g^{AB}
\end{pmatrix}\,.\label{eq:BMS}
\end{equation}
\begin{equation}
  \label{eq:96a}
  g_{AB}=r^2\bar\gamma_{AB}+rC_{AB}+\frac{1}{4} \bar\gamma_{AB}
C^C_DC^D_C+O(r^{-1}), 
\end{equation}
For simplicity, we assume here that there is no trace-free part
$D_{AB}$ at order $0$ and that the conformal factor is
time-independent, $\d_u\tilde\varphi=0$, in which case the news tensor
is simply $N_{AB}=\d_u C_{AB}$ and $f=T+\half u\tilde\psi$ with 
$T=e^{\tilde \varphi} \tilde T$. Writing
\begin{equation}
  C_{\zeta\zeta}=e^{2\tilde\varphi} c,\quad
  C_{\bar\zeta\bar\zeta}=e^{2\tilde\varphi} \bar c,\quad
  C_{\zeta\bar\zeta}=0,
\end{equation}
we have
\begin{equation}
\begin{split}
  \label{eq:30}
  \beta&=-\frac{1}{4}r^{-2}c\bar c +O(r^{-4}), \\
  U^\zeta&=-\frac{2}{r^{2}}e^{-4\tilde\varphi}\d(e^{2\tilde\varphi}\bar
  c)-\frac{2}{3r^3}\Big[ N^\zeta-4e^{-4\tilde\varphi}\bar c
\bar \d(e^{2\tilde\varphi} c)\Big]+O(r^{-4}),\\
  \frac{V}{r}&= 4e^{-2\tilde\varphi}\d\bar\d
  \tilde\varphi + r^{-1}2M +O(r^{-2}),
\end{split}
\end{equation}
which implies in particular that 
\begin{equation}
  \label{eq:13}
  r_{\rm N}=r+\frac{c\bar c}{2r}+O(r^{-3})\,.
\end{equation}
The angular momentum and mass aspects $N^\zeta=N^\zeta (u,\zeta,\bar\zeta),
M=M(u,\zeta,\bar\zeta)$ satisfy the evolution equations 
\begin{equation}
  \label{eq:3a}
  \d_uM=-\frac{1}{8} N^A_BN^B_A
 +\frac{1}{8}
  \bar \Delta \bar R 
  +\frac{1}{4}\bar D_A\bar D_C N^{CA},
\end{equation}
\begin{multline}
  \label{eq:78ter}
\d_uN_A =\d_AM+\frac{1}{4}C_A^B\d_B\bar
  R +\frac{1}{16}\d_A\big[N^B_C C^C_B\big]
-\frac{1}{4} \bar D_AC^C_BN^B_C\\
-\frac{1}{4}\bar D_B\big[C^{B}_{C}N^C_{A}-N^B_CC^C_A\big]-\frac{1}{4}
  \bar D_B \big[ \bar D^B \bar D_CC^C_A -\bar D_A \bar
  D_CC^{BC}\big].
\end{multline}

Consider now the ``eth'' operators \cite{newman:863} defined
here for a field $\eta^s$ of spin weight $s$ according to the
conventions of \cite{Penrose:1986} through 
\begin{equation}
\eth \eta^s= P^{1-s}\bar \d(P^s\eta^s),\qquad \bar \eth
\eta^s=P^{1+s}\d(P^{-s}\eta^s)\,, \qquad P=\sqrt 2
e^{-\tilde\varphi}\,,
\label{eq:34}
\end{equation}
where
$\eth,\bar\eth$ raise respectively lower the spin weight by one
unit and satisfy  
\begin{equation} 
[\bar \eth, \eth]\eta^s=\frac{s}{2} \bar R\,
  \eta^s\,.\label{eq:35}
\end{equation}
The spin weights of the various quantities are summarized in 
table \ref{t1}. Note that the $P$ used here differs from the one used in
\cite{newman:891}, which we will denote by $P_N$ below. It also no
longer denotes the particular function $\half(1+\zeta\bar\zeta)$,
contrary to the notation used in
\cite{Barnich:2010eb,Barnich:2011mi}. In the current conventions, the
particular value of $P$ adapted to the unit sphere is $\frac{1}{\sqrt
  2}(1+\zeta\bar\zeta)$. \\

In order to compare with the notation used in \cite{newman:891}, we
use $\zeta=x^3+ix^4$. With $x^{\prime
  \alpha}=u,r_{\rm N},x^3,x^4$ and
$x^\mu=u,r,\zeta,\bar\zeta$,  computing $g^{\alpha\beta}_{\rm N}(x^\prime)=-\Big(\frac{\d
  x^{\prime\alpha}}{\d x^\mu}g^{\mu\nu}_{\rm BMS}\frac{\d
  x^{\prime\beta}}{\d x^\nu}\Big)(x(x^\prime))$, where the overall
minus sign takes the change of signature into account, then gives the
following dictionary by comparing with \cite{newman:891}:
\begin{equation}
  \begin{gathered}
    \label{eq:20}
    P_N=\frac{1}{\sqrt 2}e^{-{\tilde\varphi}}=\half P \,,\quad
    \nabla= 2\bar \d\,,
\quad \mu^0=-P^2\d\bar\d\ln P=\half \bar\Delta{\tilde\varphi}=-\frac{1}{4}\bar R\,,
    \\
    \Psi^0_{2}+\bar\Psi^0_{2}=-2M-\d_u{( c \bar c)}\,,\quad \sigma^0=
    \bar c\,,
\qquad \omega^0=\bar\eth \sigma^0\,,
    \\
    \Psi^0_{1}=-P N_{\bar\zeta}-\sigma^0\eth \bar\sigma^0-\frac{3}{4}
    \eth (\sigma^0\bar\sigma^0)\,.
  \end{gathered}
\end{equation}
For convenience, let us also use
\begin{equation}
  \label{eq:53}
  \Psi^0_{3}=-\eth
  \dot{\bar\sigma}^0-\frac{1}{4}\bar\eth\bar R, \qquad 
\Psi^0_{4}=-\ddot{\bar\sigma}^0\,.
\end{equation}
In these terms, 
\begin{equation}
  \label{eq:41}
\dot \Psi^0_3=\eth \Psi^0_4,\quad 
\dot \Psi^0_{2}=\eth
\Psi^0_{3}+ \sigma^0\Psi^0_{4},\qquad  \dot \Psi^0_{1}=\eth
\Psi^0_{2}+2 \sigma^0\Psi^0_{3}\,.
\end{equation}
Indeed, the first equation holds by definition and the assumed
time-independence of $P$. The evolution equation \eqref{eq:3a} is
equivalent to the real part of the second equation. Taking into
account the on-shell relation of the NU framework,
\begin{equation}
  \label{eq:48}
  \Psi^0_2-\bar\Psi^0_2=\bar\eth^2\sigma^0-\eth^2\bar\sigma^0
+\bar\sigma^0\dot \sigma^0-\sigma^0\dot{\bar\sigma}^0\,,
\end{equation}
we find 
\begin{equation}
  \label{eq:54}
M=-\Psi^0_{2}-\sigma^0\dot{\bar\sigma}^0+\half \bar\eth^2
\sigma^0-\half\eth^2\bar\sigma^0\,,
\end{equation}
in terms of which \eqref{eq:3a} is fully equivalent to the second
equation of \eqref{eq:41} and \eqref{eq:78ter} is equivalent to the
last equation of \eqref{eq:41}, in agreement with \cite{newman:891}.

\section{Transformation laws of the NU coefficients characterizing
  asymptotic solutions}
\label{sec:transf-laws-nu}

Let $\cY=P^{-1} \bar Y$ and $\bar \cY=P^{-1}
Y$. The conformal Killing equations and the conformal factor then
become
\begin{equation}
  \label{eq:25}
  \eth \bar \cY=0=\bar\eth \cY,\qquad \psi=(\eth \cY+\bar\eth
  \bar \cY)\,.
\end{equation}
It follows for instance that 
\begin{equation}
\bar\eth \eth \cY=-\frac{\bar
  R}{2}\cY,\quad \eth^2 \psi=\eth^3\cY-\frac{1}{2}\bar\cY\eth \bar R,\quad 
\bar\eth\eth \psi=-\frac{1}{2}[\eth(\bar R\cY)+\bar\eth(\bar
R\bar\cY)]\label{eq:44}\,.
\end{equation}
Using the notation $S=(Y,\tilde T,\tilde \omega)$, we have $-\delta_S
\bar \gamma_{AB}=2\tilde\omega \bar\gamma_{AB}$ for the background
metric and
\begin{equation}
  \label{eq:32}
  [-\delta_S,\bar \eth]\eta^s=-\tilde\omega\bar\eth\eta^s+s\bar\eth
  \tilde\omega\eta^s,\quad [-\delta_S, \eth]\eta^s=
-\tilde\omega\eth\eta^s-s\eth
  \tilde\omega\eta^s\,.
\end{equation}

To work out the transformation properties of the NU coefficients
characterizing asymptotic solution space, one needs to evaluate the
subleading terms in $\cL_\xi g^{\alpha\beta}_{N}$ on-shell. This can
also be done by translating the results from the BMS gauge, which
yields
\begin{equation}
\begin{split}
  \label{eq:16b}
  -\delta_S \sigma^0 & = [f\d_u+\cY\eth+ \bar
  \cY\bar\eth+\frac{3}{2}\eth \cY-\frac{1}{2} \bar\eth \bar \cY-\tilde
  \omega] \sigma^0-\eth^2
  f\,,\\
  -\delta_S \dot\sigma^0 & = [f\d_u+ \cY\eth + \bar \cY\bar\eth+2\eth
  \cY-2\tilde\omega]\dot\sigma^0-\half \eth^2\tilde \psi\,,\\
-\delta_S\Psi^0_4&=[f\d_u+\cY\eth+\bar\cY\bar\eth+\half \eth\cY
+\frac{5}{2}\bar\eth\bar\cY-3\tilde\omega]\Psi^0_4\,,\\
-\delta_S\Psi^0_3&=[f\d_u+\cY\eth+\bar\cY\bar\eth+\eth\cY
+2\bar\eth\bar\cY-3\tilde\omega]\Psi^0_3+\eth f\Psi^0_4\,,\\
  -\delta_S
  \Psi^0_2&=[f\d_u+\cY\eth+\bar\cY\bar\eth+\frac{3}{2}\eth \cY
  +\frac{3}{2}\bar\eth \bar\cY - 3\tilde\omega]\Psi^0_2
  +2\eth f\Psi^0_3,\\
  -\delta_S \Psi^0_1&
  =[f\d_u+\cY\eth+\bar\cY\bar\eth+2\eth\cY+\bar\eth\bar\cY
  -3\tilde\omega]\Psi^0_1  +3\eth f\Psi^0_2\,.
\end{split}
\end{equation}

Following for instance the terminology in \cite{held:3145} chapter 3,
but now for general infinitesimal transformations
$\zeta^\prime=\zeta+\epsilon Y(\zeta)$, $\bar \zeta^\prime=\bar
\zeta+\epsilon \bar Y(\bar \zeta)$ instead of those associated to
linear fractional transformations on the sphere and also considering
$\bar\zeta$ as the holomorphic coordinate instead of $\zeta$, a field
$\eta$ has spin weight $s$ and conformal weight $w$ if it transforms
as
\begin{equation}
  \label{eq:30a}
 - \delta_{Y,\bar Y} \eta=\big [Y^A\d_A+\frac{s}{2}(\bar \d
  \bar Y-\d Y)-\frac{w}{2}\psi\big] \eta\,.
\end{equation}
A tensor density of rank $s\geq 0$ and weight $n$ transforms as
\begin{equation}
  \label{eq:31a}
  -\delta_{Y,\bar Y} A_{\bar\zeta\dots\bar\zeta}=\big[Y^A\d_A + s \bar \d
  \bar Y+n(\d Y+\bar \d \bar Y) \big] A_{\bar\zeta\dots\bar\zeta}\,.
\end{equation}
while for rank $s\leq0$ and weight $n$, we have
\begin{equation}
  \label{eq:39s}
  -\delta_{Y,\bar Y} A_{\zeta\dots\zeta}=\big[Y^A\d_A - s  \d
  Y+n(\d Y+\bar \d \bar Y) \big] A_{\zeta\dots\zeta}\,.
\end{equation}
It then follows that a tensor density of weights $(s,n)$ defines a
field of weights $(s,-(2n+|s|))$ and conversely, a field of weights
$(s,w)$ defines a tensor density of weights $(s,-\half(w+|s|))$. For
$s\geq 0$, this is done through $\eta=
A_{\bar\zeta\dots\bar\zeta}P^{2n+s}$ and
$A_{\bar\zeta\dots\bar\zeta}=P^w\eta$. For $s\leq 0$, we have
$\eta= A_{\zeta\dots\zeta}P^{2n-s}$ and
$A_{\zeta\dots\zeta}=P^w\eta$. Note that complex
conjugation gives rise to opposite spin weight and rank but leaves
the conformal and density weights unchanged.
Alternatively, \eqref{eq:30a} can be written as 
\begin{equation}
  \label{eq:56}
  - \delta_{\cY,\bar \cY} \eta=\big [\cY\eth+\bar\cY\bar\eth +\frac{s-w}{2}\,\eth\cY
  -\frac{s+w}{2}\,\bar\eth\bar\cY\big] \eta\,.
\end{equation}

When focusing on $T=0=\tilde\omega$ at the surface $u=0$ and on the
homogeneous part of the transformations, this gives the weights
summarized in tables \ref{t1}, \ref{t2}. These tables are extended
to the Lie algebra elements, which are passive in all our
computations, by writing $[Y,\tilde T]=-\delta_{Y,\bar Y} \tilde T$
and $[Y,Y^\prime]^A=-\delta_{Y,\bar Y} Y^{\prime A}$.

\begin{table} 
\caption{Spin and conformal weights}\label{t1}
\begin{center}
\begin{tabular}{c|cccccccc}
& $\sigma^0$  & $\dot\sigma^0$ &
  $\Psi^0_4 $&  $\Psi^0_3$ & $\Psi^0_2$ & $\Psi^0_1$ & $\cY$ & $T$  \\
\hline
s &  $2$ &  $2$  & $-2$  & $-1$ & $0$ & $1$ &  $-1$ & $0$ \\ 
w  & $- 1$  & $-2$  & $-3$ & $-3$ & $-3$ & $-3$ & $1$ & $1$ \\ 
\end{tabular}
\end{center} \end{table}
\begin{table} 
\caption{Rank and density weights}\label{t2}
\begin{center}
\begin{tabular}{c|cccccccc}
 & $P^{-1}\sigma^0$ &
  $P^{-2}\dot\sigma^0$ & $P^{-3}\Psi^0_4$ & 
$P^{-3}\Psi^0_3$ &  $P^{-3}\Psi^0_2$ & $
  P^{-3}\Psi^0_1$ & $\bar Y$ & $\tilde T$ \\
\hline
s  & $2$ & $2$ & $-2$ & $-1$ & 0 & $1$ &  $-1$  & $0$ \\ 
n  & $-\frac{1}{2}$ & 0 & $\half$ & $1$ & $\frac{3}{2}$ &
$1$ & $-1$ & $-\half$ \\  
\end{tabular}
\end{center} \end{table}

\section{Surface charge algebra}
\label{sec:surf-charge-algebra}

In this section, the local conformal rescalings are switched off ($\tilde\omega=0$) so that $f=T+\half u\psi$ and we use the
notation $s=(\cY,\bar\cY,T)$ for elements of the symmetry algebra,
which is given in these terms by $[s_1,s_2]=\hat s$ where 
\begin{equation}
\begin{gathered}
  \label{eq:57}
  \hat \cY=\cY_1\eth \cY_2 -(1\leftrightarrow 2),\qquad  \hat
  {\bar\cY}=\bar \cY_1\bar \eth \bar \cY_2 -(1\leftrightarrow 2),\\
  \hat T= (\cY_1\eth +\bar \cY_1\bar \eth)T_2-\half \psi_1 T_2 -(1\leftrightarrow 2)\,.
\end{gathered}
\end{equation}
The translation of the charges, the non-integrable piece due to the
news and the central charges computed in \cite{Barnich:2011mi} gives
here
\begin{flalign}
  Q_{s}[\cX]&=-\frac{1}{8\pi G}\int d^2\Omega^\varphi \Big[\big(
  f(\Psi^0_2+\sigma^0\dot{\bar
    \sigma}^0 )+\cY(\Psi^0_1
  +\sigma^0\eth\bar\sigma^0+\half\eth(\sigma^0\bar\sigma^0))\big)
  +{\rm c.c.}\Big],\nonumber\\
  \Theta_{s}[\delta\cX,\cX]&=\frac{1}{8 \pi G}\int d^2
  \Omega^\varphi\, f \big[\dot{\bar\sigma}^0\delta\sigma^0 +{\rm
    c.c.}\big]\,,  \label{eq:19}\\
  K_{s_1,s_2}[\cX]&= \frac{1}{8 \pi G}\int d^2 \Omega^\varphi\, \Big[
  \big(\frac{1}{4} f_1 \eth f_2 \bar\eth \bar R+\half \bar\sigma^0 f_1 \eth^2
  \psi_2 - (1\leftrightarrow 2) \big) + {\rm c.c.} \Big]\,.\nonumber
\end{flalign}
Note that one could also write the charges $Q_s[\cX]$ by allowing for
the additional terms $(\half \eth^2\bar\sigma^0-\half
\bar\eth^2\sigma^0)$ in the first parenthesis since these terms cancel
with the corresponding terms in the complex conjugate expression. Note
also that not $\Psi^0_2$ but only $\Psi^0_2+\bar\Psi^0_2$ is free data
on-shell because of the relation \eqref{eq:48}.

We recognize all the ingredients of the surface charges described in
\cite{0264-9381-1-1-005}, which in turn have been related there to
previous expressions in the literature and, in particular, to the
twistorial approach of Penrose \cite{Penrose08051982}. More precisely,
up to conventions, $Q_{0,0,T}$ agrees with Geroch's linear
super-momentum \cite{Geroch:1977aa} ${Q_{gn}}+\overline{Q}_{gn}$, as
given in equation (A1.12) of \cite{0264-9381-1-1-005}. The angular
(super-)momentum that we get is 
\begin{equation}
  Q_{\cY,0,0}=-\frac{1}{8\pi G}\int d^2\Omega^\varphi\,
  \cY\Big[\Psi^0_1
  +\sigma^0\eth\bar\sigma^0+\half\eth(\sigma^0\bar\sigma^0)
  - \frac{u}{2}\eth\big(\Psi^0_2+\bar\Psi^0_2+
  \d_u(\sigma^0\bar\sigma^0)\big)\Big]\,.
\label{eq:49}
\end{equation}
It differs from $Q_{\eta_c}$ given in equation (4) of
\cite{0264-9381-1-1-005} by the explicitly $u$-dependent term of the
second line.  It thus has a similar structure to Penrose's angular
momentum as described in equations (11), (12), and (17a) of
\cite{0264-9381-1-1-005} in the sense that it also differs by a
specific amount of linear supermomentum, but the amount is different
and explicitly $u$-dependent,
\begin{equation}
  \label{eq:50}
  Q_{\cY,0,0}= Q^{u=0}_{\cY,0,0}+\half u Q_{0,0,\eth\cY}\,.
\end{equation}

 The main result derived in \cite{Barnich:2011mi} states that 
\begin{itemize}

\item if one is allowed to integrate by parts, 
\begin{equation}
  \label{eq:27}
  \int d^2\Omega^\varphi\,\eth \eta^{-1} =0=\int
  d^2\Omega^\varphi\,\bar\eth \eta^{1},
\end{equation}
where $d^2\Omega^{\varphi}=\frac{2 d\zeta\wedge d\bar\zeta}{i P^2}$,

\item if one defines the ``Dirac bracket'' through
 \begin{equation}
   \label{eq:45}
   \{Q_{s_1},Q_{s_2}\}^*[\cX]=-\delta_{s_2}
   Q_{s_1}[\cX]+\Theta_{s_2}[-\delta_{s_1}\cX,\cX], 
 \end{equation}
\end{itemize}
then the charges define a representation of the $\mathfrak{bms}_4$
algebra, up to a field dependent central extension, 
\begin{equation}
  \label{eq:46}
  \{Q_{s_1},Q_{s_2}\}^*=Q_{[s_1,s_2]} +K_{s_1,s_2}, 
\end{equation}
where $K_{s_1,s_2}$ satisfies the generalized cocycle condition 
\begin{equation}
  \label{eq:47}
  K_{[s_1,s_2],s_3}-\delta_{s_3} K_{s_1,s_2}+{\rm cyclic} (1,2,3)=0\,.
\end{equation}
The representation theorem contained in equations \eqref{eq:46} and
\eqref{eq:47} can be verified directly in the present context by
starting from \eqref{eq:19}, \eqref{eq:48} and using the properties
\eqref{eq:35}, \eqref{eq:27} of $\eth$, the evolution equations
\eqref{eq:41}, the conformal Killing equations \eqref{eq:25}, the
$\mathfrak{bms}_4$ algebra \eqref{eq:57} and the transformation laws
\eqref{eq:16b}.\\

\noindent Several remarks are in order: 

\begin{itemize}

\item Integrations by parts are justified for regular functions on the
  sphere and thus for $\mathfrak{bms}^{\rm glob}_4$ and regular
  solutions. In the case of Laurent series more care is needed, see
  e.g.~\cite{saidi:1990xx}. 

\item For the globally well-defined $\mathfrak{bms}^{\rm glob}_4$
  algebra on the sphere, the central charge $K_{s_1,s_2}$ vanishes.

\item The non-conservation of the charges follows by taking
  $s_2=(0,0,1)$ and $s_1=s$. Indeed, since 
  $\frac{d}{du} Q_s=\frac{\d}{\d u} Q_s-\delta_{(0,0,1)} Q_s$, the
equality of the right hand sides of \eqref{eq:45}
 and \eqref{eq:46} gives 
 \begin{equation}
   \label{eq:51}
   \frac{d}{du} Q_s=-\frac{1}{8 \pi G}\int d^2
   \Omega^\varphi\, \big[\dot{\bar\sigma}^0(-\delta_s\sigma^0) +
   \frac{1}{4} \eth f\bar\eth \bar R+\half \bar\sigma^0 \eth^2
   \psi +{\rm
     c.c.}\big]
   \,.
 \end{equation}
 For $s=(0,0,1)$, this gives the standard Bondi-Sachs mass loss
 formula,
 \begin{equation}
   \label{eq:58}
   \frac{d}{du} Q_{0,0,1}= -\frac{1}{8 \pi G}\int d^2 \Omega^\varphi\,
 \big[\dot{\bar\sigma}^0\dot\sigma^0+{\rm c.c.}\big]\,.
 \end{equation}
 It also follows that the standard $\mathfrak{bms}^{\rm glob}_4$
 charges are all conserved on the sphere in the absence of news.
\end{itemize}

\section{Conclusion and perspective}

In this chapter, we have shown that the symmetry algebra of asymptotically flat
    spacetimes at null infinity in four dimensions in the sense of
    Newman and Unti is isomorphic to the direct sum of the abelian
    algebra of infinitesimal conformal rescalings with
    $\mathfrak{bms}_4$. We have worked out the
    local conformal properties of the relevant Newman-Penrose
    coefficients, constructed the surface charges and derived their
    algebra.\\

To the best of our knowledge, except for the previous analysis in the
BMS gauge, the above representation result does not exist elsewhere in
the literature. 
In the future, it should be interesting
to discuss in more detail the implication of this result, and to provide a
self-contained derivation of the $\mathfrak{bms}_4$ transformation
laws in the context of the Newman-Penrose formalism.

\newpage
\chapter{Einstein-Yang-Mills: Asymptotic symmetries}

In this chapter, asymptotic symmetries of the Einstein-Yang-Mills
    system with or without cosmological constant are explicitly worked
    out in a unified manner. In agreement with a recent conjecture,
    one finds a Virasoro-Kac-Moody type algebra not only in three
    dimensions but also in the four dimensional asymptotically flat
    case.

\section{Motivation}
\label{sec:introduction}

Even though the first discussions of asymptotic symmetries dealt with
four dimensional general relativity, both at null
\cite{Bondi:1962px,Sachs:1962wk,Sachs2} and at spatial infinity
\cite{Regge:1974zd}, most of the recent work was devoted to three
dimensions because of the occurence of a classical central charge
\cite{Brown:1986nw} that plays a key role in symmetry based
explanations \cite{Strominger:1998eq} of the entropy of the BTZ black
hole \cite{Banados:1992wn,Banados:1993gq} and in other aspects of the
AdS/CFT correspondence (see e.g.~\cite{Aharony:1999ti}, chapter 5).\\

In recent work \cite{Strominger:2013lka}, Strominger suggested to
extend the analysis for gravity in four dimensions at null infinity to
include Yang-Mills fields and established a relation to field
theoretic soft photon and graviton theorems
\cite{PhysRev.140.B516}. During these considerations, the symmetry
algebra was argued to be of Virasoro-Kac-Moody type.\\

In this chapter, we confirm this conjecture. We start by showing that the
residual symmetry algebra of a standard gauge choice adapted to the
asymptotic analysis of the Einstein-Yang-Mills system is simply the
gauge algebra in one dimension lower. The asymptotic symmetry algebra
is then obtained by a further reduction that comes from suitable
fall-off conditions on the remaining fields. Details for various
standard cases, including the flat case with asymptotics at null
infinity, are provided.\\

At this stage, one might wonder why the enhancement of the $U(1)$
electromagnetic gauge symmetry has not been discussed in previous
detailed investigations of the asymptotic properties of the
Einstein-Maxwell system
\cite{janis:902,vandeBurg06051969,Exton:1969im}. With hindsight, the
reason is that the focus was on the modifications of the equations of
motions and their solutions due to the presence of the
electro-magnetic field which had been included through its field
strength. It turns out however that a formulation in terms of gauge
potentials is required if one wants to discuss action principles and
asymptotic symmetries for both gravitational and Yang-Mills type gauge
fields in a unified manner.\\

\section{Gauge structure of the Einstein-Yang-Mills system}

The Einstein-Yang-Mills system in $d$ dimensions is described by the
action 
\begin{equation}
  \label{eq:1}
  S=\frac{1}{16\pi G}\int d^dx \sqrt{|g|}~~
  [R-2\Lambda-g_{ij}F^i_{\mu\nu}F^{j\mu\nu}], \quad \Lambda=-\frac{(d-1)(d-2)}{2l^2},
\end{equation}
where $g_{ij}$ is an invariant non-degenerate metric in a basis $T_i$
of the internal gauge algebra $\mathfrak{g}$, $F_{\mu\nu}=\partial_\mu
A_\nu-\partial_\nu A_\mu+[A_\mu,A_\nu]$ is the field strength,
$A_\mu=A^i_\mu T_i$ and the bracket denotes the Lie bracket in
$\mathfrak{g}$, $[T_i,T_j]=f^k_{ij}T_k$.\\

The complete gauge algebra consists of pairs $(\xi,\epsilon)$ of a
vector field $\xi^\mu\d_\mu$ and an internal gauge parameter
$\epsilon^i T_i$. A generating set of gauge symmetries is given by
\begin{equation}
  \label{eq:5}
  \delta_{(\xi,\epsilon)} g_{\mu\nu}=-\cL_\xi
g_{\mu\nu},\quad \delta_{(\xi,\epsilon)}A_\mu=-\cL_\xi A_\mu
+D^A_\mu\epsilon,
\end{equation}
with $D_\mu^A\epsilon=\d_\mu\epsilon+[A_\mu,\epsilon]$.\\

Let the fields be collectively denoted by
$\phi^\alpha=(g_{\mu\nu},A_\mu)$.  When the gauge parameters
$(\xi,\epsilon)$ depend only on the spacetime coordinates but not on
the fields $\phi^\alpha$, one has 
\begin{equation}
[\delta_{(\xi_1,\epsilon_1)},\delta_{(\xi_2,\epsilon_2)}]\phi^\alpha
=\delta_{(\hat\xi,\hat\epsilon)}\phi^\alpha,
\end{equation}
with $\hat\xi=[\xi_1,\xi_2]$ the Lie bracket for vector fields and
$\hat\epsilon=\xi^\mu_1\d_\mu\epsilon_2-\xi^\mu_2\d_\mu\epsilon_1
+[\epsilon_1,\epsilon_2]$. The Lie bracket for field independent gauge
parameters is given by
\begin{equation}
[(\xi_1,\epsilon_1),(\xi_2,\epsilon_2)]=
  (\hat\xi,\hat\epsilon)\label{eq:6a}.
\end{equation}

In the case of gauge parameters $(\xi,\epsilon)$ that are field
dependent, one finds instead 
\begin{equation}
[\delta_{(\xi_1,\epsilon_1)},\delta_{(\xi_2,\epsilon_2)}]\phi^\alpha
=\delta_{(\hat\xi_M,\hat\epsilon_M)}\phi^\alpha,
\end{equation}
with 
\begin{align}
  \hat\xi_M&=\hat\xi+\delta_{(\xi_1,\epsilon_1)}\xi_2-
  \delta_{(\xi_2,\epsilon_2)}\xi_1,\label{eq:103}\\
  \hat\epsilon_M&=\hat\epsilon+\delta_{(\xi_1,\epsilon_1)}
  \epsilon_2-\delta_{(\xi_2,\epsilon_2)}\epsilon_1,\label{eq:104}
\end{align}
and the Lie (algebroid) bracket for field dependent gauge parameters is thus
defined through
\begin{eqnarray}\label{eq:105}
[(\xi_1,\epsilon_1),(\xi_2,\epsilon_2)]_M=(\hat\xi_M,\hat\epsilon_M).
\end{eqnarray}

\section{Dimensional reduction through gauge fixation}
\label{sec:dimens-reduct-thro}

In terms of coordinates $x^\mu=(u,r,x^A)$, where $x^A$ are angular
variables in $d-2$ dimensions, we make the following gauge fixing
ansatz for the metric and Yang-Mills potentials:
\begin{equation}
\begin{split}
g_{\mu\nu}&=
\begin{pmatrix}
e^{2\beta}\dfrac{V}{r}+g_{CD} U^CU^D&	-e^{2\beta}	&	-g_{BC}U^C\\
-e^{2\beta}	&	0	&	0\\
-g_{AC}U^C&	0	&g_{AB}
\end{pmatrix},\label{eq:2.7}\\
A_\mu&=
\begin{pmatrix}
A_u,0,A_A
\end{pmatrix}.
\end{split}
\end{equation}
In addition, one imposes the determinant condition ${\rm det}\
g_{AB}=r^{2(d-2)}{\rm det}\bar\gamma_{AB}$, with $\bar\gamma_{AB}$ the
metric on the unit $d-2$-sphere.\\

As in the purely gravitational four dimensional case
\cite{Sachs:1962wk}, these conditions fix the gauge freedom up to some
$r$ independent functions. Indeed, the gauge transformations
\eqref{eq:5} that preserve this gauge choice, i.e., the residual gauge
symmetries, are determined by gauge parameters that have to satisfy
\begin{equation}
\cL_\xi g_{rr}=0,\hspace{0.5cm}\cL_\xi g_{rA}=0,
\hspace{0.5cm}g^{AB}\cL_\xi g_{AB}=0,\hspace{0.5cm}
-\cL_\xi A_{r}+D^A_r \epsilon=0.\label{eq:2.9}
\end{equation}
This gives the differential conditions
\begin{equation}
  \label{eq:7}
\begin{split}
 & \d_r\xi^u=0,\quad \d_r\xi^A=\d_B\xi^u g^{AB}e^{2\beta},\\ & \d_r
  (\frac{\xi^r}{r})=-\frac{1}{d-2} (\bar D_B \d_r \xi^B-\d_B\xi^u\d_r
  U^B),\quad \d_r \epsilon=\d_B\xi^u g^{AB}e^{2\beta} A_A, 
\end{split}
\end{equation}
the general solution of which is
\begin{equation}
  \begin{split}
& \xi^u =F(u,x^A),\quad \xi^A=Y^A(u,x^B)-\d_B F\int^\infty_r dr^\prime (e^{2\beta}
  g^{AB}),  \\ & \xi^r=-\frac{r}{d-2} (\bar D_B \xi^B-\d_B F
  U^B),\quad
\epsilon = E(u,x^A) - \d_B F\int^\infty_r dr^\prime
  (g^{BA}e^{2\beta} A_A),\label{eq:2.10}
  \end{split}
\end{equation}
and involves $d-1+n$ arbitrary $r$-independent functions
$F(u,x^A),Y^A(u,x^B),\\E^i(u,x^A)$.\\

At this stage, it is sufficient to impose the following fall-off
conditions on the components of the metric and the gauge potentials,
\begin{equation}
  \label{eq:6}
 e^{2\beta}g^{AB}=O(r^{-1-\epsilon})=e^{2\beta}g^{AB}A_A,\quad \ U^C
 e^{2\beta}g^{AB} =o(r^{-1})\ {\rm
   for}\ d>3.
\end{equation}
In particular, the first of these conditions guarantee that the
integrals for $\xi^A$ and $\epsilon$ in \eqref{eq:2.10} are
well-defined and that $\lim_{r\to\infty}\xi^A=Y^A$,
$\lim_{r\to\infty}\epsilon=E$.\\

Consider then the vector fields
$\xi^R=F\partial_u+Y^A\partial_A$ and the internal gauge parameter
$\epsilon^R=E^iT_i$, equipped with the Lie bracket
\begin{equation}
[(\xi^R_1,\epsilon^R_1),(\xi^R_2,\epsilon^R_2)]
=(\hat\xi^R,\hat\epsilon^R),\label{eq:7a}
\end{equation}
for field independent gauge parameters \eqref{eq:6a} of the
Einstein-Yang-Mills system in $d-1$ dimensions. 
We are now ready to state the main result of this chapter: 

{\em The Lie algebra of residual gauge parameters \eqref{eq:2.10}
  equipped with the Lie bracket $[\cdot,\cdot]_M$ of the $d$
  dimensional Einstein-Yang-Mills system is a faithful
  representation of the Lie algebra of field independent gauge
  parameters $(\xi^R,\epsilon^R)$ of the $d-1$ dimensional
  Einstein-Yang-Mills system.}

The proof for the diffeomorphism part is almost exactly the same as in
\cite{Barnich:2010eb}, except for the additional $u$ dependence in
$Y^A$, which is easily taken into account. We will thus not repeat all
details here. First, one needs to check that the result holds for
$\hat\xi^u_M$, $\hat\xi^A_M$, $r^{-1}\hat\xi^r_M$, $\hat\epsilon_M$
at $r\to\infty$. This is where the fall-off conditions \eqref{eq:6}
are needed. Note however that the fall-off condition on $U^A$ have
been considerably relaxed and, in particular, there are no conditions
for $d=3$. The rest of the proof consists in verifying that $\d_r
\hat\xi^u_M$, $\d_r \hat\xi^A_M$, $\d_r(r^{-1}\hat\xi^r_M)$,
$\d_r\hat\epsilon_M$ satisfy equations \eqref{eq:7} with
$(\xi^R,\epsilon^R)$ replaced by $(\hat\xi^R,\hat\epsilon^R)$.

\section{Fall-off conditions and asymptotic symmetry structure}
\label{sec:fall-cond-asympt}

Suppose now that in spacetime dimensions $4$ or higher, precise
fall-off conditions for the metric coefficients and gauge potentials
are given by
\begin{equation}
\begin{split}
  & \beta=o(1),\quad U^A=o(1),\quad g_{AB}dx^A
  dx^B=r^2\bar\gamma_{AB}(x^C) dx^Adx^B+o(r^2), \\
  & \frac{V}{r}=-\frac{r^2}{l^2} +o(r^2),\quad A_u=o(1),\quad
  A_B=A^{0}_B(u,x^C)+o(1). \label{eq:2}
\end{split}
\end{equation}
In the asymptotically flat $U(1)$ case in four dimensions, these
fall-off conditions are consistent with those of
\cite{janis:902,vandeBurg06051969,Exton:1969im}. They imply in
particular the conditions required in \eqref{eq:6}.\\

The gauge transformations that preserve these fall-off conditions have
to satisfy, in addition to \eqref{eq:2.9}, the supplementary
conditions
\begin{equation}
\begin{split}
  &\cL_\xi g_{ur}=o(1),\hspace{0.5cm}\cL_\xi
  g_{uA}=o(r^2),\hspace{0.5cm}
  \cL_\xi g_{AB}=o(r^2),\hspace{0.5cm}\cL_\xi g_{uu}=o(r^2),\\
  &-\cL_\xi A_u+D^A_u \epsilon=o(1),\hspace{0.5cm} -\cL_\xi A_B+D^A_B
  \epsilon=O(1).\label{eq:2.21}
\end{split}
\end{equation}
They are equivalent to the following differential equations on the
$(\xi^R,\epsilon^R)$
\begin{equation}
\begin{split}
  \label{eq:2.22}
&  \d_u F=\frac{1}{d-2}\Psi, \quad \d_u
Y^A\bar\gamma_{AB}=\frac{1}{l^2} \d_B F, \\
& \cL_Y\bar\gamma_{AB} =
\frac{2}{d-2} \Psi \bar\gamma_{AB},\quad \d_u
E=\frac{1}{l^2}\d^B F A^{0}_B, 
\end{split}
\end{equation}
with $\Psi=\bar D_B Y^B$, the general solution of which is
\begin{equation}
\begin{split}
 & F =f(x^A)+\dfrac{1}{d-2}\int_0^u du^\prime~~\Psi,\quad
  Y^A=y^A(x^B)+\dfrac{1}{l^2}\int^u_0 du^\prime
  (\bar\gamma^{AB}\d_B F),  \\
 & E = e(x^A)+\dfrac{1}{l^2}\int^u_0 du^\prime(\bar\gamma^{AB}\d_A F
  A_B^0).\label{eq:4.5}
\end{split}
\end{equation}
Let us denote by $a_B$ the values of $A^{0}_B$ at $u=0$ and consider
time independent conformal Killing vectors of the $d-2$ sphere,
\begin{equation}
  \label{eq:4}
\cL_y \bar\gamma_{AB}=\frac{2}{d-2}\psi\bar\gamma_{AB},
\end{equation} 
with $\psi=\bar D_B y^B$. In addition, in the case of a non-vanishing
cosmological constant, the vectors $\d^Af=\bar\gamma^{AB}\d_Bf$ are
also required to be conformal Killing vectors of the $d-2$ sphere, as
follows by differentiating the third of \eqref{eq:2.22} with respect
to $u$ and setting $u=0$. A second derivative with respect to $u$ at
$u=0$ then implies that $\d^A\psi$ are also conformal Killing vectors
of the $d-2$ sphere. This can be continued for higher order
derivatives.\\

In terms of these quantities, the asymptotic symmetry structure is
described through the brackets
\begin{equation}
\begin{split}
  \label{eq:3}
  & \hat f=\frac{1}{d-2}f_1\psi_2+y^A_1\d_A f_2-(1\leftrightarrow
  2),\quad \hat y^A=\frac{1}{l^2} f_1 \d^A f_2+ y^B_1\d_B
  y^A_2-(1\leftrightarrow 2),\\
  & \hat e=\frac{1}{l^2} f_1 \d^A f_2 a_A +y^A_1\d_A
  e_2-(1\leftrightarrow 2)+[e_1,e_2].
\end{split}
\end{equation}
On account of the explicit field dependence in $\hat e$, one has to
use the Lie algebroid bracket in order to check the Jacobi identity
for $e$ with $\delta_{y,e} a_A=-\cL_y a_A+ D^a_A e$ in the case of a
non-vanishing cosmological constant. More generally, it is implicitly
understood that each time an element depends explicitly on the fields,
the Lie algebroid bracket has to be used.\\

By the same reasoning as before, one then shows that the asymptotic
symmetry structure is represented at infinity for all values of $u$
through the Lie algebroid bracket that involves the dependence on
$A^0_B$, and then also in the bulk spacetime through the result of the
previous section.\\

The gauge theory part of the asymptotic symmetry structure consists of
elements of the form$(0,0,e)$. It is is a non-abelian ideal that
contains an arbitrary $\mathfrak g$-valued function on the $d-2$
sphere. In that sense, it is a generalisation of a loop algebra where
the base space is a higher dimensional sphere instead of a circle.\\

The quotient of the total structure by this ideal is the spacetime
part. It can be described by elements of the form $(f,y^A,0)$ with
brackets determined by the first line of \eqref{eq:3}. In the case of
vanishing cosmological constant, elements of the form $(f,y^A,0)$ form
a subalgebra that acts on the gauge theory ideal.

\section{Explicit description of asymptotic symmetry structure in
  particular cases}
\label{sec:expl-descr-part}

\subsection{Dimensions 4 and higher, anti-de Sitter case}
\label{sec:dimensions-5-or}

For $d\geq 4$ and $l\neq 0$, the space-time part of the asymptotic
symmetry structure is isomorphic to $\mathfrak{so}(d-1,2)$, the algebra
of exact Killing vectors of $d$-dimensional anti-de Sitter space, in
agreement with the analysis in \cite{Henneaux:1985ey}. \\

Indeed, in the coordinates we are using, the anti-de Sitter metric is
given by
\begin{equation}
g_{\mu\nu}=\begin{pmatrix}
-\frac{r^2}{l^2}-1&	-1	&	0\\
-1	&	0	&	0\\
0 &	0	&r^2\bar\gamma_{AB}
\end{pmatrix}.\label{eq:ads}
\end{equation} 
Besides the conditions 
\begin{gather}
  \label{eq:4a}
  \bar\xi^u=\bar F(u,x),\quad \bar\xi^A=\bar Y^A(u,x)-\frac{1}{r}\d^A
  \bar F,\quad
  \bar\xi^r=\frac{1}{d-2}(-r\bar \Psi+\bar \Delta \bar F),\\
  \d_u \bar F=\frac{1}{d-2}\bar \Psi,\quad \d_u \bar
  Y^A=\frac{1}{l^2}\d^A \bar F,
\end{gather}
where $\bar Y^A$ and $\d^A \bar F$ are conformal Killing vectors of
$\bar\gamma_{AB}$, which correspond to an asymptotic Killing vector
evaluated for the anti-de Sitter metric, an exact Killing vector
$\bar\xi=\bar\xi^u\d_u +\bar\xi^r\d_r +\bar \xi^A\d_A$ has also to
satisfy the additional conditions
\begin{equation}
  \label{eq:8}
  \d_B \bar F=-\frac{1}{d-2}\d_B\bar\Delta \bar F, \quad  
\bar \Psi=-\frac{1}{d-2}\bar\Delta \bar \Psi.
\end{equation}
The latter are automatically satisfied for conformal Killing vectors
$\bar Y^A,\d^A \bar F$ of the unit $d-2$ sphere.\\

Even though it is not needed for this proof, one can also check
directly that, if $y^A$ are conformal Killing vectors of the $d-2$
sphere, then the requirement that $\d^A\psi$ are also conformal
Killing vectors is automatically satisfied if $d\neq 4$, while for
$d=4$, this reduces the local conformal algebra in $2$ dimensions to
the globally well-defined algebra $\mathfrak{so}(3,1)$ on the $2$
sphere.

\subsection{Dimensions 5 and higher, flat case}
\label{sec:dimensions-5-higher}

For $l\to\infty$, the asymptotic symmetry structure of field
independent parameters $(f,y^A,e)$ simplifies. The subalgebra
$(0,y^A,0)$ of conformal Killing vectors of the $d-2\geq 3$ sphere
represents the Lorentz algebra $\mathfrak{so}(d-1,1)$. It acts both on the
abelian ideal $(f,0,0)$ of arbitrary functions on the sphere, 
representing supertranslations, and on the gauge theory ideal. \\

Stronger fall-off conditions motivated by the Einstein equations of
motions have been considered in \cite{Tanabe:2011es}. They require
$\d^A F$ to be conformal Killing of the $d-2$ sphere. In turn this
requires both $\d^A f$ and $\d^A\psi$ to be conformal Killing
vectors. Again, by comparing with the conditions satisfied by exact
Killing vectors of Minkowski space-time, the only additional
conditions are \eqref{eq:8}, which are automatically satisfied for
conformal Killing vectors $Y^A,\d^A F$. This shows that the additional
conditions reduce super to standard translations so that the spacetime
part of the asymptotic structure becomes the Poincar\'e algebra
$\mathfrak{iso}(d-1,1)$.

\subsection{4 dimensional flat case}
\label{sec:4-dimens-asympt}

In $4$ dimensions, it is useful to introduce stereographic coordinates
$\zeta=\cot{\frac{\theta}{2}}e^{i\phi}$ and its complex conjugate, so
that $\bar\gamma_{AB} dx^Adx^B=2P_S^{-2}d\zeta d\xbar\zeta$ with
$P_{S}=\frac{1}{\sqrt 2}(1+\zeta\xbar\zeta)$. The covariant derivative
on the $2$ surface is then encoded in the operator
\begin{equation}
\eth \eta^s= P^{1-s}_S\bar \d(P^s_S\eta^s),\qquad \xbar \eth
\eta^s=P^{1+s}\d(P^{-s}\eta^s)\,, 
\label{eq:34}
\end{equation}
where $\eth,\bar\eth$ raise respectively lower the spin weight $s$ by
one unit and satisfy
\begin{equation} 
[\bar \eth, \eth]\eta^s=\frac{s}{2}  R_S\,
  \eta^s\,,\label{eq:35}
\end{equation}
with $ R_S=4P^2_S\d \bar \d \ln P_S=2$. \\

Let $\cY=P^{-1}_S y^{\bar\zeta}$ and $\xbar \cY=P^{-1}_S y^\zeta$ be of spin
weights $-1$ and $1$ respectively. The conformal Killing equations and
the conformal factor then become
\begin{equation}
  \label{eq:25}
  \eth \xbar \cY=0=\xbar\eth \cY,\qquad \psi=(\eth \cY+\xbar\eth
  \xbar \cY)\,.
\end{equation}
It follows for instance that $\xbar\eth \eth \cY=-\cY$, $\eth^2
\psi=\eth^3\cY$, $\xbar\eth\eth \psi=-\psi$. \\

In order to describe the asymptotic symmetry structure there are then
two options. \\

The first is to require well-defined functions on the $2$-sphere. This
amounts to restricting oneself to the conformal Killing vectors that
satisfy $\eth^3\cY=0=\xbar\eth^3\xbar\cY$ and require that the
functions $f,e^a$ that occur in \eqref{eq:3} (with $l\to\infty$) can
be expanded in spherical harmonics. The asymptotic symmetry algebra is
then the semi-direct sum of the globally well-defined
$\mathfrak{bms}^{\rm glob}_4$ algebra \cite{Bondi:1962px,Sachs2} with
a globally well-defined ``sphere'' algebra.\\

Alternatively \cite{Barnich:2009se,Barnich:2010eb}, one admits Laurent
series and expands $y^{\zeta}\d_\zeta$ in terms of
$l_m=-\zeta^{m+1}\d_\zeta$, $y^{\bar\zeta}\d_{\bar\zeta}$ in terms of
$\bar l_m$, $f$ in terms of $t_{m,n}=P_S^{-1}\zeta^m\bar\zeta^n$ and
$e$ in terms of $j_i^{m,n}=T_i\zeta^m\bar\zeta^n$. In these terms, the
non-vanishing brackets of the asymptotic symmetry algebra become
\begin{align}
[l_l,t_{m,n}]&=\left(\dfrac{l+1}{2}-m\right)t_{m+l,n},
\quad [l_m,l_n]=(m-n)~~l_{m+n},\\
[\bar l_l,t_{m,n}]&=\left(\dfrac{l+1}{2}-n\right)t_{m,n+l},
\quad [\bar l_m,\bar l_n]=(m-n)~~\bar l_{m+n},\\
[l_l,j^{m,n}_i]&=-m~~j^{m+l,n}_i,\quad
[\bar l_l,j^{m,n}_i]=-n~~j^{m,n+l}_i,\\\label{eq:3.67}
[j^{l,p}_i,j^{m,n}_j]&= f_{ij}^k~~j_k^{l+m,p+n}. 
\end{align}

\subsection{3 dimensional anti-de Sitter case}
\label{sec:3-dimensional-anti}

On the metric components, we use the same fall-off conditions as in
\ref{sec:fall-cond-asympt}. Note that the determinant condition
requires $g_{\phi\phi}=r^2$ and that the fall-off conditions allow for
$\ln r$ terms both in $g_{uu}$ and $g_{u\phi}$. The spacetime part of
the asymptotic symmetry structure is then described by two copies of
the conformal algebra \cite{Brown:1986nw},
$F\d_u+Y\d_\phi=Y^+(x^+)\d_++Y^-(x^-)\d_-$, where
$x^\pm=\frac{u}{l}\pm \phi$.\\

In order to accommodate the charged and rotating black hole solution
\cite{Martinez:1999qi}, the fall off conditions on the gauge
potentials can be chosen as $A_+=O(\ln r)$, while one simultaneously
requires $A_-=o(1)$.  Alternatively, one could also exchange the
r\^ole of $+$ and $-$.  Requiring $-\cL_\xi A_+ +D^A_+ \epsilon=O(\ln
r)$ gives no conditions, while $-\cL_\xi A_- +D^A_- \epsilon=o(1)$
leads to $\d_-E=0$. In this case, there is no explicit field
dependence and the asymptotic symmetry structure simplifies as
compared to the higher dimensional case.\\

When expanding $Y^+\d_+,Y^-\d_-,E$ in terms of modes,
$l^{\pm}_m=e^{imx^\pm}\d_\pm,j^m_i=T_ie^{imx^+}$, the non-vanishing
brackets of the asymptotic symmetry algebra are explicitly given by
\begin{equation}
i[l^\pm_m,l^\pm_n]=(m-n) l^\pm_{m+n},\quad i[l^+_m,j^n_i]=-n
j^{m+n}_i,\quad i[j^m_i,j^n_i]=if_{ij}^k j^{m+n}_k. 
\end{equation}

\subsection{3 dimensional flat case}
\label{sec:3-dimensional-flat}

In this case, the spacetime part of the asymptotic symmetry structure
is the $\mathfrak{bms}_3$ algebra described by
$F\d_u+Y\d_\phi=[f(\phi)+uy(\phi)]\d_u+y\d_\phi$. For the gauge
potentials, one may then choose $A_u=o(1)$, $A_\phi=O(\ln r)$. 
Requiring $-\cL_\xi A_u+D_u^A \epsilon=o(1)$ leads to
$\d_u E=0$, while $-\cL_\xi A_\phi+D_\phi^A \epsilon=O(\ln r)$
gives no conditions.\\

When expanding $F\d_u+Y\d_\phi,E$ in terms of modes,
$l_m=e^{im\phi}\d_\phi+uim\phi e^{im\phi}\d_u$, $t_m=e^{im\phi}\d_u$,
$j^m_i=T_i e^{im\phi}$, the non vanishing brackets of the asymptotic
symmetry algebra are given by
\begin{equation}
\begin{split}
  \label{eq:10}
 & i[l_m,l_n]=(m-n) l_{m+n},\quad i[l_m,t_n]=(m-n) t_{m+n},\\
 & i[l_m,j^n_i]=-n j^{m+n}_i,\quad i[j^m_i,j^n_j]=if_{ij}^k j^{m+n}_k.
\end{split}
\end{equation}

\section{Discussion and outlook}
\label{sec:discussion}

The gauge fixing and fall-off conditions that we have considered have
been mainly dictated by the desire to yield the usual asymptotic
symmetry structure, at least for the spacetime part, while otherwise
being as relaxed as possible. As partly already discussed in the text,
additional more restrictive conditions motivated by finiteness of
associated conserved currents or their integrability for example can
further reduce the asymptotic symmetry structure, in particular also
in the Yang-Mills part.\\

On the other hand, one may wonder how far these conditions can be
relaxed even further. From a holographic point of view, the role of
the gauge fixing conditions considered in section
\ref{sec:dimens-reduct-thro} is to fix the dependence in $r$ of the
gauge parameters. This can be achieved in various ways. In the
Newman-Unti gauge \cite{newman:891} for instance, one can require
$g_{ur}=-1$ instead of the determinant condition, leading to another
integration function in $\xi^r$, that may or may not be fixed through
additional conditions, see e.g.~\cite{Barnich:2011ty}. One may also relax
conditions \eqref{eq:6}. For instance at fixed but finite $r$ no such
conditions are needed. Even though one will then not get the symmetry
structure of the Einstein-Yang-Mills system in one dimension lower,
the resulting structure will still be well-defined.\\

Similarly, except for the fall-off conditions on $g_{AB}$, the role
of the other conditions in section \ref{sec:fall-cond-asympt} is to
fix the time dependence of the gauge parameters and thus of the
symmetry structure of the dual boundary theory. In the present set-up,
the fall-off conditions on $g_{AB}$ are the only ones that constrain
the dependence of the symmetry structure, or more precisely of
$Y^A,F$, on the spatial coordinates $x^A$. In other words, relaxing
this condition leads to ``superrotations'' that, like
supertranslations and the Yang-Mills gauge parameters, have an
arbitrary $x^A$ dependence.\\

With the symmetries under control, the next stage is to work out
asymptotic solutions. This will be done in the next chapter, for simplicity first in
three, and then in four dimensions, in the case of the Einstein-Maxwell system.
Once this is done, the
symmetry transformations of the fields characterizing asymptotic
solutions can be computed.

\newpage
\chapter{Einstein-Maxwell theory: equations of motion}
In chapter five, the Einstein-Yang-Mills system was studied in $d$ dimensions from the point of view of asymptotic symmetries. The next step in the asymptotic analysis is to work out the solution space of the theory. This is the goal of this chapter. For simplicity, the analysis is carried out in the case of Einstein-Maxwell theory, in three and four spacetime dimensions. The first section generalizes Bianchi identities  \eqref{3:33} to the case of Einstein-Maxwell equations of motion. In sections 6.2 and 6.3, the equations of motion are solved (under the assumption of a perturbative expansion in inverse power or $r$) in three and four dimensions.

\section{Identitites for equations of motion}
The Einstein-Maxwell system in $d$ dimensions ($d>2$) with a cosmological constant is described by the action
\begin{eqnarray}
S=\dfrac{1}{16\pi}\int d^dx\sqrt{-g}(R-2\Lambda-F_{\mu\nu}F^{\mu\nu}),\hspace{1cm}\Lambda=-\dfrac{(d-1)(d-2)}{2l^2},\label{eq:action}
\end{eqnarray}
where $F_{\mu\nu}=\partial_\mu A_\nu-\partial_\nu A_\mu$ is the $U(1)$ field strengh. Equations of motion, obtained by varying the action with respect to the fields $(g,A)$, read
\begin{align}
 R_{\mu\nu}-\dfrac{1}{2}Rg_{\mu\nu}+\Lambda g_{\mu\nu}&=
2\left(F_{\mu\sigma}F_\nu^{\phantom\nu\sigma}-\dfrac{1}{4}F^2g_{\mu\nu}\right)\equiv 8\pi T_{\mu\nu},\label{eq:3.1}\\
F^{\mu\nu}_{\phantom{\mu\nu};\mu}&=0.\label{eq:3.2}
\end{align}
Contracting equation \eqref{eq:3.1} gives $R=\frac{-2}{d-2}(8\pi T-d\Lambda)$, where $T\equiv T^\mu_{\phantom\mu\mu}$, so that equations of motion \eqref{eq:3.1} simplify to
\begin{eqnarray}\label{eq:eomcoolchap6}
R_{\mu\nu}-\dfrac{2}{d-2}\Lambda g_{\mu\nu}=8\pi\left(T_{\mu\nu}-\dfrac{T}{d-2}g_{\mu\nu}\right).
\end{eqnarray}
In four dimensions it follows from the definition of $T_{\mu\nu}$ that  $T=0$, and equations of motion \eqref{eq:eomcoolchap6} become $ R_{\mu\nu}-\frac{2}{d-2}\Lambda g_{\mu\nu}= 8\pi T_{\mu\nu}$. Note that this is no longer true in $d\neq 4$ where in general $T\neq0$.\\

Suppose that equations of motion \eqref{eq:3.2} are satisfied. Then, the associated electromagnetic energy momentum tensor is conserved, $T^{\mu\nu}_{\phantom{\mu\nu};\mu}=0$ \cite{Felsager}. This fact, and the contracted Bianchi identitites  in $d$ dimensions for the metric, allows one to organize the field equations of the Einstein-Maxwell system in a very convenient way, as in the pure gravity case. A general expression for the equations of motion is proven, valid in $d$ dimensions with and without a cosmological constant. This expression is used explicitly in sections 6.2 and 6.3 when solving equations of motion. Let the equations of motion $R_{\mu\nu}-\frac{2}{d-2}\Lambda g_{\mu \nu}-8\pi\left(T_{\mu\nu}-\frac{T}{d-2}g_{\mu\nu}\right)$ be denoted by $E_{\mu\nu}$ for briefty.  Then, one has\footnote{This result is explicitly used in four dimensions in \cite{vandeBurg06051969}, but the proof was not given. In this thesis, the result is proven in $d$ dimensions and includes the presence of a cosmological constant.}
\begin{theorem}
\begin{eqnarray}
g^{\alpha\epsilon}\left(E_{\mu\alpha,\epsilon}-\dfrac{1}{2}E_{\alpha\epsilon,\mu}-\Gamma^{\sigma}_{\alpha\epsilon}E_{\mu\sigma}\right)=0.\label{eq:bianchi}
\end{eqnarray}
\end{theorem}
The proof goes as follow:
\begin{itemize}
\item
In the case of a vanishing cosmological constant,\\  $E_{\mu\nu}=R_{\mu\nu}-8\pi\left(T_{\mu\nu}-\frac{T}{d-2}g_{\mu\nu}\right)$ and one has\\
\begin{align*}
0=&g^{\alpha\epsilon}(G_{\mu\alpha}-8\pi T_{\mu\alpha})_{;\epsilon}=\\
&=g^{\alpha\epsilon}(R_{\mu\alpha;\epsilon}-\dfrac{1}{2}R_{\alpha\epsilon;\mu}-8\pi T_{\mu\alpha;\epsilon})\\
&=g^{\alpha\epsilon}\left[(R_{\mu\alpha}-8\pi T_{\mu\alpha})_{,\epsilon}-\dfrac{1}{2}R_{\alpha\epsilon,\mu}-\Gamma^\sigma_{\alpha\epsilon}(R_{\mu\sigma}-8\pi T_{\mu\sigma})+8\pi \Gamma^\sigma_{\mu\epsilon}T_{\sigma\alpha}\right]\\
&=g^{\alpha\epsilon}\left[(R_{\mu\alpha}-8\pi T_{\mu\alpha})_{,\epsilon}-\dfrac{1}{2}(R_{\alpha\epsilon}-8\pi T_{\alpha\epsilon})_{,\mu}-\Gamma^\sigma_{\alpha\epsilon}(R_{\mu\sigma}-8\pi T_{\mu\sigma})-4\pi T_{\alpha\epsilon;\mu}\right]\\
&=g^{\alpha\epsilon}[E_{\mu\alpha,\epsilon}-\dfrac{8\pi}{d-2}(Tg_{\mu\alpha})_{,\epsilon}-\dfrac{1}{2}E_{\alpha\epsilon,\mu}+\dfrac{4\pi}{d-2}(Tg_{\alpha\epsilon})_{,\mu}-\Gamma^{\sigma}_{\alpha\epsilon}E_{\mu\sigma}+\notag\\
&\hspace{1cm}+\dfrac{8\pi}{d-2}\Gamma^\sigma_{\alpha\epsilon}T_{\mu\sigma}-4\pi T_{\alpha\epsilon;\mu}]\\
&=g^{\alpha\epsilon}\left[E_{\mu\alpha,\epsilon}-\dfrac{8\pi}{d-2}(Tg_{\mu\alpha})_{;\epsilon}-\dfrac{1}{2}E_{\alpha\epsilon,\mu}+\dfrac{4\pi}{d-2}(Tg_{\alpha\epsilon})_{;\mu}-\Gamma^{\sigma}_{\alpha\epsilon}E_{\mu\sigma}-4\pi T_{\alpha\epsilon;\mu}\right]\\
&=g^{\alpha\epsilon}\left[E_{\mu\alpha,\epsilon}-\dfrac{1}{2}E_{\alpha\epsilon,\mu}-\Gamma^{\sigma}_{\alpha\epsilon}E_{\mu\sigma}\right]+\dfrac{T_{;\mu}}{d-2}\left(-8\pi+4\pi d-4\pi (d-2)\right)\\
&=g^{\alpha\epsilon}\left(E_{\mu\alpha,\epsilon}-\dfrac{1}{2}E_{\alpha\epsilon,\mu}-\Gamma^{\sigma}_{\alpha\epsilon}E_{\mu\sigma}\right).
\end{align*}

\item This result can be generalized in the case of gravity with a non-vanishing cosmological constant. Indeed, with $E_{\mu\nu}=R_{\mu\nu}-\dfrac{2}{d-2}\Lambda g_{\mu\nu}-8\pi\left(T_{\mu\nu}-\dfrac{T}{d-2}g_{\mu\nu}\right)$ a similar computation as before shows that
\begin{align*}
&g^{\alpha\epsilon}(G_{\mu\alpha}-8\pi T_{\mu\alpha})_{;\epsilon}=
g^{\alpha\epsilon}\left(E_{\mu\alpha,\epsilon}-\dfrac{1}{2}E_{\alpha\epsilon,\mu}-\Gamma^{\sigma}_{\alpha\epsilon}E_{\mu\sigma}\right).
\end{align*}
This concludes the proof of \eqref{eq:bianchi}.
\end{itemize}

\section{Einstein-Maxwell system in three dimensions}

In three spacetime dimensions, the following gauge fixing ansatz for the metric and gauge potential is made, in terms of coordinates $(u,r,\phi)$,
\begin{align}
g_{\mu\nu}&=
\begin{pmatrix}
e^{2\beta}\dfrac{V}{r}+r^2U^2	&	-e^{2\beta}	&	-r^2U\\
-e^{2\beta}	&	0	&	0\\
-r^2U&	0	&r^2
\end{pmatrix},
g^{\mu\nu}=
\begin{pmatrix}
0	&	-e^{-2\beta}	&	0\\
-e^{-2\beta}	&	-\dfrac{V}{r}e^{-2\beta}	&	-Ue^{-2\beta}\\
0	&	-Ue^{-2\beta}	&	\dfrac{1}{r^2}
\end{pmatrix},\label{eq:2.1}\\
A_\mu&=
\begin{pmatrix}
A_u,0,A_\phi
\end{pmatrix},\label{eq:2.2}
\end{align}
in agreement with the gauge ansatz considered in the asymptotic symmetries analysis of the Einstein-Yang-Mills system in chapter 5, when restricted to the case of abelian gauge fields. The boundary conditions of these fields will be introduced in section 6.2.2.

\subsection{Integration procedure}
First, equations of motion \eqref{eq:3.2} must be solved in order to use the result \eqref{eq:bianchi}.
The ansatz  \eqref{eq:2.1} implies the following relations
\begin{eqnarray}
\Gamma^u_{ur}+\Gamma^r_{rr}+\Gamma^\phi_{r\phi}=\dfrac{1}{r}+2\beta_r,\hspace{0.2cm}
\Gamma^u_{u\phi}+\Gamma^r_{r\phi}+\Gamma^\phi_{\phi\phi}=2\beta_\phi,\hspace{0.2cm}
\Gamma^\phi_{\phi u}+\Gamma^u_{uu}+\Gamma^r_{r u}=2\beta_u,\notag
\end{eqnarray}
and moreover let $\mathcal F^{\mu\nu}$ be defined by $F^{\mu\nu}=e^{-2\beta}\mathcal F^{\mu\nu}$.\\
The field equations \eqref{eq:3.2} then become
\begin{align}
F^{u\sigma}_{\phantom{u\sigma};\sigma}&=e^{-2\beta}\left(
\mathcal F^{ur}_{\phantom{ur},r}+\dfrac{1}{r}\mathcal F^{ur}+\mathcal F^{u\phi}_{\phantom{u\phi},\phi}\right)\label{eq:maxu},\\
F^{\phi\sigma}_{\phantom{u\sigma};\sigma}&=e^{-2\beta}\left(
\mathcal F^{\phi u}_{\phantom{\phi u},u}+\dfrac{1}{r}\mathcal F^{\phi r}+\mathcal F^{\phi r}_{\phantom{\phi r},r}\right)\label{eq:maxphi},\\
F^{r\sigma}_{\phantom{u\sigma};\sigma}&=e^{-2\beta}\left(
\mathcal F^{ru}_{\phantom{ru},u}+\mathcal F^{r\phi}_{\phantom{r\phi},\phi}\right)\label{eq:maxr}.
\end{align}

Suppose that the two following equations of motion (called main equations) are satisfied,
\begin{eqnarray}\label{eq:maxmain}
F^{u\sigma}_{\phantom{u\sigma};\sigma}=0,\hspace{1cm} F^{\phi\sigma}_{\phantom{u\sigma};\sigma}=0.
\end{eqnarray}
Then, by dropping the $e^{-2\beta}$, taking the $r$--derivative of \eqref{eq:maxr} and by combining with the $u$--derivative of \eqref{eq:maxu} and the $\phi$--derivative of \eqref{eq:maxphi}, one finds the following differential equation for $(\mathcal F^{ru}_{\phantom{ru},u}+\mathcal F^{r\phi}_{\phantom{r\phi},\phi})\equiv D$,
\begin{align}
\partial_rD&=
-\dfrac{1}{r}D\notag\\
\dfrac{1}{r}\partial_r(rD)&=0
\hspace{1cm}\Rightarrow\hspace{1cm}
D=e^{2\beta}F^{r\sigma}_{\phantom{r\sigma};\sigma}=\dfrac{\text{h}(u,\phi)}{r}.\label{eq:maxsupp}
 \end{align}
where $h(u,\phi)$ is an arbitrary function of its arguments.
The equation of motion $D=0$ restricts the function $h$ to vanish.\\

When equations \eqref{eq:3.2} are solved, relations \eqref{eq:bianchi} allow one to solve the equations of motion \eqref{eq:3.1} in a very convenient way.\\
First, suppose that the three following equations of motion (main equations), are satisfied\footnote{In three dimensions: $E_{\mu\nu}=R_{\mu\nu}-2\Lambda g_{\mu\nu}-8\pi\left(T_{\mu\nu}-Tg_{\mu\nu}\right)$.}
\begin{eqnarray}
E_{rr}=0,\hspace{1cm}E_{r\phi}=0,\hspace{1cm}E_{\phi\phi}=0.\label{eq:main}
\end{eqnarray}
Then, the identities \eqref{eq:bianchi} reduce to
\begin{align}
\mu=r:\hspace{1cm}&-g^{\alpha\epsilon}\Gamma^u_{\alpha\epsilon}E_{ur}=0,\label{6:bianchi1}\\
\mu=\phi:\hspace{1cm}&g^{ur}(E_{\phi u,r}-E_{ur,\phi})-g^{\alpha\epsilon}\Gamma^u_{\alpha\epsilon}E_{u\phi}=0,\label{6:bianchi2}\\
\mu=u:\hspace{1cm}&
g^{ur}E_{uu,r}+g^{rr}E_{ur,r}+g^{r\phi}(E_{ur,\phi}+E_{u\phi,r})+g^{\phi\phi}E_{u\phi,\phi}-\notag\\
&-g^{\alpha\epsilon}\Gamma^u_{\alpha\epsilon}E_{uu}-g^{\alpha\epsilon}\Gamma^r_{\alpha\epsilon}E_{ur}-g^{\alpha\epsilon}\Gamma^\phi_{\alpha\epsilon}E_{\phi u}=0.\label{6:bianchi3}
\end{align}

From the gauge fixing of the metric \eqref{eq:2.1}, we get $g^{\alpha\epsilon}\Gamma^u_{\alpha\epsilon}=\frac{e^{-2\beta}}{r}$ and therefore equation \eqref{6:bianchi1} implies
\begin{eqnarray}
E_{ur}=0\label{eq:trivial}.
\end{eqnarray}
Equation of motion $E_{ur}=0$ is called the trivial equation because it is automatically satisfied as a  consequence of the main equations \eqref{eq:main}.
This result  implies that \eqref{6:bianchi2} becomes
\begin{align}
-e^{-2\beta}\left(\partial_r+\dfrac{1}{r}\right)E_{\phi u}&=0\notag\\
-\dfrac{e^{-2\beta}}{r}\partial_r(rE_{\phi u})&=0\hspace{1cm}\Rightarrow\hspace{1cm}
E_{\phi u}=\dfrac{\text{f}(u,\phi)}{r}\label{eq:supppl1}.
\end{align}
where $f(u,\phi)$ is an arbitrary function of its arguments.
This equation (called supplementary equation) implies that the equations of motion $E_{\phi u}=0$ restricts that the function $f(u,\phi)$ vanishes.
Finally, when equations of motion $E_{ur}=0$ and $E_{u\phi}=0$ are satisfied, \eqref{6:bianchi3} becomes
\begin{align}
-e^{-2\beta}\left(\partial_r+\dfrac{1}{r}\right)E_{uu}&=0\notag\\
-\dfrac{e^{-2\beta}}{r}\partial_r(rE_{uu})&=0\hspace{1cm}\Rightarrow\hspace{1cm}
E_{uu}=\dfrac{\text{g}(u,\phi)}{r}\label{eq:supppl2}.
\end{align}
where $g(u,\phi)$ is an arbitrary function of its arguments.\\

In summary, the following hierarchy can be introduced between the equations of motion, for computational convenience:
\begin{itemize}
\item 5 main equations: $E_{rr}=0,\,F^{u\sigma}_{\phantom{u\sigma};\sigma}=0,\,E_{r\phi}=0,\,E_{\phi\phi}=0,\,F^{\phi\sigma}_{\phantom{\phi\sigma};\sigma}=0$,
\item 1 trivial equation: $E_{ur}=0$ (automatically satisfied),
\item 3 supplementary equations: $F^{r\sigma}_{\phantom{r\sigma};\sigma}=0,\,E_{u\phi}=0,\,E_{uu}=0.$
\end{itemize}
As will be seen in the next section, the main equations produce arbitrary integration functions (with respect to $r$). Some of these functions will be set to zero in order to preserve the gauge fixing conditions imposed on the metric and gauge field, while the time evolution of the remaining constants will be constrained by supplementary equations.

\subsubsection*{Remark on main equations}
In the case of pure gravity in four dimensions, there is a further splitting \cite{Sachs:1962wk} between main equations that contain time derivative (called the standard equations) and those that do not (called hypersurface equation), see chapter 2. This splitting does not occur for Einstein-Maxell set-up.



\subsection{General solution when $\Lambda =0$}
For simplicity, let us consider the case of a vanishing cosmological constant.
Besides the gauge fixing \eqref{eq:2.1}\eqref{eq:2.2}, some additional fall-off conditions are imposed on the coefficients of the metric field,
\begin{eqnarray}
\beta=\mathcal O(r^{-1}),\hspace{1cm}U=\mathcal O\left(\dfrac{\ln r}{r^{2}}\right),\label{fo1}
\end{eqnarray}
As in the four dimensional case, the function $V$ is completely determined by the other boundary conditions, due to the equations of motion.

\subsubsection{Case 1: $A_u=O(r^{-1})$ and $A_\phi=O(ln~r)$}
In addition to the fall-off conditions for the metric field \eqref{fo1}, the following boundary conditions are imposed on the gauge potential,
\begin{eqnarray}
A_u=\mathcal O(r^{-1}),\hspace{1cm}A_\phi=\alpha_\phi \ln r+\mathcal O(1),\label{fo2}
\end{eqnarray}
in agreement with the boundary conditions considered in chapter 5 in the case of the Einstein-Yang-Mills system.\\

One has the following result concerning the solution of the equations of motion for the fields $g$ and $A$,
\begin{theorem}
If one prescribes $A_\phi$ and the the gauge fixing conditions \eqref{fo1} and \eqref{fo2} on hypersurface $u=u_0$, then the 6 main equations can be integrated and determine completely $\beta,U,V,A_{u}$ up to 2 arbitrary integration functions: $M(u,\phi), N(u,\phi)$ and determine also the $u$--derivative of $A_{\phi}$.   The time evolution of integration functions $M(u,\phi),N(u,\phi)$ is determined by the supplementary equations. 
\end{theorem}

The proof goes as follows.
\footnote{
 Christoffel symbols associated with the metric \eqref{eq:2.1} were computed in \cite{Barnich:2010eb}, up to one misprint in $\Gamma^r_{uu}$ (the last term here below):
\begin{eqnarray*}
\Gamma^r_{uu}=\dfrac{V}{r}\beta_{u}-\dfrac{V_{u}}{2r}+\dfrac{V^2}{r^2}\beta_{r}+\dfrac{V}{2r^2}V_{r}-\dfrac{V^2}{2r^3}+e^{-2\beta}VU^2+re^{-2\beta}UVU_{r}+\notag\\
+\dfrac{UV}{r}\beta_{\phi}+\dfrac{U}{2r}V_{\phi}+r^2e^{-2\beta}U^2U_{\phi},
\end{eqnarray*}
where indices on functions indicate derivatives. The Ricci tensors
$R_{rr},R_{r\phi}$ and $R_{\phi\phi}$ were also computed in \cite{Barnich:2010eb}, up to one misprint in $R_{r\phi}$ (first term here below):
\begin{eqnarray*}
R_{r\phi}=-\beta_{\phi r}+\dfrac{\beta_\phi}{r}-r^2e^{-2\beta}\beta_r U_r+\dfrac{3}{2}re^{-2\beta}U_r+\dfrac{r^2}{2}e^{-2\beta}U_{rr}.
\end{eqnarray*}
}
\begin{itemize}
\item The equation of motion $R_{rr}=8\pi (T_{rr}-Tg_{rr})$ gives
\begin{eqnarray}
\beta_r=\dfrac{1}{r}(F_{r\phi})^2,\hspace{1cm}\Rightarrow \beta=-\int_r^\infty dr~~\dfrac{1}{r'}(F_{r\phi})^2.
\end{eqnarray}
Note that $F_{r\phi}$ is a known function (since $A_\phi$ is a initial data and $A_r=0$), and that there is no integration function due to the fall-off condition $\beta=O(r^{-1})$.\\

\item Using \eqref{eq:maxu}, the equation $F^{u\sigma}_{\phantom{u\sigma};\sigma}=0$ becomes
\begin{eqnarray}
\mathcal F^{ur}_{\phantom{ur},r}+\dfrac{1}{r}\mathcal F^{ur}+\mathcal F^{u\phi}_{\phantom{u\phi},\phi}=0.\label{625}
\end{eqnarray}
Let us define $E\equiv e^{-2\beta}(F_{ur}-UF_{r\phi})$ so that
$\mathcal F^{ur}=-E$ and
$\mathcal F^{u\phi}=-\frac{1}{r^2}F_{r\phi}$.
Equation \eqref{625} is a first order differential equation for $E$. Writing $g\equiv-\frac{1}{r^2}(F_{r\phi})_{,\phi}$ one finds
\begin{eqnarray}
E_{,r}+\dfrac{E}{r}=g\hspace{0.5cm}\Rightarrow
E=\dfrac{Q(u,\phi)-\int_r^\infty dr'~~(r'g)}{r},
\end{eqnarray}
with $Q(u,\phi)$ an integration function. Boundary conditions \eqref{fo1}\eqref{fo2} imply $E=\mathcal O(r^{-2})$, therefore requiring $Q(u,\phi)$ to be zero. The solution for $E$ is thus
\begin{eqnarray}\label{e2.8}
E=\dfrac{1}{r}\left(\int_r^\infty dr'~~\frac{\partial_\phi F_{r\phi}}{r'}\right).
\end{eqnarray}

\item 
By defining $n\equiv\dfrac{r^2}{2}e^{-2\beta}U_r$, equation $R_{r\phi}=8\pi (T_{r\phi}-Tg_{r\phi})$ is
\begin{eqnarray}
R_{r\phi}=-\beta_{\phi r}+\dfrac{\beta_\phi}{r}+\left(\partial_r+\dfrac{1}{r}\right)n,
\end{eqnarray}
and becomes a first order differential equation for $n$. Writing $h \equiv -2F_{r\phi}E+\beta_{r\phi}-\dfrac{\beta_\phi}{r}$, one finds
\begin{eqnarray}
n_{,r}+\dfrac{n}{r}=h\hspace{0.6cm}\Rightarrow
n=\dfrac{N(u,\phi)-\int_r^\infty dr'~~(r'h)}{r}\label{eq:n},
\end{eqnarray}
with $N(u,\phi)$ an integration function.
From \eqref{eq:n}, the definition of $n$ and the boundary condition \eqref{fo1}, one finds
\begin{eqnarray}\label{628}
U=\int_{r}^\infty dr' ~\dfrac{2e^{2\beta}}{r'^2}\left(\dfrac{N}{r'}+\int_{r'}^\infty d\tilde r(2\tilde r F_{\tilde r\phi}E-\tilde r\beta_{\tilde r\phi}+\beta_\phi)\right),\label{eq:N}
\end{eqnarray}
the leading of which being $U= -\frac{N(u,\phi)}{r^2}+\mathcal O(r^{-3})$.
Results \eqref{e2.8} and \eqref{628} determine $A_u$ uniquely, by using the definition of $E$.\\

\item Equation $R_{\phi\phi}=8\pi (T_{\phi\phi}-Tg_{\phi\phi})$ is a differential equation for $V$,
\begin{align}
V_{r}-\dfrac{V}{r}=2r^2e^{2\beta}E^2-2rU_{\phi}+2e^{2\beta}\beta_{\phi\phi}-r^2U_{\phi r}+\notag\\
+2e^{2\beta}(\beta_\phi)^2+\dfrac{e^{-2\beta}}{2}r^4(U_r)^2\label{629}.
\end{align}
Let the auxiliary quantities $q$ and $W$ be defined by the relations
\begin{align}
W&\equiv \dfrac{V}{r},\notag\\
q&\equiv
2r^2e^{2\beta}E^2-2rU_{\phi}+2e^{2\beta}\beta_{\phi\phi}-r^2U_{\phi r}+2e^{2\beta}(\beta_\phi)^2+\dfrac{e^{-2\beta}}{2}r^4(U_r)^2.\notag
\end{align}
Then the equation of motion \eqref{629} reads
\begin{align}
rW_{,r}&=q\hspace{1cm}\Rightarrow\hspace{1cm}
\dfrac{V}{r}=\left[M(u,\phi)-\int_r^\infty~~dr' \dfrac{q}{r'}\right],
\end{align}
with $M(u,\phi)$ an integration function.\\


\item Finally, by using \eqref{eq:maxphi} and the definition of $E$, the last main equation $F^{\phi\sigma}_{\phantom{\phi\sigma};\sigma}=0$ becomes
\begin{eqnarray}\label{631}
\mathcal F^{\phi u}_{\phantom{\phi u},u}+\dfrac{1}{r}\mathcal F^{\phi r}+\mathcal F^{\phi r}_{\phantom{\phi r},r}=0
\end{eqnarray}
with: $\mathcal F^{\phi r}=-UE+\frac{1}{r^2}\left(F_{u\phi}+\frac{V}{r}F_{r\phi}\right)$ and $\mathcal F^{u\phi}=-\frac{1}{r^2}F_{r\phi}$.
Equation of motion \eqref{631} thus becomes
\begin{eqnarray}\label{632}
F_{r\phi,u}=r^2\left(\partial_r+\dfrac{1}{r}\right)\left[UE-\dfrac{1}{r^2}\left(F_{u\phi}+\dfrac{V}{r}F_{r\phi}\right)\right]
\end{eqnarray} 
which is a differential equation governing the $u$--dependance of the initial data $A_{\phi}$. Using the fact that $(\partial_r+1/r)X=1/r\partial_r(rX)$, and the general identity $\left(\alpha\partial_r-\frac{1}{r}\right)X=\alpha~~ r^{1/\alpha}~~ \partial_r\left(r^{-1/\alpha}X\right)$, equation \eqref{632} reduces to
\begin{eqnarray}
2\sqrt r~~\partial_u\partial_r\left(\dfrac{ A_\phi}{\sqrt r}\right)=r\partial_r\left(rUE-\dfrac{V}{r^2}\partial_{r}A_{\phi}+\dfrac{1}{r}\partial_\phi A_u\right)
\label{eq.16}
\end{eqnarray}
So far, no assumption has been made about the development of the initial data $A_\phi$ in terms of $r$. However, in order to extract more information from equation \eqref{eq.16}, one assumes that all the data can be expanded in inverse power of $r$, as in the case of pure gravity in four dimensions. In particular, $A_\phi$ is assumed to be of the form
\begin{eqnarray}
A_\phi=\alpha_{\phi}(u,\phi)\ln r+A_\phi^0(u,\phi)+\mathcal O(r^{-1}),\label{azerty}
\end{eqnarray}
with only one logarithmic term, in the leading term of the expansion.\\

Due to the derivative $\partial_u\partial_r$ in the left hand side of \eqref{eq.16}, the $u$-derivative of all terms in the expansion of $A_\phi$ are determined by expression \eqref{eq.16}, except the one proportional to $\sqrt r$. The $u$-derivative of this term is not determined by the above equation, and must be given as part of the initial data.\\

However, there is no such a term in the expansion \eqref{azerty} of $A_\phi$, due to the boundary condition \eqref{fo2}. Solving equation \eqref{eq.16} at order $\mathcal O(\ln r)$ and $\mathcal O(r^{-1})$ gives
\begin{eqnarray}
\partial_u \alpha_\phi=0,\hspace{1cm}\partial_u A_\phi^0=0.
\end{eqnarray}
\end{itemize}

The solution to the 3 supplementary equations goes as follows.
\begin{itemize}
\item The last supplementary equation $D=\frac{h}{r}$ can be computed using \eqref{eq:maxsupp}. The equation of motion $D=0$ does not lead to any new constraint.
\item The equation of motion $E_{uu}=0$ yields
\begin{eqnarray}
E_{uu}=-\dfrac{\partial_u M}{2r}.
\end{eqnarray}
This gives differential equation on the integration constant $M$,
\begin{eqnarray}
\partial_u M=0\hspace{1cm}\Rightarrow\hspace{1cm}M=M(\phi).
\end{eqnarray}

\item For the supplementary equation $E_{u\phi}$, one finds
\begin{eqnarray}
E_{u\phi}=\dfrac{1}{r}\left(\dfrac{\partial_\phi M}{2}-\partial_u N\right).
\end{eqnarray}
 The integration constant $N$ therefore satisfies
\begin{eqnarray}
\partial_u N=\dfrac{1}{2}\partial_\phi M\hspace{1cm}\Rightarrow\hspace{1cm}N(u,\phi)=\Xi(\phi)+\dfrac{u}{2}\partial_\phi M(\phi).
\end{eqnarray}
\end{itemize}


\subsubsection{Case 2: $A_u=\mathcal O(ln~r)$ and $A_\phi=\mathcal O(r^{-1})$}
In this second case, the boundary conditions imposed on the gauge potential are given by
\begin{eqnarray}
A_u=\mathcal O(\ln r),\hspace{1cm}A_\phi=\mathcal O(r^{-1}).\label{fal20}
\end{eqnarray}
The integration of the equations of motion is the same as for case 1 and will not be reproduced here. However, the boundary conditions \eqref{fal20} of the gauge potential are different and affect the solution. Let us assume that all the data can be expanded in inverse power of $r$,
\begin{eqnarray}
A_{\phi}=\dfrac{A_\phi^0}{r}+ \mathcal O(r^{-2}).
\end{eqnarray}
Then, the six main equations yield
\begin{align}
\beta&=\mathcal O(r^{-4}),\\
E&=\dfrac{Q(u,\phi)}{r}+\mathcal O(r^{-3}),\hspace{1cm}\Rightarrow F_{ur}=\dfrac{Q}{r}+\mathcal O(r^{-3})\notag,\\
&\phantom{=\dfrac{Q(u,\phi)}{r}+O(r^{-3}),}\hspace{1cm}\Rightarrow A_{u}=-Q\ln {r}+\mathcal O(r^{-2}),\\
U&=-\dfrac{N}{r^2}+\mathcal O(r^{-3}),\\
\dfrac{V}{r}&=2 Q^2\ln r+M(u,\phi)+\mathcal O(r^{-1}).
\end{align}
The equation of motion \eqref{eq.16} restricts the form of the solution, due to the perturbative expansion. Solving this equation at order $\mathcal O(\ln r), \mathcal O(r^{-1})$ implies
either $Q=0$ or the conditions $A_\phi=0$ and $\partial_\phi Q=0$.
The second choice implies that the gauge potential is  of the form $A_r=0=A_\phi$ and is called the monopole solution, in analogy with the case studied in Minkowskian gravity \cite{janis:902}.

\subsubsection{Monopole solution}
The explicit form of the monopole solution is given by fields \eqref{eq:2.1},\eqref{eq:2.2} with
\begin{align}
A_\phi&=0,\\
\beta&=0,\\
E&=\dfrac{Q(u,\phi)}{r}=F_{ur},\hspace{1cm}\Rightarrow A_u=-Q(u,\phi)\ln r\\
U&=-\dfrac{N(u,\phi)}{r^2}\\
\dfrac{V}{r}&=2Q^2(u,\phi)\ln r+M(u,\phi)-\dfrac{N^2(u,\phi)}{r^2}
\end{align}
The main condition \eqref{eq.16} imposes $\partial_\phi Q=0$ and $N=0$. Finally, the three supplementary equations restrict $M$ to be constant.\\

In summary, the metric and gauge field for the monopole solution read
\begin{eqnarray}
g_{\mu\nu}=
\begin{pmatrix}
2Q^2(u)\ln r +M 	&	1	&	0	\\
1				&	0	&	0	\\
0				&	0	&	r^2	
\end{pmatrix},
\hspace{1cm}
A_\mu=(-Q(u)\ln r,0,0),
\end{eqnarray}
and this solution is static.

\subsection{Comment on a possible rotation trick}
The integration procedure described in the previous section does not seem to produce a solution that is both charged and rotating.
In the case of a negative cosmological constant, such a solution exists and is the charged rotating BTZ, derived implicitly in \cite{Martinez:1999qi,Clement}. It could be interesting to find an explicit form for this solution.\\

Usually, one of the motivations for studying lower dimensional gravity is the fact that it is more trackable than four dimensional gravity. From this perspective, three dimensional gravity plays the role of  a toy model to get insight in the four dimensional case.  To get an explicit expression for the charged rotating BTZ black hole, one first idea could be to apply the inverse logic and use the four dimensional case to get a hint of the three dimensional solution.\\

In four dimensions, the rotating charged black hole (Kerr-Newman \cite{Newman}) was found by using a rotation trick, due to Newman and Janis \cite{Janis}, applied to the Reissner-Nordstrom metric. The original Newman-Janis trick is only efficient to produce the form of the metric; the form of the gauge field must be found by direct integration of the fields equations. However, a recent extension \cite{Keane} of the Newman-Janis' algorithm allows one to obtain the form of the gauge field without having to solve the equations of motion, by adding a null rotation in the original algorithm of Newman-Janis.\\

Interestingly, the Newman-Janis' rotation algorithm can be adapted to be applied in the case of gravity in three dimensions, as noted by Kim \cite{Kim}. One natural question one can wonder is whether the modified algorithm can also be adapted to three dimensions, by using the prescription of  Kim. A positive answer to this question will produce the solution of interest, i.e. the charged rotating BTZ black hole in an explicit form. This question is investigated in appendix B, but the answer appears to be negative.\\

\section{Einstein-Maxwell system in four dimensions}

The starting point is the following action for Einstein-Maxwell theory in four dimensions without cosmological constant,
\begin{eqnarray}
S=\dfrac{1}{16\pi}\int d^4x\sqrt{-g}\left(R-\dfrac{1}{4}F^2\right),\label{eq:action}
\end{eqnarray}
with $F^2=F_{\mu\nu}F^{\mu\nu}$ and $F_{\mu\nu}=\partial_\mu A_\nu-\partial_\nu A_\mu$. Equations of motion are
\begin{align}
R_{\mu\nu}&=8\pi T_{\mu\nu},\label{eq:eomcool}\\
F^{\mu\nu}_{\phantom{\mu\nu};\mu}&=0.\label{6:eq:3.2}
\end{align}
The following gauge fixing ansatz for the metric and gauge potential is made, in terms of coordinates $(u,r,x^A)$,
\begin{align}
g_{\mu\nu}&=
\begin{pmatrix}
\dfrac{V}{r}e^{2\beta}+g_{AB}U^AU^B	&-e^{2\beta}	&-g_{BC}U^C\\
-e^{2\beta}&0	&0\\
-g_{AC}U^C&0&g_{AB}
\end{pmatrix},\label{ladh}\\
A_\mu&=
\begin{pmatrix}
A_u,0,A_A
\end{pmatrix},
\end{align}
in agreement with the gauge ansatz considered in the asymptotic symmetries analysis of the Einstein-Yang-Mills system in chapter 5, when restricted to the case of abelian gauge fields. The boundary conditions of these fields are
\begin{align}
\beta&=\mathcal O(r^{-2})=U^A,\hspace{1cm}g_{AB}=r^2\bar\gamma_{AB}+\mathcal O(r),\label{18h19}\\
A_u&=\mathcal O(r^{-1}),\hspace{1cm}A_A=\mathcal O(1),\label{18h20}
\end{align}
where $\bar\gamma_{AB}$ is the metric of the two-sphere. Moreover, the determinant of $g_{AB}$ is fixed and given to be det$(g_{AB})=r^4$det$(\bar\gamma_{AB})$. The fall-off condition of $V$ is a consequence of boundary conditions \eqref{18h19},\eqref{18h20}, due to the equations of motion.

\subsection{Integration procedure}
First, equations \eqref{6:eq:3.2} must be solve in order to use identities \eqref{eq:bianchi}. One has the
\begin{theorem}
 If the 3 equations of motion $F^{u\sigma}_{\phantom{\mu\sigma};\sigma}=0,F^{A\sigma}_{\phantom{A\sigma};\sigma}=0$ are satisfied on-shell, 
then
 the following relation is valid off-shell: $e^{2\beta}F^{r\sigma}_{\phantom{\mu\sigma};\sigma}=\frac{h(u,x^A)}{r^2}$.
\end{theorem}
The proof is the following. 
Suppose that the three following equations are satisfied (the main equations)
\begin{eqnarray}
F^{u\sigma}_{\phantom{\mu\sigma};\sigma}=0,\hspace{1cm}F^{A\sigma}_{\phantom{A\sigma};\sigma}=0.\label{mainEM}
\end{eqnarray}
Then, a short computation shows that
\begin{align}
\partial_r(F^{r\sigma}_{\phantom{\mu\sigma};\sigma})=-\left(2\beta_{,r}+\dfrac{2}{r}\right)\left(F^{r\sigma}_{\phantom{\mu\sigma};\sigma}\right)-\partial_uF^{u\sigma}_{\phantom{\mu\sigma};\sigma}-\partial_AF^{A\sigma}_{\phantom{\mu\sigma};\sigma}-\Gamma_{AE}^EF^{A\sigma}_{\phantom{\mu\sigma};\sigma},
\end{align}
or equivalently, taking into account the main equations \eqref{mainEM},
\begin{align}
\left(\partial_r+\dfrac{2}{r}\right)\left(e^{2\beta}F^{r\sigma}_{\phantom{\mu\sigma};\sigma}\right)&=0\notag\\
\dfrac{1}{r^2}\partial_r\left(r^2 e^{2\beta}F^{r\sigma}_{\phantom{\mu\sigma};\sigma}\right)&=0\hspace{1cm}\Rightarrow\hspace{1cm}e^{2\beta}F^{r\sigma}_{\phantom{\mu\sigma};\sigma}=\dfrac{h(u,x^A)}{r^2}.
\end{align}
This means that the equation of motion $F^{r\sigma}_{\phantom{\mu\sigma};\sigma}=0$ is automatically satisfied at all orders, except at order $\mathcal O(r^{-2})$, due to main equations \eqref{mainEM}. This concludes the proof.\\

The identities \eqref{eq:bianchi} were derived in $d$ dimensions.
By taking on $d=4$, one has the
\begin{theorem} If the 6 equations of motion $E_{rr}=0,E_{rA}=0,E_{AB}=0$ are satisfied on-shell, then the 4 following relations are valid off-shell: $E_{ur}=0$ and $E_{A u}=\dfrac{f_A(u,x^C)}{r^2}$, $E_{uu}=\dfrac{g(u,x^C)}{r^2}$.
\end{theorem}
The proof goes as follow.
Suppose that the 6 following equations of motion (main equations) are satisfied,
\begin{eqnarray}
E_{rr}=0,\hspace{1cm}E_{rA}=0,\hspace{1cm}E_{AB}=0,\label{6:eq:main}
\end{eqnarray}
then the identities \eqref{eq:bianchi} reduce to
\begin{align}
\mu=r:\hspace{1cm}&-g^{\alpha\epsilon}\Gamma^u_{\alpha\epsilon}E_{ur}=0,\label{eq:bianchi1}\\
\mu=A:\hspace{1cm}&g^{ur}(E_{A u,r}-E_{ur,A})-g^{\alpha\epsilon}\Gamma^u_{\alpha\epsilon}E_{uA}=0,\label{eq:bianchi2}\\
\mu=u:\hspace{1cm}&
g^{ur}E_{uu,r}+g^{rr}E_{ur,r}+g^{rA}(E_{ur,A}+E_{uA,r})+g^{AB}E_{uA,B}-\notag\\
&-g^{\alpha\epsilon}\Gamma^u_{\alpha\epsilon}E_{uu}-g^{\alpha\epsilon}\Gamma^r_{\alpha\epsilon}E_{ur}-g^{\alpha\epsilon}\Gamma^A_{\alpha\epsilon}E_{A u}=0.\label{eq:bianchi3}
\end{align}
From the form of the BMS metric \eqref{ladh}, one has $g^{\alpha\epsilon}\Gamma^u_{\alpha\epsilon}=g^{AB}\Gamma^u_{AB}=2\frac{e^{-2\beta}}{r}$. Equation  \eqref{eq:bianchi1} thus implies
\begin{eqnarray}
E_{ur}=0.\label{6:eq:trivial}
\end{eqnarray}
Equation $E_{ur}=0$ is called the trivial equation.\\
This result in \eqref{eq:bianchi2} implies
\begin{align}
-e^{-2\beta}\left(\partial_r+\dfrac{2}{r}\right)E_{Au}&=0\notag\\
-\dfrac{e^{-2\beta}}{r^2}\partial_r(r^2E_{Au})&=0\hspace{1cm}\Rightarrow\hspace{1cm}
E_{A u}=\dfrac{f_A(u,x^C)}{r^2}\label{6:eq:supppl1},
\end{align}
for some integration function $f_A(u,x^C)$.\\
Finally, equations $E_{ur}=0$ and $E_{u\phi}=0$ in \eqref{eq:bianchi3} gives
\begin{align}
-e^{-2\beta}\left(\partial_r+\dfrac{2}{r}\right)E_{uu}&=0\notag\\
-\dfrac{e^{-2\beta}}{r^2}\partial_r(r^2E_{uu})&=0\hspace{1cm}\Rightarrow\hspace{1cm}
E_{uu}=\dfrac{\text{fct}(u,x^A)}{r^2}\label{6:eq:supppl2},
\end{align}
with $g(u,x^C)$ an integration function.
This concludes the proof.\\


In summary, equations of motion \eqref{eq:eomcool} \eqref{6:eq:3.2} can be organized as follow, for computational convenience:
\begin{itemize}
\item 9 main:
$E_{rr}=0,\,F^{u\sigma}_{\phantom{u\sigma};\sigma}=0,\,E_{rA}=0,\,E_{AB}=0,\,F^{A\sigma}_{\phantom{A\sigma};\sigma}=0$,
\item 1 trivial: $E_{ur}=0$ (automatically satisfied),
\item 4 supplementary: $E_{uu}=0,\,E_{uA}=0,\,e^{2\beta}F^{r\sigma}_{\phantom{r\sigma};\sigma}=0.$
\end{itemize}
As we will see in the next section, main equations produce arbitrary integration functions (with respect to $r$), some of these functions will be set to zero in order to preserve boundary conditions, while the time evolution of the remaining constants will be constrained by supplementary equations.

\subsection{General solution}
The rest of this section is devoted to the proof  of the following result:
\begin{theorem}
If one prescribes $g_{AB}$ and  $A_A$ on hypersurface $u=u_0$, then:\\
The 9 main equations can be integrated and determine $\beta,U^A,V,F_{ur},F_{uA}$ up to arbitrary integration constants (with respect to $r$): $M(u_0,x^B), N^A(u_0,x^B),\\ Q(u_0,x^B)$, and determine also the $u$--derivative of $F_{rA}$ and of $g_{AB}$, up to the first subleading terms in $g_{AB}$ and up to the leading term in $A_A$.
 Trivial equation is then satisfied, and finally
  the supplementary equations determine the $u$-derivative of integration functions [$M(u_0,x^B),N^A(u_0,x^B),Q(u_0,x^B)$].\\
\end{theorem}
The proof is the following. First, the main equations are integrated:
\begin{itemize}
\item Equation of motion $E_{rr}$ gives
\begin{align}
\beta_r&=\dfrac{1}{4r^3}K^A_BK^B_A+\dfrac{r}{8}(\partial_rA_A)(\partial_rA_B)g^{AB},\\ \beta&=-\int_r^\infty\left(\dfrac{1}{4r^3}K^A_BK^B_A+\dfrac{r}{8}(\partial_rA_A)(\partial_rA_B)g^{AB}\right),
\end{align}
with $k ^A_B=\dfrac{\delta^A_B}{r}+\dfrac{K^A_B(r)}{r^2}, k^A_B\equiv g^{AC}(\half \partial_rg_{BC})$. There is no integration function, due to the boundary condition $\beta=O(r^{-2})$. The function $\beta$ is known in terms of initial data $g_{AB}$ and $A_A$.\\

\item The equation $F^{u\sigma}_{\phantom{u\sigma};\sigma}=0$ yields
\begin{eqnarray}
F^{u\sigma}_{\phantom{u\sigma};\sigma}&=e^{-2\beta}\left[\left(\partial_r+\dfrac{2}{r}\right)\mathcal F^{ur}+\left(\partial_A+^{(2)}\Gamma^E_{AE}\right)\mathcal F^{uA}\right],
\end{eqnarray}
with $\mathcal F^{\mu\nu}\equiv e^{2\beta}\mathcal F^{\mu\nu}.$
The equation of motion reduces to
\begin{align}
\partial_r A_u+U^A\partial_rA_A&=\dfrac{e^{2\beta}}{r^2}\left[Q(u,x^A)-\int_r^\infty r^2(\partial_A+^{(2)}\Gamma^E_{AE})g^{AB}\partial_rA_Bdr\right],\label{E}
\end{align}
with $Q(u,x^A)$ an integration function.\\

\item
As in the pure gravity case\cite{Barnich:2010eb}, let us introduce the auxiliary quantity $n_A\equiv\dfrac{e^{-2\beta}}{2}g_{AB}\partial_rU^B$. The equation $E_{rA}=0$ implies\\
\begin{align}
&\partial_r(r^2n_A)=r^2\left(\partial_r-\dfrac{2}{r}\right)\beta_A-^{(2)}D_B K^B_A+\notag\\
&+\dfrac{e^{-2\beta}}{2} (\partial_{r}A_A)\left(\partial_rA_u+(\partial_rA_B)U^B\right)+ \dfrac{\partial_{r}A_B}{2}(\partial_{A}A_C-\partial_CA_A)g^{BC}.\label{18h29}
\end{align}
This expression for $n_A$ is known in a closed form, due to equation \eqref{E} for $\left(\partial_rA_u+(\partial_rA_B)U^B\right)$. By using the definition of $n_A$, equation \eqref{18h29} can be integrated twice to give
\begin{eqnarray}
U^B=-\int_r^\infty dr'\left[\dfrac{2}{r^2}e^{2\beta}g^{AB}\left(N_A-\int_{r'}^\infty dr'' L_A\right)\right],
\end{eqnarray}
where $N_A$ in an integration function\footnote{The indice of $N_A$ is raised with the metric $\bar\gamma^{AB}$.}, and where $L_A$ denotes the right hand side of equation \eqref{18h29}.
The knowledge of $U^B$ allows one to get the solution for $A_u$, by using equation \eqref{E}.\\

\item The trace part equation $g^{AB}E_{AB}$ gives
\begin{align}
\partial_r V&-\left\{
e^{2\beta}r^2\left(\Delta\beta+\partial^\alpha\beta\partial_\alpha\beta+n^\alpha n_\alpha-\half ^{(2)}R\right)-\dfrac{1}{2r^2}\partial_r\left[r^4\left(D_AU^A\right)\right]\right\}\notag\\
&=F_{ur}F_{Ar}g^{ur}g^{Ar}+\half (g^{ur})^2(F_{ur})^2+\half F_{Ar}g^{Br}(F_{Br}g^{Ar}+F_{BC}g^{BC}).\end{align}
This equations determine the function $V$ up to an arbitrary integration function $2M(u,x^C)$.\\
The traceless equation $g^{AD}E_{AB}=0$ gives the time evolution of all the terms in the expansion of $g_{AB}$ in inverse power of $r$, except for the first subleading. The time derivative of this subleading must therefore be given as part of the initial data, and is called the news function.\\

\item The equation $F^{A\sigma}_{\phantom{u\sigma};\sigma}=0$ is a differential equation for $F_{rA}$:
\begin{align}
&\partial_u\left[F_{rB}g^{AB}\right]+\notag\\
&\dfrac{1}{r}\partial_r\left[r^2U^A(F_{ru}g^{ru}+F_{rB}g^{rB})+r^2g^{AB}(F_{uB}+\dfrac{V}{r}F_{Br}+F_{BC}U^C)\right]+\notag\\
&+\left(\partial_B+^{(2)}\Gamma^E_{EB}\right)(F_{rC}g^{CB}U^C+F_{rC}g^{AC}U^B-e^{2\beta}g^{AC}g^{BD}F_{CD})=0,
\end{align}
and determine the time evolution of $\partial_r A_A$ in terms of known functions. Note that the time evolution of the leading term of $A_A$ is thus not determined by this equation, and must be given.
\end{itemize}

Let us assume that the angular part of the metric and the gauge potential can be expanded in inverse power of $r$,
\begin{align}
g_{AB}(u,r,x^C)&=r^2\bar\gamma_{AB}(x^C)+rC_{AB}(u,x^C)+\mathcal O(1),\label{18h43}\\
A_A(u,r,x^C)&=C_A(u,x^C)+\dfrac{E_A(u,x^C)}{r}+\mathcal O(r^{-2}),\label{18h44}
\end{align}
where indices on functions $C_{AB},C_A,E_A$ are raised with the metric $\bar\gamma^{AB}$. The time derivative of $C_{AB}$ in \eqref{18h43} and $C_A$ in \eqref{18h44} must be given as part of the initial data and are the news functions.\\
Due to the perturbative expansion, solution for $\beta,U^B,A_u,V$ become
\begin{align}
\beta&=-\dfrac{1}{32r^2} C^{AB}C_{AB}+\mathcal O(r^{-3}),\\
U^B&=-\dfrac{1}{2r^2}\bar D_C C^{BC}-\dfrac{2}{3r^3}\left[N^B(u,x^C)-\half C^{AB}\bar D_CC^C_A\right]+\mathcal O(r^{-4}),\\
A_u&=-\dfrac{Q(u,x^A)}{r}-\dfrac{\bar D_AE^A}{2r^2}+\mathcal O(r^{-2}),\\
\dfrac{V}{r}&=-1+\dfrac{2M(u,x^C)}{r}+\mathcal O(r^{-2}).
\end{align}
Functions $\beta,U^B,V$ have contributions from the gauge potential at order $\mathcal O(r^{-4}),\mathcal O(r^{-4}),\mathcal O(r^{-2})$, respectively.\\

In summary, the integration of the main equations  determine functions $\beta,V,U^A,A_u$ in terms of the initial free data $g_{AB}, A_B$ up to arbitrary functions ($M,N^A,Q$) and to news functions $\partial_u C_{AB},\partial_u C_A$.\\

The final step in the integration of the equations of motion is to solve the supplementary equations. These equations determine the time evolution of integration functions $M,N^A,Q$ and yield
\begin{align}
\partial_u M(u,x^A)&=\dfrac{1}{4}\bar D_A\bar D_B \partial_u C^{AB}-\dfrac{1}{8}\partial_uC^{AB}\partial_u C_{AB}-\dfrac{1}{4}\partial_uC_A\partial_uC^A,\\
&\notag\\
\partial_uN_A(u,x^B)&=\partial_AM+\dfrac{1}{16}\partial_A(C_{BC}\partial_uC^{BC})-\dfrac{1}{4}(\partial_uC_{BC})\partial_AC^{BC}-\notag\\
&-\dfrac{1}{4}\bar D_B(\bar D^B\bar D^CC_{CA}-\bar D_A\bar D_CC^{BC})-\notag\\
&-\dfrac{1}{4}\bar D_B(C^{CB}\partial_u C_{AC}-C_{AC}\partial_uC^{CB})\notag\\
&-\dfrac{1}{2}(Q \partial_uC_A+(\partial_AC_B-\partial_BC_A)\partial_uC^C),\\
&\notag\\
\partial_uQ(u,x^A)&=\bar D_A(\partial_uC^A).
\end{align}

\subsection{On-shell variation of the electromagnetic solution space}
As a first application, let us compute the action of the asymptotic symmetries (found in chapter 5) on the functions $Q,C_A,\partial_u C_A$, parametrizing the electromagnetic part of the solution space. This is done by computing the on-shell variation\footnote{
$-\delta_{(\xi,\epsilon)} A_\mu=\mathcal L_\xi A_\mu+\partial_\mu\epsilon.$
}
 of the field $A_u,A_A$ at leading order. One finds
\begin{align}
-\delta Q&=(f\partial_u+Y^A\partial_A+\bar D_AY^A) Q ~+~ (C_A~~\partial^A\psi+\partial_u C_A~~\partial^Af),\\
-\delta C_A&=(f\partial_u+Y^B\partial_B)C_A~+~(C_B\partial_A Y^B+\partial_A E),\label{18h57}
\end{align}
where functions $f,Y,E$ characterize the asymptotic symmetries of the Einstein-Maxell system, see chapter 5 for more detail.\\
Taking the time derivative of \eqref{18h57} yields\footnote{Recall from chapter 5 that $\partial_u E=0$.}
\begin{eqnarray}
-\delta (\partial_u C_A)=\left(f\partial_u+Y^B\partial_B+\dfrac{1}{2}\bar D_BY^B\right)(\partial_u C_A)+(\partial_u C_B) \partial_A Y^B.
\end{eqnarray}

\section{Conclusions et perspectives}
In this chapter, the solution space of the Einstein-Maxwell system has been worked out, in three and four spacetime dimensions. During the integration of the equations of motion, a perturbative expansion in inverse power of the radial coordinate was performed. It could be interesting to use more general perturbative expansion, for instance with logarithmic terms. The next step in the asymptotic analysis is  to work out the
holographic current algebra, including potential central
extensions.

\chapter{Conclusions}

In the past, asymptotic methods developed for studying the symmetry structure of gravity 
\cite{Brown:1986nw} have allowed to anticipate  the AdS-CFT correspondence \cite{Maldacena} and also to reproduce microscopically, in a completely non-stringy way, the entropy of black-holes that are locally AdS$_3$ \cite{Strominger:1998eq}. Due to the richness of these results, it is of interest to try to extend these results to other cases than the one of AdS gravity.\\

Moreover, the study of asymptotic symmetries sheds light on the symmetry enhancement phenomenon, according to which the symmetry algebra at the boundary of the spacetime differs from the rigid symmetry algebra of the bulk theory. In gravitational theories, 
the enhancement phenomenon is particularly interesting when the asymptotic symmetry algebra is realized by the Virasoro algebra, as first observed in the case of gravity with a negative cosmological constant, when expanded around anti-de Sitter space in three dimensions with suitable boundary conditions \cite{Brown:1986nw}.
This enhancement to a Virasoro algebra can be generalized
 to four spacetime dimensions, in the case of 
asymptotically flat spacetimes at null infinity \cite{Barnich:2009se}.
The novel feature of this generalization is that local singularities are allowed in the symmetry algebra. A natural question one could wonder about is to find other examples of enhancement to a Virasoro algebra.\\

In view of these motivations, the symmetry structure of gravity at null infinity was studied further in this thesis, in the case of pure gravity in four spacetime dimensions, and also in the case of gravity coupled to matter in $d$ spacetime dimensions with and without a cosmological constant.\\

Firstly, we have shown that the symmetry enhancement from Lorentz to Virasoro algebra also occurs for asymptotically flat spacetimes  defined in the sense of Newman-Unti \cite{newman:891}. In this set-up, the asymptotic symmetry algebra was derived, focusing on the local conformal properties, and was shown to be given by the direct sum of the bms algebra with the abelian algebra of infinitesimal conformal rescalings. As a first application, the transformation laws of the Newman-Penrose coefficients characterizing the solution space of the Newman-Unti approach were worked out, focusing on the inhomogeneous terms that contain the information about central extensions of the theory. These transformations laws make the conformal structure particularly transparent, and constitute the main original result of the thesis. We have also constructed the dictionary between BMS and Newman-Unti gauges, constructed the surface charges and derived their algebra. In future, it should be interesting to provide a self-contained derivation of the transformations laws in the context of the Newman-Penrose formalism.\\

Secondly, we have worked out the asymptotic symmetries of the Einstein-Yang-Mills system with or
without cosmological constant in a unified manner in $d$ dimensions. In
agreement with a recent conjecture \cite{Strominger:2013lka}, we have found a Virasoro-Kac-Moody type algebra
not only in three dimensions but also in the four dimensional asymptotically
flat case.\\

The gauge fixing and fall-off conditions that we have considered have
been mainly dictated by the desire to yield the usual asymptotic
symmetry structure (at least for the spacetime part) while otherwise
being as relaxed as possible. As partly already discussed at the end of chapter five,
additional more restrictive conditions motivated by finiteness of
associated conserved currents or their integrability can
further reduce the asymptotic symmetry structure, in particular also
in the Yang-Mills part.\\

With the symmetries under control, the next stage of the asymptotic analysis of the Einstein-Yang-Mills system is to work out
asymptotic solutions. This has been done in chapter six for simplicity first in
three, and then in four dimensions, in the case of the Einstein-Maxwell system under the assumption of a perturbative expansion in inverse power of the radial coordinate.
Once this is done, the
symmetry transformations of the fields characterizing asymptotic
solutions can be computed. It could be interesting to use more general perturbative expansion, for instance with logarithmic terms and analyze its consequences. The next step in the asymptotic analysis will be to work out the
holographic current algebra, including potential central extensions.\\

In this thesis, we have frequently assumed that the physical quantities could be expanded in inverse power of the radial coordinate. This assumption was motivated by the fact of having the solution space under control but this assumption can be relaxed to admit other types of perturbative expansions, like for instance with logarithmic terms (polyhomogeneity, see \cite{Chrusciel:1993hx}).
\\

Very recently \cite{Strominger:2013jfa}, a correspondence was found between the supertranslations part of the BMS symmetry group
and the Weinberg' soft theorems. This observation links gravitational symmetries in four dimensions and S-matrix physics and therefore provides an additional motivation for studying asymptotic symmetries.
It could be interesting to try to connect the results developed in this thesis to that approach. For instance, one can wonder about wether the local conformal rescalings present in the Newman-Unti symmetry algebra are also a symmetry of the S-matrix, in the same spirit as for the supertranslations transformations \cite{He:2014laa}.

\appendix

\newpage
\chapter{Conformal Killing vectors in $d$ dimensions}
In this appendix, some properties of conformal Killing vectors are established in $d$ spacetime dimensions.

\section{$X_A$ conformal Killing vector field}

Let $X_A$ be a conformal Killing vector of the sphere in $(d-2)$ dimensions\footnote{In this chapter (as in the rest of this thesis) the spacetime dimension $d$ is assumed to be $d\geq3$, and capital latin indices $A,B,\dots$ denote the coordinates on the $d-2$ sphere $\bar\gamma_{AB}$.},
\begin{eqnarray}\label{A:1}
\bar D_AX_B+\bar D_BX_A=\dfrac{2}{d-2}\psi \bar \gamma_{AB},
\end{eqnarray}
where the metric of the sphere is denoted by $\bar\gamma_{AB}$ and associated covariant derivative by $\bar D_A$, and
with the conformal factor $\psi=\bar D_A\bar X^A$.
Recall that the Riemann curvature tensor associated with the metric $\bar \gamma_{AB}$ is defined by the commutator of the covariant derivatives $\bar D_C$ acting on a vector field $U^A$,
\begin{eqnarray}\label{A:2}
[\bar D_A,\bar D_B]U^C=\bar R_{AB\phantom CD}^{\phantom{AB}C}~~U^D.
\end{eqnarray}
In the case of a maximally symmetric spacetime (here the $(d-2)$-sphere, but a similar expression also exists for de Sitter or Anti-de Sitter  spaces) the Riemann tensor is given explicitly by
\begin{eqnarray}\label{A:3}
\bar R_{AB\phantom CD}^{\phantom{AB}C}=\kappa(\bar \gamma_A^{\phantom AC}\bar \gamma_{BD}-\bar \gamma_{AD}\bar \gamma_B^{\phantom BC}),
\end{eqnarray}
where $\kappa=1$ in the case of the sphere (and $\kappa=\frac{-1}{l^2}$ for AdS).\\
Finally,  one also have the following relation for any vector field $U^A$
\begin{align}
[\bar D_A,\bar D_B]\bar D_CU_D=\bar \gamma_{CA}\bar D_BU_D-\bar \gamma_{CB}\bar D_AU_D+\bar \gamma_{DA}\bar D_CU_B-\bar \gamma_{DB}\bar D_CU_A\label{A:6}.
\end{align}
The proof follows from \eqref{A:2} and \eqref{A:3}. The relation \eqref{A:6} implies in particular that $[\bar D_C,\bar D_B]\bar D^BU^C=0$.\\

\noindent After having reviewed these definitions, some properties of the conformal Killing vector can be derived:
\begin{itemize}
\item One can write the commutator \eqref{A:2} three times (by permuting the indices $A,B,C$) with the vector field  $U^C$ replaced by the conformal Killing vector field $X_C$ and using the conformal Killing equation \eqref{A:1} to rewrite the second term of all these commutators. Finally, by using the definition of the curvature tensor \eqref{A:3}, one obtains the first property satisfied by any conformal Killing vector $X_C$ of the $(d-2)$-sphere $\bar \gamma_{AB}$,
\begin{align}\label{A:4}
\bar D_A\bar D_BX_C~&=~\dfrac{1}{d-2}~(\bar \gamma_{CA}\bar D_B\psi+\bar \gamma_{CB}\bar D_A\psi-\bar \gamma_{AB}\bar D_C\psi)-\notag\\
&-\bar \gamma_{AB}X_C+\bar \gamma_{AC}X_D.
\end{align}

\item By contracting the indices $A$ and $B$ of \eqref{A:4} (and writing $\bar \Delta=\bar D_A\bar D^A$ the Laplacian of the $(d-2)$-sphere), one gets the second property,
\begin{eqnarray}\label{A:5}
\bar \Delta X_C~=~\dfrac{4-d}{d-2}~\bar D_C\psi+(3-d)X_C.
\end{eqnarray}
In four spacetime dimensions, $\bar\gamma_{AB}$ is a two-sphere and \eqref{A:5} implies that $\bar \Delta X_C=-X_C$.\\

\item Finally, by taking the covariant derivative $\bar D_C$ of \eqref{A:5} and using the property \eqref{A:6} on the left hand side, one obtains the last property,
\begin{eqnarray}\label{A:7}
\bar \Delta\psi=-(d-2)\psi.
\end{eqnarray}
In four space-time dimensions, this relation reduces to $\bar \Delta\psi=-2\psi$.\\
\end{itemize}

\section{$X_A=\partial_A f$ conformal Killing vector field}
Suppose that the conformal Killing vector $X_A$ is a gradient field, i.e. the derivative of a scalar field $f$, $X_A=\partial_A f$. Then, one has $\bar D_AX_B=\bar D_BX_A$ because of the symmetry of the Christoffel symbols,  $\bar D_AX_B=\bar D_A\partial_Bf
=\bar D_B\partial_Af=\bar D_BX_A$.
The conformal Killing equation \eqref{A:1} then becomes
\begin{eqnarray}\label{A:9}
\bar D_AX_B=\dfrac{1}{d-2}\psi \bar \gamma_{AB},
\end{eqnarray}
with the conformal factor $\psi\equiv \bar D_AX^A=\bar D_A\partial^A f=\bar \Delta f$.\\
Taking the covariant derivative $\bar D_C$ of \eqref{A:9} gives
\begin{eqnarray}
\bar D_C\bar D_A\partial_B f=\dfrac{1}{d-2}(\bar D_C\bar \Delta f)~~\bar \gamma_{AB}.
\end{eqnarray}
Using the property \eqref{A:3} and contracting the indices $B$ and $C$ gives the relation,
\begin{eqnarray}\label{A:11}
\partial_A f=-\dfrac{1}{d-2}\bar D_A\bar \Delta f,
\end{eqnarray}
which in four spacetime dimensions reduces to $\partial_Af=-\dfrac{1}{2}\bar D_A\bar \Delta f$.

\section{Property of $\partial_Af$  a conformal killing vector}
In this last section, a theorem is proved relating the conformal Killing character of a vector $X_C$ and the derivative of its conformal factor $\partial_A\psi$ (recall $\psi=\bar D_CX^C$). More precisely, we will prove the
\begin{theorem}\label{theoA4}
In $d\neq4$, if $X^C$ is a conformal Killing vector of $\bar \gamma_{AB}$, then $\partial_A \psi$ is also a conformal Killing vector of $\bar \gamma_{AB}$.
\end{theorem}

\textit{Proof.}
By hypothesis $X^C$ is a conformal Killing vector in $d$ dimensions, so the relations \eqref{A:1}, \eqref{A:5} and \eqref{A:7} can be used,
\begin{align}
\bar D_AX_B+\bar D_BX_A&=\dfrac{2}{d-2}\psi \bar \gamma_{AB}\label{A:12},\\
\bar \Delta X_C&=\dfrac{4-d}{d-2}\bar D_C\psi+(3-d)X_C\label{A:13},\\
\bar \Delta\psi&=-(d-2)\psi\label{A:14},
\end{align}
with $\psi=\bar D_AX^A$ the conformal factor of $X^C$.\\
From \eqref{A:13}, we get
\begin{eqnarray}\label{15}
\partial_B\psi=\dfrac{d-2}{4-d}(\bar \Delta X_B-(3-d)X_B),
\end{eqnarray}
which is valid since we are in $d\neq 4$, by hypothesis.\\
To prove that $\partial_A\psi$ is a conformal Killing vector of $\bar \gamma_{AB}$, we have to show that $\partial_A\psi$ satisfies
\begin{eqnarray}\label{A:16}
\bar D_A\partial_B\psi+\bar D_B\partial_A\psi=\dfrac{2}{d-2}(\bar D_C\partial^C\psi)\bar \gamma_{AB}.
\end{eqnarray}
Using the conformal Killing equation \eqref{A:1}, the property \eqref{15} and the following relation (consequence of \eqref{A:6})
\begin{eqnarray}
\bar D_B\bar \Delta X_A=
\bar \Delta \bar D_BX_A+2\bar \gamma_{AB}\psi-2\bar D_AX_B-(d-3)\bar D_BX_A,
\end{eqnarray}
the left hand side of \eqref{A:16} becomes
\begin{eqnarray}\label{A:18}
\bar D_A\partial_B\psi+\bar D_B\partial_A\psi~=~\dfrac{2}{4-d}~[\bar \Delta\psi\bar \gamma_{AB}~+~2(d-3)\psi\bar \gamma_{AB}].
\end{eqnarray}
Finally, the relation \eqref{A:14} allows to simplify the right hand side of equation \eqref{A:18} to
\begin{eqnarray}\label{A:19}
\dfrac{2}{4-d}[\bar \Delta\psi\bar \gamma_{AB}+2(d-3)\psi\bar \gamma_{AB}]=\dfrac{2}{d-2}\bar \Delta\psi\bar \gamma_{AB}.
\end{eqnarray}
Relations \eqref{A:18} and \eqref{A:19} give
\begin{eqnarray}\label{19}
\bar D_A\partial_B\psi+\bar D_B\partial_A\psi=\dfrac{2}{d-2}\bar \Delta\psi\bar \gamma_{AB},
\end{eqnarray}
which concludes the proof.\\

\noindent So we have shown that when one has a conformal Killing vector in $d\neq4$ spacetime dimensions, then the derivative of its conformal factor is also a conformal Killing vector.





\newpage
\chapter{Newman-Janis rotation algorithm applied to BTZ}

In this appendix, the following question is investigated: in three spacetime dimensions, can one find the metric of the charged rotating BTZ black hole by applying the Newman-Janis rotation trick to the static charged BTZ black hole? This question is answered in this appendix, but appears to be negative.

\section{Introduction}
\label{sec:introduction}

The Newman-Janis rotation algorithm is a trick that allows, in some cases, to produce a rotating black-hole metric, starting from a non-rotating one. It has been proposed in the sixties by Newman and Janis \cite{Janis} who re-derived the Kerr black-hole metric starting from the Schwarzschild black hole. This algorithm successively consists of writting the non-rotating black-hole metric in a suitable Newman-Penrose complex null tetrad, then allowing the radial coordinate to become complex and performing a complex change of variables (this is the main part of the rotation trick), and finally reducing the radial coordinate to its real value\footnote{See \cite{d'Inverno:1992rk} for a pedagogical presentation.}. The use of this algorithm allowed  to find \cite{Newman} the rotating version of the Reissner-Nordstrom black hole, by applying the same complex coordinates transformation, but in the Einstein-Maxwell set-up. The Newman-Janis rotation algorithm can be schematized in the following way (see section B.2 for more detail),\\

\begin{centering}
\begin{tikzpicture}
\node (A) at (0,-1) {$A_\mu$};
\node (F) at (1.1,-1) {$F_{\mu\nu}$};
\node (phi) at (3.8,-1) {$(\phi_0,\phi_1,\phi_2)$};
\node (phi') at (6.8,-1) {$(\phi'_0,\phi'_1,\phi'_2)$};
\node (F') at (9.8,-1) {$F'_{\mu\nu}$};
\node (A') at (11.05,-1) {$A'_\mu$};
\node (trick) at (5.5,0.25) {Rotation trick};
\node (trick) at (5.3,-0.75) {Rot.};
\node (trick) at (5.3,-1.22) {trick};
\node (g) at (0,0) {$g^{\mu\nu}$};
\node (l) at (2.1,0) {$(l^\mu,n^\mu,m^\mu,\bar m^\mu$)};
\node (l') at (8.6,0) {$(l'^\mu,n'^\mu,m'^\mu,\bar m'^\mu$)};
\node (g') at (11.05,0) {$g'^{\mu\nu}$};
\draw[->,>=latex] (g) -- (l);
\draw[->,>=latex]  (l) -- (l');
\draw[->,>=latex]  (l') -- (g');
\draw[->,>=latex] (A) -- (F);
\draw[->,>=latex] (F) -- (phi);
\draw[->,>=latex] (phi) -- (phi');
\draw[->,>=latex] (phi') -- (F');
\draw[->,>=latex] (F') -- (A');
\draw (2.1,-1) -- (l.south);
\draw [fill] (2.1,-1) circle [radius=0.05];
\draw (8.6,-1) -- (l'.south);
\draw [fill] (8.6,-1) circle [radius=0.05];
\end{tikzpicture}
\end{centering}

Interestingly, the Newman-Janis' trick can be applied in three spacetime dimensions as well, as noted by Kim in \cite{Kim}. The procedure he proposed in this lower dimensional case is the following: first, a dimensional continuation to four dimensions is made from the three dimensional non-rotating solution, then the rotation algorithm of Newman-Janis is applied to this four dimensional metric, and finally a dimensional reduction to three dimensions is performed on the resulting metric. In \cite{Kim}, Kim illustrated this procedure in the case of the uncharged non-rotating BTZ black hole to successfully obtain the uncharged rotating BTZ solution.\\

The main drawback of the Newman-Janis' algorithm in the Einstein-Maxwell set-up is that it does not provide the right expression for the gauge field (i.e. $A'$ in the schema on previous page does not satisfies the equations of motion).
 As already stressed in \cite{Newman}, the form of the Kerr-Newman gauge field was found by integrating the field equations of motion.
This incompleteness of the rotation algorithm in the case of the Maxwell gauge field is by no mean peculiar to four dimensions, but is also present in higher dimensions \cite{Higher}.\\

Very recently however, it has been proposed by Keane \cite{Keane} to extend the original Newman-Janis' algorithm by adding to it a null rotation that has the effect (among other things)  of provinding the right Maxwell field in an algorithmic way, contrary to the original algorithm.
This extended algorithm comes from the nice observation that the Newman-Janis' algorithm does not preserve the principal null directions\footnote{See \cite{Wald}, section (7.3) for more detail.} of the initial and final tetrads, but that this issue can be solved by applying a null rotation of class I\footnote{See \cite{Chandrasekhar}, page 53 for detail.} on the final rotating tetrad. 
The extended algorithm of Keane can be schematized as follow (see section B.2 for further detail):\\

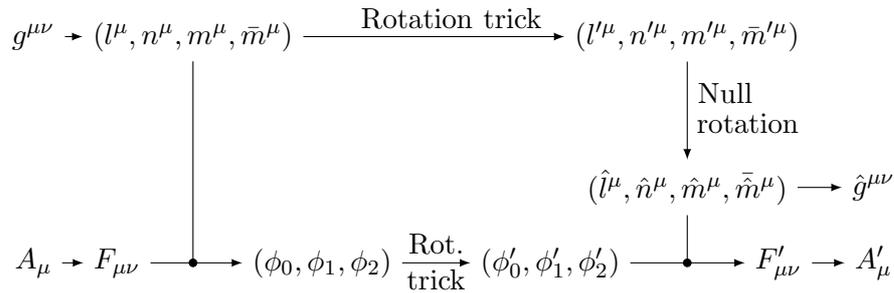
\begin{centering}
\begin{tikzpicture}
\node (A) at (0,-3) {$A_\mu$};
\node (F) at (1.1,-3) {$F_{\mu\nu}$};
\node (phi) at (3.8,-3) {$(\phi_0,\phi_1,\phi_2)$};
\node (phi') at (6.8,-3) {$(\phi'_0,\phi'_1,\phi'_2)$};
\node (hatF) at (9.8,-3) {$F'_{\mu\nu}$};
\node (hatA) at (11.05,-3) {$A'_\mu$};
\node (trick) at (5.5,0.25) {Rotation trick};
\node (trick) at (5.3,-2.75) {Rot.};
\node (trick) at (5.3,-3.22) {trick};
\node (null) at (9.1,-0.75) {Null};
\node (null) at (9.4,-1.15) {rotation};
\node (g) at (0,0) {$g^{\mu\nu}$};
\node (l) at (2.1,0) {$(l^\mu,n^\mu,m^\mu,\bar m^\mu$)};
\node (l') at (8.6,0) {$(l'^\mu,n'^\mu,m'^\mu,\bar m'^\mu$)};
\node (lhat) at (8.6,-2) {$(\hat l^\mu,\hat n^\mu,\hat m^\mu,\bar{\hat m}^\mu$)};
\node (g') at (11.05,-2) {$\hat g^{\mu\nu}$};
\draw[->,>=latex] (g) -- (l);
\draw[->,>=latex]  (l) -- (l');
\draw[->,>=latex]  (l') -- (lhat);
\draw[->,>=latex]  (lhat) -- (g');
\draw[->,>=latex] (A) -- (F);
\draw[->,>=latex] (F) -- (phi);
\draw[->,>=latex] (phi) -- (phi');
\draw (phi') -| (lhat);
\draw[->,>=latex] (hatF) -- (hatA);
\draw (2.1,-3) -- (l.south);
\draw [fill] (2.1,-3) circle [radius=0.05];
\draw[->,>=latex] (8.6,-3) -- (hatF);
\draw [fill] (8.6,-3) circle [radius=0.05];
\end{tikzpicture}
\end{centering}\\

\noindent And now, the fields $g'_{\mu\nu}$ and $\hat A_\mu$ solve the equations of motion.\\

This extended Newman-Janis algorithm gives the hope that it can be applied in three dimensions as well, by following the prescription of Kim. In the rest of this appendix, the algorithm of Newman-Janis-Kim-Keane is applied to the case of the charged unrotating BTZ black hole to attempt a derivation of the charged rotating BTZ black hole. However, the final fields obtained by this algorithm do not satisfy the equations of motion $D_\mu F^{\mu\nu}=0$. The conclusion is that the four dimensional extended algorithm of Keane\cite{Keane} together with the three dimensional prescription of Kim \cite{Kim} are not sucessful in the search of the charged rotating BTZ black hole.\\

It should be interesting to try to adapt this approach in higher dimensions. In \cite{Higher} for instance, the author uses the rotation algorithm of Newman-Janis to recover rotating black holes from non-rotating ones, in even higher dimensions. The hope given by the extended algorithm of Keane \cite{Keane} is to be able to find an exact analytical expression for the Kerr-Newman black hole in higher dimensions.

\section{Trick of Newman-Janis-Kim-Keane for BTZ}
\label{sec:2}
In this section, the procedure of Newman-Janis-Kim-Keane \cite{Janis,Kim,Keane} is applied in the case of the static charged BTZ black hole.
The starting point is the following action in three spacetime dimensions, 
\begin{eqnarray}
S=\dfrac{1}{16\pi G}\int d^3x\sqrt{-g}\left(R-{2}\Lambda-\frac{1}{4}F^{2}\right),
\end{eqnarray}
where $F^2=F_{\mu\nu}F^{\mu\nu}, F_{\mu\nu}=\partial_\mu A_\nu-\partial_\nu A_\mu$ and $\Lambda=\frac{-1}{l^2}$. Equations of motion read
\begin{eqnarray}
R_{\mu\nu}-\half R g_{\mu\nu}-\frac{1}{l^2}g_{\mu\nu}=-\frac{1}{8}F^2g_{\mu\nu}+\half F_{\mu\sigma}F_{\nu}^{\phantom\nu\sigma},\hspace{1cm}D_\mu F^{\nu\mu}=0.
\end{eqnarray}
The charged static BTZ black hole \cite{BTZ} is given by the following fields,
\begin{align}
ds^2&=-\left(\dfrac{r^2}{l^2}-M-\dfrac{Q^2}{4}\ln r^2\right) dt^2+\left(\dfrac{r^2}{l^2}-M-\dfrac{Q^2}{4}\ln r^2\right)^{-1} dr^2+r^2 d\varphi^2,\label{2.1}\\
A&=-Q\ln r~~dt\hspace{1cm}\Rightarrow
F_{\mu\nu}=
\begin{pmatrix}
0&\frac{Q}{r}&0\\
-\frac{Q}{r}&0&0\\
0&0&0
\end{pmatrix}
.\label{B:4}
\end{align}
In view of applying the rotation trick, it is advantageous to change the signature of the metric \eqref{2.1},
\begin{align}
ds^2&= \left(\dfrac{r^2}{l^2}-M-\dfrac{Q^2}{4}\ln r^2\right)dt^2-\left(\dfrac{r^2}{l^2}-M-\dfrac{Q^2}{4}\ln r^2\right)^{-1} dr^2-r^2 d\varphi^2.\label{21}
\end{align}
This metric is now a solution to $R_{\mu\nu}-\half R g_{\mu\nu}+\frac{1}{l^2}g_{\mu\nu}=\frac{1}{8}F^2g_{\mu\nu}-\half F_{\mu\sigma}F_{\nu}^{\phantom\nu\sigma}$.\\

In four spacetime dimensions, the Newman-Janis' trick consists of a complex coordinates transformation performed on the complex null tetrad associated with a metric beforehand written in Eddington-Finkelstein coordinates. To apply this trick in three dimensions, Kim \cite{Kim} proposed to perform a dimensional continuation of the three spacetime solution \eqref{B:4}\eqref{21}, by assuming the three-dimensional solutions are the $\theta=\pi/2$ slice of four-dimensional fields,
\begin{align}
A_\mu&=(-Q\ln r,0,0,0),\\
ds^2&=\left(\dfrac{r^2}{l^2}-M-\dfrac{Q^2}{4}\ln r^2\right) dt^2-\left(\dfrac{r^2}{l^2}-M-\dfrac{Q^2}{4}\ln r^2\right)^{-1} dr^2-r^2 d\Omega^2,\label{2.5}
\end{align}
where $d\Omega^2=d\theta^2+\sin^2\theta d\varphi^2$. It is important to stress that the four-dimensional metric \eqref{2.5} is not supposed to be a solution to the Einstein's field equations in four dimensions. It is only required the slice $\theta=\pi/2$ to be a solution to the field equations in three dimensions.\\
Then, Eddington-Finkelstein coordinates are introduced,
\begin{align}
dt=du+\frac{dr}{\left(\dfrac{r^2}{l^2}-M-\dfrac{Q^2}{4}\ln r^2\right)},
\end{align}
so that the metric \eqref{2.5} and the $U(1)$ field strength  become
\begin{align}
ds^2&=\left(\dfrac{r^2}{l^2}-M-\dfrac{Q^2}{4}\ln r^2\right) du^2+2dudr-r^2 (d\theta^2+\sin^2\theta d\varphi^2)\label{2.7},\\
F_{\mu\nu}&=
\begin{pmatrix}
0&\frac{Q}{r}&0&0\\
-\frac{Q}{r}&0&0&0\\
0&0&0&0\\
0&0&0&0
\end{pmatrix},\label{B.11}
\end{align}
and the contravariant form of the metric is
\begin{eqnarray}\label{2.9}
g^{\mu\nu}=
\begin{pmatrix}
0&1&0&0\\
1&\left(\dfrac{r^2}{l^2}-M-\dfrac{Q^2}{4}\ln r^2\right)&0&0\\
0&0&\frac{-1}{r^2}&0\\
0&0&0&\frac{-1}{r^2\sin^2\theta}
\end{pmatrix}.
\end{eqnarray}

\subsubsection{Newman-Penrose formalism in very short}
\label{sec:Newman-Penrose formalism in very short}
The rotation algorithm is based on the Newman-Penrose formalism \cite{newman:566}, that can be quickly summarized as follow: To a metric $g_{\mu\nu}$ with signature $(+,-,-,-)$ is constructed a complex\footnote{The complex conjugation is denoted by a bar.} null tetrad $e_{a\mu}=(l_\mu,n_\mu,m_\mu,\bar m_\mu$) defined by the relation
\begin{eqnarray}
g_{\mu\nu}=e_{a\mu}e_{b\nu}\eta^{ab},
\end{eqnarray}
where $\eta^{ab}$ represents the normalization of the tetrad vectors, given by $l^\mu n_\mu=1=-m^\mu\bar m_\mu$, or equivalently in terms of $\eta_{ab}$,
\begin{align}
e_{a\mu}e_b^{\phantom b\mu}&=\eta_{ab}=
\begin{pmatrix}
0&1&\phantom{-}0&\phantom{-}0\\
1&0&\phantom{-}0&\phantom{-}0&\\
0&0&\phantom{-}0&-1\\
0&0&-1&\phantom{-}0
\end{pmatrix}
,~~ \eta^{ac}\eta_{cb}=\delta^{a}_{\phantom ab}.\label{ortho}
\end{align}
The explicit relation between the complex null tetrad vectors and the metric is therefore
\begin{eqnarray}\label{metric}
g_{\mu\nu}=l_\mu n_\nu+l_\nu n_\mu-m_\mu\bar m_\nu-m_\nu\bar m_\mu.
\end{eqnarray}
In the Newman-Penrose formalism,
the six components of the Maxwell tensor $F_{\mu\nu}$ are represented by three complex scalars, defined by
\begin{eqnarray}\label{B.15}
\phi_0=F_{\mu\nu}l^\mu m^\nu,~~
\phi_1=\dfrac{1}{2}F_{\mu\nu}(l^\mu n^\nu+\bar m^\mu m^\nu),~~
\phi_2=F_{\mu\nu}\bar m^\nu n^\nu.
\end{eqnarray}
In terms of the Newman-Penrose scalars, the Maxwell tensor reads
\begin{align}
F_{\mu\nu}=-&\phi_0(n_\mu\bar m_\nu-n_\nu\bar m_\mu)-\bar\phi_0(n_\mu m_\nu-n_\nu m_\mu)\notag\\
+&(\phi_1+\bar\phi_1)(n_\mu l_\nu-n_\nu l_\mu)+(\phi_1-\bar\phi_1)(m_\mu \bar m_\nu-m_\nu \bar m_\mu)\label{fmunu}\\
-&\phi_2(m_\mu l_\nu-m_\nu l_\mu)-\bar\phi_2(\bar m_\mu l_\nu-\bar m_\nu l_\mu)\notag.
\end{align}
\subsection{Algorithm of Newman-Janis-Kim for $g_{\mu\nu}$}
Complex null tetrad vectors associated with the metric \eqref{2.9} are given by
\begin{align}
l^\mu&=(0,1,0,0), \hspace{1cm}n^\mu=(1,-\frac{1}{2}\left(\dfrac{r^2}{l^2}-M-\dfrac{Q^2}{4}\ln r^2\right),0,0),\label{2.11}\\
m^\mu&=\frac{1}{\sqrt{2}r}\left(0,0,1,\frac{i}{\sin\theta}\right),\hspace{0.5cm}\hspace{1cm}\bar m^\mu=\frac{1}{\sqrt{2}r}\left(0,0,1,\frac{-i}{\sin\theta}\right).\label{2.12}
\end{align}
This complex null tetrad is the starting point of the rotation trick, which consists in two different parts. Firstly, the $r$ coordinate becomes  complex, so the tetrad vectors \eqref{2.11}\eqref{2.12} can be re-written as
\begin{align}
l^\mu&=(0,1,0,0),\hspace{1cm}n^\mu=(1,-\frac{1}{2}\left(\dfrac{r\overline r}{l^2}-M-\dfrac{Q^2}{4}\ln r\overline r\right),0,0),\label{2.14}\\
m^\mu&=\frac{1}{\sqrt{2}~~\overline r}\left(0,0,1,\frac{i}{\sin\theta}\right),\hspace{1cm}\bar m^\mu=\frac{1}{\sqrt{2}~~r}\left(0,0,1,\frac{-i}{\sin\theta}\right).\label{2.15}
\end{align}
Secondly, the following complex coordinate transformation is performed (with $a$ being a constant real parameter),
\begin{align}
r'=r+ia\sin\theta,\hspace{1cm}u'=u-ia\cos\theta,\hspace{1cm}a\in\mathcal R.
\end{align}
Under this coordinates transformation, the four complex null tetrad vectors \eqref{2.14}\eqref{2.15} become (the prime is omitted)
\begin{align}
l^\mu&=(0,1,0,0),\label{B22}\\
n^\mu&=(1,-\frac{1}{2}\left(\dfrac{r^2+a^2\cos^2\theta}{l^2}-M-\dfrac{Q^2}{4}\ln (r^2+a^2\cos^2\theta)\right),0,0),\\
m^\mu&=\frac{1}{\sqrt{2}~~(r+ia\cos\theta)}\left(ia\sin\theta,-ia\sin\theta,1,\frac{i}{\sin\theta}\right),\\
\bar m^\mu&=\frac{1}{\sqrt{2}~~(r-ia\cos\theta)}\left(-ia\sin\theta,ia\sin\theta,1,\frac{-i}{\sin\theta}\right),\label{B25}
\end{align}
from which the metric is known, by using \eqref{metric}.\\

The three-dimensional metric is found by  considering the $\theta=\pi/2$ slice of this metric
\begin{eqnarray}
g^{\mu\nu}=
\begin{pmatrix}
-\frac{a}{r^2}&1+\frac{a^2}{r^2}&\frac{-a}{r^2}\\
1+\frac{a^2}{r^2}&-\left(\dfrac{r^2}{l^2}-M-\dfrac{Q^2}{4}\ln r^2\right)-\frac{a^2}{r^2}&\frac{a}{r^2}\\
\frac{-a}{r^2}&\frac{a}{r^2}&\frac{-1}{r^2}
\end{pmatrix},
\end{eqnarray}
which can be inverted to give the line element
\begin{align}
ds^2=\left(\dfrac{r^2}{l^2}-M-\dfrac{Q^2}{4}\ln r^2\right)(du-ad\varphi)^2+2(du-ad\varphi)(dr+ad\varphi)-r^2d\varphi^2.\label{2.25}
\end{align}
This metric can be re-written using Boyer-Lindquist coordinates, by making the following coordinates transformation,
\begin{align}
dt&=du+\dfrac{(a^2+r^2)~dr}{a^2+r^2\left(\dfrac{r^2}{l^2}-M-\dfrac{Q^2}{4}\ln r^2\right)},\\
d\phi&=d\varphi+\dfrac{a ~dr}{a^2+r^2\left(\dfrac{r^2}{l^2}-M-\dfrac{Q^2}{4}\ln r^2\right)}.
\end{align}
In Boyer-Lindquist coordinates, the metric \eqref{2.25} becomes
\begin{align}\label{2.23}
ds^2=&\left(\dfrac{r^2}{l^2}-M-\dfrac{Q^2}{4}\ln r^2\right) dt^2-\dfrac{r^2 dr^2}{a^2+r^2\left(\dfrac{r^2}{l^2}-M-\dfrac{Q^2}{4}\ln r^2\right)}+\notag\\
&+2a\left[1-\left(\dfrac{r^2}{l^2}-M-\dfrac{Q^2}{4}\ln r^2\right)\right]dtd\phi-\notag\\
&-
\left[r^2+2a^2-a^2\left(\dfrac{r^2}{l^2}-M-\dfrac{Q^2}{4}\ln r^2\right)\right]d\phi^2.
\end{align}
Finally, this metric can be cast in the ADM form \cite{BTZ}, using the following coordinates transformation
\begin{eqnarray}
\tilde t=t-a\phi.
\end{eqnarray}
The metric in ADM form reads
\begin{align}
ds^2=&\left[\frac{a^2}{r^2}+\left(\frac{r^2}{l^2}-M-\dfrac{Q^2}{4}\ln r^2\right)\right]d\tilde t^2
-\dfrac{ dr^2}{\frac{a^2}{r^2}+\left(\dfrac{r^2}{l^2}-M-\dfrac{Q^2}{4}\ln r^2\right)}
-\notag\\
&-r^2\left(d\phi-\frac{a}{r^2} d\tilde t\right)^2.\label{227}
\end{align}
When $Q=0$, one recovers the rotating BTZ black hole that is solution to $R_{\mu\nu}-\half R g_{\mu\nu}+\frac{1}{l^2}g_{\mu\nu}=0$, in agreement with \cite{Kim}\footnote{In \cite{Kim}, there is a sign misprint in the last term of the equation of motion, just after equation (12). This change of sign comes from the signature of the metric, which has been changed in order to apply the Newman-Janis algorithm.}.

\subsection{Keane's extended algorithm for $A_\mu$}
The three complex electromagnetic scalars \eqref{B.15} associated with field strength \eqref{B.11} and tetrad \eqref{2.11}\eqref{2.12} are
\begin{eqnarray}\label{B33}
\phi_0=0,\hspace{1cm}\phi_1=-\dfrac{Q}{2r},\hspace{1cm}\phi_2=0.
\end{eqnarray}
Following the algorithm, the rotation trick of Newman-Janis is applied and \eqref{B33} become
\begin{eqnarray}\label{B34}
\phi'_0=0,\hspace{1cm}\phi'_1=-\dfrac{Q}{2(r-ia\cos\theta)},\hspace{1cm}\phi'_2=0.
\end{eqnarray}
Note that one recovers \eqref{B:4} when the dimensional reduction is perfomed by taking $\theta=\pi/2$. The gauge field associated with  \eqref{B34} and the metric \eqref{2.25} do not solve the equations of motions.\\

The final step in the extended algorithm of Keane is to perfom a null rotation on the tetrad vectors. This null rotation
 preserves the tetrad vector $l_\mu$ (called null rotation of class I in \cite{Chandrasekhar}) and is parametrized by a complex parameter $\alpha$. It acts on the tetrad vectors and complex scalars as follow,
\begin{align}
\hat l^\mu&= l^\mu,~~\hat m^\mu= m^\mu+\alpha l^\mu,~~\hat n^\mu= n^\mu+\alpha\bar m^\mu+\bar\alpha m^\mu+\alpha\bar\alpha l^\mu,\label{null1}\\
\hat \phi_0&=\phi_0,~~\hat \phi_1= \phi_1+\bar\alpha\phi_0,~~\hat \phi_2=\phi_2+2\bar\alpha\phi_1+\bar\alpha^2\phi_0.\label{null2}
\end{align}
Of course, both the metric tensor $g_{\mu\nu}$ and the Maxwell tensor $F_{\mu\nu}$ are invariant under the null rotation \eqref{null1}\eqref{null2}. However $F_{\mu\nu}$ is no longer invariant if only one of the two transformations between \eqref{null1} \eqref{null2} is performed on it, see \eqref{fmunu}.\\

Under \eqref{null1} with parameter $\alpha=\dfrac{i a\sin\theta}{\sqrt 2(r-ia\cos\theta)}$, the tetrad vectors \eqref{B22}\eqref{B25} become
\begin{align}
\hat l^\mu&=(0,1,0,0),\label{B37}\\
\hat m^\mu&=\dfrac{1}{\sqrt2(r+ia\cos\theta)}\left(ia\sin\theta,0,1,\dfrac{i}{\sin\theta}\right),\\
\hat n^u&=1+\dfrac{a^2\sin^2\theta}{r^2+a^2\cos^2\theta},\hspace{1cm}\hat n^\theta=0,\hspace{1cm}
\hat n^\phi=\dfrac{a}{r^2+a^2\cos^2\theta},\\
\hat n^r&=-\dfrac{1}{2}\left(\dfrac{r^2+a^2\cos^2\theta}{l^2}-M-\dfrac{Q^2}{4}\ln (r^2+a^2\cos^2\theta)\right)
+\dfrac{a^2\sin^2\theta}{r^2+a^2\cos^2\theta},\label{B40}
\end{align}
so that the Maxwell field $F^{\mu\nu}$ associated with scalars \eqref{B34} and tetrad \eqref{B37}\eqref{B40} is known, due to equation \eqref{fmunu}.  The dimensional reduction is done by taking $\theta=\frac{\pi}{2}$, and the non-vanishing components of the field strength are given by
\begin{align}
&F^{ur}=-\dfrac{Q}{r}\left(1+\dfrac{a^2}{r^2}\right),\hspace{1cm}F^{r\phi}=\dfrac{aQ}{r^3},\\
\Leftrightarrow\hspace{1cm}&F_{ur}=\dfrac{Q}{r},\hspace{1cm}F_{r\phi}=\dfrac{aQ}{r}.\label{B42}
\end{align}

The rotation algorithm of Newman-Janis, modified by Keane, adapted to three dimensions by Kim, therefore leads to the expressions \eqref{2.25} and \eqref{B42} for the metric and gauge field, respectively. These fields are solutions to $R_{\mu\nu}-\half R g_{\mu\nu}+\frac{1}{l^2}g_{\mu\nu}=\frac{1}{8}F^2g_{\mu\nu}-\half F_{\mu\sigma}F_{\nu}^{\phantom\nu\sigma}$. Unfortunalety, they are not solutions to $D_\mu F^{\mu\nu}=0$, due to
\begin{eqnarray}\label{B43}
D_\mu F^{\mu u}=\dfrac{2a^2Q}{r^4},\hspace{1cm}D_\mu F^{\mu \phi}=\dfrac{2aQ}{r^3}.
\end{eqnarray}
Equations \eqref{B43} are zero either for $Q=0$ (i.e. uncharged case), or for $a=0$ (i.e. unrotating case).

\subsection{Modification of Keane's algorithm in $4d$}
The algorithm of Keane \cite{Keane} performs a null rotation on the tetrad vectors, while leaving invariant the complex electromagnetic scalars. In this section, we simply note that one can slightly modify this algorithm, by performing the null rotation on the electromagnetic complex scalars, while leaving the rotating tetrad invariant. The new scheme is the following,\\

\begin{centering}
\begin{tikzpicture}
\node (A) at (0,-1) {$A_\mu$};
\node (F) at (1.1,-1) {$F_{\mu\nu}$};
\node (phi) at (3.8,-1) {$(\phi_0,\phi_1,\phi_2)$};
\node (phi') at (6.8,-1) {$(\phi'_0,\phi'_1,\phi'_2)$};
\node (phihat) at (6.8,-3) {$(\hat\phi_0,\hat\phi_1,\hat\phi_2)$};
\node (hatF) at (9.8,-3) {$\hat F_{\mu\nu}$};
\node (hatA) at (11.05,-3) {$\hat A_\mu$};
\node (trick) at (5.5,0.25) {Rotation trick};
\node (trick) at (5.3,-0.75) {Rot.};
\node (trick) at (5.3,-1.22) {trick};
\node (null) at (7.25,-1.75) {Null};
\node (null) at (7.25,-2.15) {rot.};
\node (g) at (0,0) {$g^{\mu\nu}$};
\node (l) at (2.1,0) {$(l^\mu,n^\mu,m^\mu,\bar m^\mu$)};
\node (l') at (8.6,0) {$(l'^\mu,n'^\mu,m'^\mu,\bar m'^\mu$)};
\node (g') at (11.05,0) {$g'^{\mu\nu}$};
\draw[->,>=latex] (g) -- (l);
\draw[->,>=latex]  (l) -- (l');
\draw[->,>=latex]  (l') -- (g');
\draw[->,>=latex] (A) -- (F);
\draw[->,>=latex] (F) -- (phi);
\draw[->,>=latex] (phi) -- (phi');
\draw[->,>=latex] (phi') -- (phihat);
\draw (phihat) -| (l');
\draw[->,>=latex] (hatF) -- (hatA);
\draw (2.1,-1) -- (l.south);
\draw [fill] (2.1,-1) circle [radius=0.05];
\draw[->,>=latex] (8.6,-3) -- (hatF);
\draw [fill] (8.6,-3) circle [radius=0.05];
\end{tikzpicture}
\end{centering}\\

\noindent This modified algorithm is equivalent to the one proposed by Keane. In the four dimensional case, the rotation trick applied to the Reissner-Nordstrom fields yields $(\phi_0=0,\phi_1\neq 0,\phi_2=0)$, while a null rotation of parameter $\alpha=\frac{i a \sin\theta}{\sqrt 2(r-ia\cos\theta)}$ produces  $(\hat\phi_0=0,\hat\phi_1\neq 0,\hat\phi_2=2\bar\alpha\hat\phi_1)$ as it should, see equation (6) of \cite{Newman}.

\newpage
\chapter{Boundary conditions from Kerr metric}

\section{Introduction}
In this appendix, the relation between BMS boundary conditions and those obtained by acting with the Minkowski Killing vectors on the Kerr metric is investigated.
The strategy is similar to the case of asymptotically anti-de Sitter \cite{Henneaux:1985tv}, and is the following:
\begin{itemize}
\item Start with Minkowski in spherical coordinates $(u,r,\theta,\phi)$ and find the exact Killing vectors of it. The simplest way is to consider the Minkoswki metric in cartesian coordinates ($t,x,y,z$) and then translate the exact Killing vector from cartesian coordinates to spherical coordinates.

\item The Kerr metric can be written in retarded null coordinates ($u,r,\theta,\phi$). When $m=0$, it reduces to the Minkowski coordinates written in unusual coordinates. The change of coordinates that brings the Kerr metric when $m=0$ to the Minkowski metric in spherical coordinates is found.

\item The change of coordinates found in previous item is then performed on the general ($m\neq0$) Kerr metric but it appears to be quite heavy. Only the leading order in the radial coordinate is really needed. Then, the Kerr metric can be written as the sum of Minkoski metric plus a deviation tensor,
\begin{eqnarray}
g_{\mu\nu}^{\text{Kerr}}=g_{\mu\nu}^{\text{Mink}}+k_{\mu\nu}.
\end{eqnarray}
This is the asymptotic expansion of the Kerr metric in the new coordinates system.

\item The exact Killing vectors of the Minkowski metric act on this asymptotic expansion of the Kerr metric and  generate boundary conditions, $h_{\mu\nu}$,
\begin{eqnarray}
\delta_{\xi_{\text{Mink}}}~~g_{\mu\nu}^{\text{Kerr}}=h_{\mu\nu},\hspace{1cm}
\end{eqnarray}

\item The last step is to find asymptotic Killing vectors preserving the boundary conditions $h_{\mu\nu}$. This means solving the asymptotic Killing equation
\begin{eqnarray}
\mathcal L_{\xi}g_{\mu\nu}=\mathcal O(h_{\mu\nu})\label{Cquatre},
\end{eqnarray}
for any asymptotically flat metric $g$. Relation \eqref{Cquatre} is also valid in the case of the Minkowski metric.
\end{itemize}

\section{Killing vectors of Minkoswki in spherical coordinates}

The isometries of the Minkowski metric in cartesian coordinates $ds^2=-dt^2+dx^2+dy^2+dz^2$  are given by the solutions of the Killing equation,
\begin{eqnarray}
\xi_{\mu;\nu}+\xi_{\nu;\mu}=0\hspace{1cm}\leftrightarrow\hspace{1cm}\xi_{\mu,\nu}+\xi_{\nu,\mu}=0,\label{Kerr1.2}
\end{eqnarray}
where the covariant derivatives can be replaced by partial derivatives, because $\Gamma_{\mu\nu}^\sigma=0$ in cartesian coordinates.
The most general solutions to \eqref{Kerr1.2} are found by considering polynomials in the coordinates $x^\sigma$:
\begin{itemize}
\item $\xi_\mu=a_\mu$ is a solution (with $a_\mu$ constant);
\item $\xi_\mu=b_{\mu\nu}x^\nu$ is a solution (with $b_{\mu\nu}=-b_{\nu\mu}$);
\item $\xi_\mu=c_{\mu\nu\rho}x^\nu x^\rho$ is also a solution for $c_{\mu\nu\rho}=-c_{\nu\mu\rho}$, but the symmetry on the last two indices and the antisymmetry on the first two reduce $c_{\mu\nu\rho}$ to be zero: $c_{\mu\nu\rho}=c_{\mu\rho\nu}=-c_{\rho\mu\nu}=-c_{\rho\nu\mu}=c_{\nu\rho\mu}=c_{\nu\mu\rho}=-c_{\mu\nu\rho}=0$.
\item There is no higher order polynomial in the coordinate $x^\sigma$ that is solution to \eqref{Kerr1.2}, due to the same symmetry argument as for the quadratic term.
\end{itemize}
The components of the most general isometry vector field of the Minkowski metric can therefore be written as
\begin{eqnarray}
\xi_t&=a_0+b_{01}x+b_{02}y+b_{03}z=-\xi^t,\\
\xi_x&=a_1+b_{10}t+b_{12}y+b_{13}z=\xi^x,\label{Kerr1.6}\\
\xi_y&=a_2+b_{20}t+b_{21}x+b_{23}z=\xi^y,\\
\xi_z&=a_3+b_{30}t+b_{31}x+b_{32}y=\xi^z,\label{Kerr1.7}
\end{eqnarray}
and depend on ten arbitrary constants ($a_\mu,b_{\mu\nu}$) therefore giving rise to ten different Killing vectors \footnote{The following notation is introduced: $\xi_0$ will denotes the vector field $\xi^\mu\partial_\mu$ such that in $\xi^\mu$ the constant $a_0$ is set to one, and all other constants are set to zero (i.e. $\xi_0$ is the span of $a_0$) and similarly for the other constants.}
\begin{align}
\xi_0&=-\dfrac{\partial}{\partial t},\hspace{1cm}\xi_1=-\dfrac{\partial}{\partial x},\hspace{1cm}\xi_2=-\dfrac{\partial}{\partial y},\hspace{1cm}\xi_3=-\dfrac{\partial}{\partial z},\\
\xi_{01}&=x\dfrac{\partial}{\partial t}-t\dfrac{\partial}{\partial x},\hspace{1cm}\xi_{02}=y\dfrac{\partial}{\partial t}-t\dfrac{\partial}{\partial y},\hspace{1cm}\xi_{03}=z\dfrac{\partial}{\partial t}-t\dfrac{\partial}{\partial z},\\
\xi_{12}&=y\dfrac{\partial}{\partial x}-x\dfrac{\partial}{\partial y},\hspace{1cm}\xi_{13}=z\dfrac{\partial}{\partial x}-x\dfrac{\partial}{\partial z},\hspace{1cm}\xi_{23}=z\dfrac{\partial}{\partial y}-y\dfrac{\partial}{\partial z}.
\end{align}
 $\xi_0$ represents invariance of the spacetime under time translation, $(\xi_1,\xi_2,\xi_3)$ represent invariance under space translations, $(\xi_{01},\xi_{02},\xi_{03})$ represent invariance under spacetime rotations (boosts), and $(\xi_{12},\xi_{13},\xi_{23})$ represent invariance under space rotations. \\

Spherical coordinates with a retarded null time coordinate $u$ are introduced,
\begin{align}
t&=u+r,\hspace{1cm}x=r\sin\theta\cos\phi,\label{1.3}\\
y&=r\sin\theta\sin\phi,\hspace{1cm}z=r\cos\theta,\label{1.4}
\end{align}
so that the Minkowski line element becomes
$
ds^2=-du^2-2dudr+r^2(d\theta^2+\sin^2\theta d\phi^2),\label{1.7}
$
and the isometry vector field \eqref{Kerr1.6}\eqref{Kerr1.7} becomes in this coordinates system
\begin{eqnarray}
\tilde \xi_\mu=\dfrac{\partial x^\sigma}{\partial\tilde x^\mu}\xi_\sigma,\hspace{1cm}\tilde x^\mu=(u,r,\theta,\phi),\hspace{1cm}\tilde\xi^\mu=g^{\mu\nu}\tilde\xi_\nu.
\end{eqnarray}
They can be written explicitly (droping the tilde on them)\\
\begin{align}
 \xi^u=&-(a_0+a_1\sin\theta\cos\phi+a_2\sin\theta\sin\phi+a_3\cos\theta)-\notag\\
-&u(b_{10}\sin\theta\cos\phi+b_{20}\sin\theta\sin\phi+b_{30}\cos\theta)\label{xiu},\\
&\notag\\
\xi^r=&-r(b_{01}\sin\theta\cos\phi+b_{02}\sin\theta\sin\phi+b_{03}\cos\theta)+\notag\\
+&((a_1+ub_{10})\sin\theta\cos\phi+(a_2+ub_{20})\sin\theta\sin\phi+(a_3+ub_{30})\cos\theta),\\
&\notag\\
\xi^\theta=&(b_{10}\cos\theta\cos\phi+b_{20}\cos\theta\sin\phi-b_{30}\sin\theta+b_{13}\cos\phi+b_{23}\sin\phi)+\notag\\
+&\dfrac{1}{r}\left[(a_1+ub_{10})\cos\theta\cos\phi+(a_2+ub_{20})\cos\theta\sin\phi-(a_3+ub_{30})\sin\theta\right],\\
&\notag\\
\xi^\phi=&\frac{1}{\sin\theta}\left(-b_{10}\sin\phi+b_{20}\cos\phi-b_{12}\sin\theta-b_{13}\sin\phi\cos\theta+b_{23}\cos\phi\cos\theta\right)+\notag\\
+&\dfrac{1}{r}\dfrac{1}{\sin\theta}\left(-(a_1+ub_{10})\sin\phi+(a_2+ub_{20})\cos\phi)\right)\label{xiphi}.\\
\notag
\end{align}
\newpage
This isometry vector field still depends on ten constants, and the ten Killing vectors can be written as
\begin{align}
\xi_{0}&=-\dfrac{\partial}{\partial u},\\
\xi_{1}&=-\sin\theta\cos\phi\dfrac{\partial}{\partial u}+\sin\theta\cos\phi\dfrac{\partial}{\partial r}+\dfrac{1}{r}\left(\cos\theta\cos\phi \dfrac{\partial}{\partial \theta}-\dfrac{\sin\phi}{\sin\theta}\dfrac{\partial}{\partial \phi}\right),\\
\xi_{2}&=-\sin\theta\sin\phi\dfrac{\partial}{\partial u}+\sin\theta\sin\phi \dfrac{\partial}{\partial r}+\dfrac{1}{r}\left(\cos\theta\sin\phi\dfrac{\partial}{\partial\theta}+\dfrac{\cos\phi}{\sin\theta}\dfrac{\partial}{\partial \phi}\right),\\
\xi_3&=-\cos\theta\dfrac{\partial}{\partial u}+\cos\theta\dfrac{\partial}{\partial r}-\dfrac{1}{r}\sin\theta\dfrac{\partial}{\partial\theta},\\
\xi_{01}&=u\sin\theta\cos\phi\dfrac{\partial}{\partial u}-(r+u)\sin\theta\cos\phi\dfrac{\partial}{\partial r}-\notag\\
&\hspace{3cm}-\left(1+\dfrac{u}{r}\right)\cos\theta\cos\phi\dfrac{\partial}{\partial\theta}+\left(1+\dfrac{u}{r}\right)\dfrac{\sin\phi}{\sin\theta}\dfrac{\partial}{\partial\phi},\\
\xi_{02}&=u\sin\theta\sin\phi\dfrac{\partial}{\partial u}-(r+u)\sin\theta\sin\phi\dfrac{\partial}{\partial r}-\notag\\
&\hspace{3cm}-\left(1+\dfrac{u}{r}\right)\cos\theta\sin\phi\dfrac{\partial}{\partial\theta}-\left(1+\dfrac{u}{r}\right)\dfrac{\cos\phi}{\sin\theta}\dfrac{\partial}{\partial\phi},\\
\xi_{03}&=u\cos\theta\dfrac{\partial}{\partial u}-(r+u)\cos\theta\dfrac{\partial}{\partial r}+\left(1+\dfrac{u}{r}\right)\sin\theta\dfrac{\partial}{\partial\theta},\\
\xi_{12}&=-\dfrac{\partial}{\partial\phi}\\
\xi_{13}&=\cos\phi\dfrac{\partial}{\partial\theta}-\sin\phi\dfrac{\cos\theta}{\sin\theta}\dfrac{\partial}{\partial\theta},\\
\xi_{13}&=\sin\phi\dfrac{\partial}{\partial\theta}+\cos\phi\dfrac{\cos\theta}{\sin\theta}\dfrac{\partial}{\partial\theta}.
\end{align}\\
Let us introduce some notation, in order to make a comparison later on with the BMS vector field:
\begin{align}
f(u,x^A)&\equiv \xi^u,\hspace{2cm}x^A=(\theta,\phi),\\
T(x^A)&\equiv-(a_0+a_1\sin\theta\cos\phi+a_2\sin\theta\sin\phi+a_3\cos\theta),\label{1.13}\\
Y^\theta(x^A)&\equiv(b_{10}\cos\theta\cos\phi+b_{20}\cos\theta\sin\phi-b_{30}\sin\theta+b_{13}\cos\phi+b_{23}\sin\phi),\label{1.15}\\
Y^\phi(x^A)&\equiv\frac{\left(-b_{01}\sin\phi+b_{20}\cos\phi-b_{12}\sin\theta-b_{13}\sin\phi\cos\theta+b_{23}\cos\phi\cos\theta\right)}{\sin\theta}\label{1.16}\\
g_{AB}&=r^2\bar \gamma^{AB},\hspace{1cm}\bar \gamma_{AB}dx^Adx^B=d\theta^2+\sin^2\theta d\phi^2.
\end{align}\\
Let $\bar D_A (\bar \Delta)$ denotes the covariant derivative (Laplacian) with respect to the two-sphere $\bar \gamma_{AB}$, and  using $\Gamma_{\theta\phi}^\phi=\dfrac{\cos\theta}{\sin\theta}$ a short computation shows that
\begin{align}
\bar D_AY^A&=\partial_\theta Y^\theta+\partial_\phi Y^\phi+\Gamma_{\theta\phi}^\phi Y^\theta\notag\\
&=-2b_{10}\sin\theta\cos\phi-2b_{20}\sin\theta\sin\phi-2b_{30}\cos\theta\label{1.18},\\
\bar \Delta f&=\dfrac{1}{\sin\theta}\partial_A(g^{AB}\sin\theta\partial_B)f\notag\\
&=2(a_1+ub_{10})\sin\theta\cos\phi+2(a_2+ub_{20})\sin\theta\sin\phi+2(a_3+ub_{30})\cos\theta.\label{1.19}
\end{align}
It follows from these notations that $Y^A$ defined by \eqref{1.15},\eqref{1.16} is a conformal Killing vector with respect to the two-sphere $\bar \gamma_{AB}$,
\begin{eqnarray}
Y^C\partial_C \bar \gamma_{AB}+\bar \gamma_{CB}\partial_AY^C+\bar \gamma_{AC}\partial_BY^C=(D_CY^C)\bar \gamma_{AB}.\label{C36}
\end{eqnarray}
Let us finally introduce the notation $\psi=\bar D_CY^C$ for the conformal factor. Using $\Gamma_{\theta\theta}^\phi=0=\Gamma_{\theta\theta}^\theta$, we find that $\partial_A\psi$ is also a conformal Killing vector with respect to the two-sphere $\gamma_{AB}$,
\begin{eqnarray}
\bar D_A\partial_B\psi+\bar D_B\partial_A\psi=\bar \Delta\psi\gamma_{AB}.\label{C37}
\end{eqnarray}
Note that in $d\neq4$, relation \eqref{C36} always implies \eqref{C37}, due to theorem A.1 of appendix A.
The definitions \eqref{1.13}-\eqref{1.16} and properties \eqref{1.18}-\eqref{1.19} allow to rewrite the components of the isometry vector field of Minkowski metric in spherical coordinates as
\begin{align}
\xi^u&\equiv f= T+\dfrac{u}{2}\bar D_AY^A,\label{kerr:1.38}\\
\xi^r&=-\dfrac{r}{2}\bar D_AY^A+\dfrac{1}{2}\bar \Delta f,\\
\xi^A&=Y^A-\partial_B f\dfrac{\gamma^{AB}}{r}\hspace{0.5cm}\text{i.e.}\hspace{0.5cm}\xi^\theta=Y^\theta-\dfrac{\partial_\theta f}{r},
\xi^\phi=Y^\phi-\dfrac{\partial_\phi f}{r\sin^2\theta}\label{kerr:1.40},
\end{align}
with $Y^A$ and $\partial_B\psi$ both conformal Killing vectors.\\

\subsection{Kerr metric}

The Kerr metric is  (with the notation $\rho^2=r^2+a^2\cos^2\theta$)
\begin{align}
ds^2&=-\left(1-\dfrac{2mr}{\rho^2}\right)du^2-2dudr-2a\sin^2\theta\dfrac{2mr}{\rho^2}dud\phi-2a\sin^2\theta drd\phi+\notag\\
&+\rho^2d\theta^2+\sin^2\theta\left(a^2+r^2+a^2\sin^2\theta\dfrac{2mr}{\rho^2}\right)d\phi^2\\
g_{\mu\nu}&=
\begin{pmatrix}
-\left(1-\dfrac{2mr}{\rho^2}\right)&-1&0&-a\sin^2\theta\left(\dfrac{2mr}{\rho^2}\right)\\
-1&0&0&a\sin^2\theta\\
0&0&\rho^2&0\\
-a\sin^2\theta\left(\dfrac{2mr}{\rho^2}\right)&a\sin^2\theta&0&\sin^2\theta\left(r^2+a^2+a^2\sin^2\theta\dfrac{2mr}{\rho^2}\right)
\end{pmatrix}\label{1.25}.
\end{align}
When $a=0$, the metric \eqref{1.25} reduces to the Schwarschild one \eqref{Kerr1.42}. When $m=0$ instead, the Kerr metric \eqref{1.25} reduces to
\begin{eqnarray}
ds^2 =-du^2-2dudr-2a\sin^2\theta dr d\phi+\rho^2d\theta^2+\sin^2\theta (a^2+r^2)d\phi^2\label{kerr1.4},
\end{eqnarray}
which is also the Minkowski metric, but written in unusual coordinates. To see it, let us define new coordinates ($U,R,\Theta,\Phi$) by the implicit relations
\begin{align}
U&=u+r-R,\hspace{1cm}R\cos\Theta=r\cos\theta,\label{Kerr1.42}\\
R^2&=r^2+a^2\sin^2\theta,\hspace{1cm}\Phi=\phi+\arctan\left(\dfrac{r}{a}\right),\label{1.43}
\end{align}
 so that the metric \eqref{kerr1.4} becomes
\begin{eqnarray}
ds^2=-dU^2-2dUdR+R^2(d\Theta^2+\sin^2\Theta d\Phi^2).
\end{eqnarray}
The explicit coordinates transformation \eqref{Kerr1.42}-\eqref{1.43} is given by
\begin{align}
u&=U-\sqrt{\dfrac{R^2-a^2+\sqrt{(R^2-a^2)^2+4a^2R^2\cos^2\Theta}}{2}}+R,\\
r&=\sqrt{\dfrac{R^2-a^2+\sqrt{(R^2-a^2)^2+4a^2R^2\cos^2\Theta}}{2}},\\
\theta&=\arccos\left(\dfrac{\sqrt2 R\cos\Theta}{\sqrt{R^2-a^2+\sqrt{(R^2-a^2)^2+4a^2R^2\cos^2\Theta}}}\right),\\
\phi&=\Phi-\arctan\left(\sqrt{\dfrac{R^2-a^2+\sqrt{(R^2-a^2)^2+4a^2R^2\cos^2\Theta}}{2a^2}}\right).
\end{align}
The perturbative expansion of this coordinates transformation in inverse power of $R$ is given by
\begin{align}
u&=U+\dfrac{a^2\sin^2\Theta}{2R}+\dots,\\
r&=R-\dfrac{a^2\sin^2\Theta}{2R}+\dots,\\
\theta&=\Theta+O(R^{-3}),\\
\phi&=\Phi-\arctan\left(\dfrac{R}{a}-\dfrac{a^2\sin^2\Theta}{2R}+\dots\right).
\end{align}

Let us compute the Kerr metric \eqref{1.25} (when $m\neq0$) in this new coordinates system, up to the leading order in $1/R$. First, note the the Kerr metric in Boyer-Lindquist coordinates can be written as the sum of the Minkowski metric ($ds_0^2=-du^2-2dudr+r^2d\Omega^2$) plus a deviation tensor ($k_{\mu\nu}dx^\mu dx^\nu$),
\begin{align}
ds^2_{Kerr}=&-\left(1-\dfrac{2mr}{\rho^2}\right)du^2-2dudr-2a\sin^2\theta\dfrac{2mr}{\rho^2}dud\phi-2a\sin^2\theta drd\phi+\\
&+\rho^2d\theta^2+\sin^2\theta\left(a^2+r^2+a^2\sin^2\theta\dfrac{2mr}{\rho^2}\right)d\phi^2,\\
=&~ds_0^2+k_{\mu\nu}dx^\mu dx^\nu.
\end{align}
with the non-vanishing $k_{\mu\nu}$ given by
\begin{align}
k_{uu}&=\dfrac{2mr}{r^2+a^2\cos^2\theta},\\
k_{u\phi}&=-a\sin^2\theta\dfrac{2mr}{r^2+a^2\cos^2\theta},\\
k_{\phi\phi}&=a^2\sin^4\theta\dfrac{2mr}{r^2+a^2\cos^2\theta}.
\end{align}
In order to have the Kerr metric in coordinates ($U,R,\Theta,\Phi$), let us compute the deviation tensor $k_{\mu\nu}$ in the new coordinates system, up to the leading order in $R$,
\begin{align}
k_{UU}&=\dfrac{2m}{R}+O(R^{-3}),\label{kerr1.59}\\
k_{U\Phi}&=-a\sin^2\Theta\dfrac{2m}{R}+O(R^{-3}),\\
k_{\Phi\Phi}&=a^2\sin^4\Theta\dfrac{2m}{R}+O(R^{-3}),\\
k_{\Theta\Theta}&=a^4\sin^2\Theta\cos^2\Theta\dfrac{2m}{R^3}+O(R^{-5}),\\
k_{RR}&=\dfrac{3}{2}a^4\sin^4\Theta\dfrac{m}{R^5}+O(R^{-3}),\\
k_{R\Theta}&=a^4\sin^3\Theta\cos\Theta\dfrac{m}{R^4}+O(R^{-6}),\\
k_{U\Theta}&=a^2\sin\Theta\cos\Theta\dfrac{m}{R^2}+O(R^{-4}),\\
k_{UR}&=a^2\sin^2\Theta\dfrac{m}{R^3}+O(R^{-5}),\\
k_{\Phi\Theta}&=-a^3\sin\Theta\cos\Theta\dfrac{2m}{R^2}+O(R^{-4}),\\
k_{R\Phi}&=-a^3\sin^4\Theta\dfrac{m}{R^3}+O(R^{-5}).\label{kerr1.68}
\end{align}
We can observe that when $m=0$, all the components of the deviation tensor $k_{\mu\nu}$ are zero, as it should. These deviation vectors define the asymptotic form of the Kerr metric in the coordinates ($U,R,\Theta,\Phi$).

\subsection{Boundary conditions for asymptotically flat spacetimes}
In this subsection, we act with the Killing vectors of Minkowski metric, given in components by \eqref{xiu}-\eqref{xiphi}, on the asymptotic Kerr metric, defined by \eqref{kerr1.59}-\eqref{kerr1.68}. This procedure produces boundary conditions $h_{\mu\nu}$,
\begin{eqnarray}
\delta_{\xi^{Mink}}\left(g_{\mu\nu}^{Kerr}\right)=h_{\mu\nu},
\end{eqnarray}
where $\delta_\xi (g_{\mu\nu}^{Kerr})=\mathcal L_{\xi}(g_{\mu\nu}^{Kerr})$. 
These boundary conditions are (capital coordinates are replace by small ones for conveniance)
\begin{align}
h_{rr}&=\mathcal O(r^{-5}),\hspace{1cm}h_{r\theta}=\mathcal O(r^{-3}),\hspace{1cm}h_{r\phi}=\mathcal O(r^{-3}),\notag\\
h_{\theta\theta}&=\mathcal O(r^{-1}),\hspace{1cm}h_{\theta\phi}=\mathcal O(r^{-1}),\hspace{1cm}h_{\phi\phi}=\mathcal O(r^{-1}),\label{1.27}\\
h_{ur}&=\mathcal O(r^{-3}),\hspace{1cm}h_{u\theta}=O(r^{-1}),\hspace{1cm}h_{u\phi}=\mathcal O(r^{-1}),\hspace{1cm}h_{uu}=\mathcal O(r^{-1})\notag.
\end{align}

\subsection{Asymptotic symmetries}
Asymptotic symmetries are the most general gauge transformation that preserve the boundary conditions \eqref{kerr1.59}-\eqref{kerr1.68}. The gauge parameter of  such transformation is such that
\begin{eqnarray}
\mathcal L_{\xi}g=\mathcal O(h_{\mu\nu})\label{kerr1.71},
\end{eqnarray}
with $h_{\mu\nu}$ given by \eqref{1.27}.
Equation \eqref{kerr1.71} is valid for any metric with boundary conditions \eqref{1.27}, so in particular the Minkowski metric can be used. Solving \eqref{kerr1.71} for the Minkowski metric gives,
\begin{eqnarray}
\delta_{\xi}g_{\mu\nu}^{\text{Mink}}=\mathcal O(h_{\mu\nu}).
\end{eqnarray}
The solutions are
\begin{itemize}
\item $g_{rr}:\hspace{1cm}\xi^2=f(u,x^A)+O(r^{-4})$,
\item $g_{rA}:\hspace{1cm}\xi^A=Y^A-\dfrac{\partial^Af}{r}+O(r^{-4})$,
\item $g_{AB}:\hspace{1cm}\xi^r=-\dfrac{r}{2}D_CY^C+\dfrac{1}{2}\Delta f+O(r^{-2})$, and $Y^A,\partial_Af$ are conformal Killing of $\gamma_{AB}$,
\item $g_{ur}:\hspace{1cm}\partial_u f=\dfrac{1}{2}D_CY^C$,
\item $g_{uA}:\hspace{1cm}\partial_u Y^C=0$ and $-\partial_A f-\dfrac{1}{2}\Delta f=0$,
\item $g_{uu}:\hspace{1cm}D_CY^C+\dfrac{1}{2}\Delta D_CY^C=0.$
\end{itemize}
The two last relations are automatically satisfied for any conformal Killing vectors, see equation \eqref{A:11} of Appendix A for more detail.
This symmetry algebra is not the BMS symmetry algebra, but the Poincar\'e one, see \eqref{kerr:1.38}-\eqref{kerr:1.40}.\\

\backmatter

\newpage
\def\cprime{$'$}
\providecommand{\href}[2]{#2}\begingroup\raggedright\endgroup

\cleardoublepage

\newpage
\cleardoublepage
\thispagestyle{empty}
\cleardoublepage

\end{document}